\documentstyle[epsfig]{mn}

\newif\ifAMStwofonts
\ifoldfss
  \ifCUPmtlplainloaded \else
    \NewTextAlphabet{textbfit} {cmbxti10} {}
    \NewTextAlphabet{textbfss} {cmssbx10} {}
    \NewMathAlphabet{mathbfit} {cmbxti10} {} % for math mode
    \NewMathAlphabet{mathbfss} {cmssbx10} {} %  "   "    "
  \fi
  \ifAMStwofonts
    \ifCUPmtlplainloaded \else
      \NewSymbolFont{upmath} {eurm10}
      \NewSymbolFont{AMSa} {msam10}
      \NewMathSymbol{\upi}     {0}{upmath}{19}
      \NewMathSymbol{\umu}     {0}{upmath}{16}
      \NewMathSymbol{\upartial}{0}{upmath}{40}
      \NewMathSymbol{\leqslant}{3}{AMSa}{36}
      \NewMathSymbol{\geqslant}{3}{AMSa}{3E}

      \let\leq=\leqslant 
      \let\geq=\geqslant 
    \fi
  \fi
\fi % End of OFSS

\ifnfssone
  \newmathalphabet{\mathit}
  \addtoversion{normal}{\mathit}{cmr}{m}{it}
  \addtoversion{bold}{\mathit}{cmr}{bx}{it}
  \newmathalphabet{\mathbfit} % math mode version of \textbfit{..}
  \addtoversion{normal}{\mathbfit}{cmr}{bx}{it}
  \addtoversion{bold}{\mathbfit}{cmr}{bx}{it}
  \newmathalphabet{\mathbfss} % math mode version of \textbfss{..}
  \addtoversion{normal}{\mathbfss}{cmss}{bx}{n}
  \addtoversion{bold}{\mathbfss}{cmss}{bx}{n}
  \ifAMStwofonts
    \ifCUPmtlplainloaded \else
      \UseAMStwoboldmath
      \makeatletter
      \new@mathgroup\upmath@group
      \define@mathgroup\mv@normal\upmath@group{eur}{m}{n}
      \define@mathgroup\mv@bold\upmath@group{eur}{b}{n}
      \edef\UPM{\hexnumber\upmath@group}
      \new@mathgroup\amsa@group
      \define@mathgroup\mv@normal\amsa@group{msa}{m}{n}
      \define@mathgroup\mv@bold\amsa@group{msa}{m}{n}
      \edef\AMSa{\hexnumber\amsa@group}
      \makeatother
      \mathchardef\upi="0\UPM19
      \mathchardef\umu="0\UPM16
      \mathchardef\upartial="0\UPM40
      \mathchardef\leqslant="3\AMSa36
      \mathchardef\geqslant="3\AMSa3E

      \let\leq=\leqslant 
      \let\geq=\geqslant 
    \fi
  \fi
\fi % End of NFSS release 1

\ifnfsstwo
  \DeclareMathAlphabet{\mathbfit}{OT1}{cmr}{bx}{it}
  \SetMathAlphabet\mathbfit{bold}{OT1}{cmr}{bx}{it}
  \DeclareMathAlphabet{\mathbfss}{OT1}{cmss}{bx}{n}
  \SetMathAlphabet\mathbfss{bold}{OT1}{cmss}{bx}{n}
  \ifAMStwofonts
    \ifCUPmtlplainloaded \else
      \DeclareSymbolFont{UPM}{U}{eur}{m}{n}
      \SetSymbolFont{UPM}{bold}{U}{eur}{b}{n}
      \DeclareSymbolFont{AMSa}{U}{msa}{m}{n}
      \DeclareMathSymbol{\upi}{0}{UPM}{"19}
      \DeclareMathSymbol{\umu}{0}{UPM}{"16}
      \DeclareMathSymbol{\upartial}{0}{UPM}{"40}
      \DeclareMathSymbol{\leqslant}{3}{AMSa}{"36}
      \DeclareMathSymbol{\geqslant}{3}{AMSa}{"3E}

      \let\leq=\leqslant 
      \let\geq=\geqslant 
    \fi
  \fi
\fi % End of NFSS release 2

\ifCUPmtlplainloaded \else
  \ifAMStwofonts \else % If no AMS fonts
    \def\upi{\pi}
    \def\umu{\mu}
    \def\upartial{\partial}
  \fi
\fi

\title[HST Imaging of QDOT ULIRGs]
  {HST/WFPC2 Imaging of the QDOT Ultraluminous Infrared Galaxy Sample}
\author[D. Farrah et al.]
  {D.~Farrah$^1$, M.~Rowan-Robinson$^1$, S.~Oliver$^2$, S.~Serjeant$^{3}$, K.~Borne$^4$, 
   \newauthor A.~Lawrence$^5$, R.A.~Lucas$^6$, H.~Bushouse$^6$ and L.~Colina$^7$\\
 $^1$Astrophysics Group, Blackett Laboratory, Imperial College, Prince Consort Road, London SW7 2BW, UK\\
 $^2$Astronomy Centre, University of Sussex, Falmer, Brighton BN1 9QJ, UK\\
 $^3$Unit for Space Sciences \& Astrophysics, School of Physical Sciences, University of Kent at Canterbury,\\
     Canterbury, Kent, CT2 7NZ, UK\\
 $^4$Raytheon ITSS, NASA GSFC, Greenbelt, MD, USA\\
 $^5$Institute for Astronomy, University of Edinburgh, Royal Observatory, Blackford Hill, Edinburgh, EH9 3HJ, UK\\
 $^6$Space Telescope Science Institute, Baltimore, Maryland, USA\\ 
 $^7$Instituto de Fisica de Cantabria, Santander, Spain}
\date{Received 2001 May 15}
\pagerange{\pageref{01}--\pageref{16}}
\pubyear{2001}

\def\LaTeX{L\kern-.36em\raise.3ex\hbox{a}\kern-.15em
    T\kern-.1667em\lower.7ex\hbox{E}\kern-.125emX}

\begin{document}

\label{firstpage}

\maketitle

\begin{abstract}
We present {\em HST} WFPC2 {\em V} band imaging for 23 Ultraluminous Infrared Galaxies 
taken from the QDOT redshift survey. The fraction of sources observed to be interacting 
is 87\%. Most of the merging systems show a number of compact `knots', whose colour and 
brightness differ substantially from their immediate surroundings. Colour maps for nine 
of the objects show a non-uniform colour structure. Features include blue regions located 
towards the centres of merging systems which are likely to be areas of enhanced star 
formation, and compact red regions which are likely to be dust shrouded starbursts or AGN. 
The host galaxies of the QSOs in the sample were found to be either interacting systems 
or ellipticals. Our data shows no evidence that ULIRGs are a simple transition stage 
between galaxy mergers and QSOs. We propose an alternative model for ULIRGs based 
on the morphologies in our sample and previous N-body simulations. Under this model 
ULIRGs as a class are much more diverse than a simple transition between galaxy merger 
and QSO. The evolution of IR power source and merger morphology in ULIRGs is driven solely 
by the local environment and the morphologies of the merger progenitors.
\end{abstract}


\section{Introduction}

The existence of galaxies that emit more radiation in the rest-frame infrared than at all other wavelengths 
combined was established with the first extragalactic mid-IR observations \cite{low,rie}.
The Infrared Astronomical Satellite ({\em IRAS}), launched in 1983, was the first observatory with sufficient 
sensitivity to detect these Luminous Infrared Galaxies in large numbers \cite{soi,row,san}. Luminous Infrared 
Galaxies (LIRGs) are broadly defined as any object with $L_{ir}>10^{11}L_{\odot}$\footnote{where $L_{ir}$ spans 
8-1000$\mu$m and is computed using the flux in all four {\em IRAS} bands according to the 
prescription given by Perault (1987).}. These objects become the dominant extragalactic 
population at luminosities above $10^{11}L_{\odot}$ in the local ($z<0.5$) Universe, 
outnumbering Seyferts, Quasi-Stellar Objects (QSOs) and Starburst Galaxies of comparable 
bolometric luminosity. 

\begin{table*}
\begin{minipage}{175mm}
\caption{Ultraluminous QDOT Galaxies Observed By {\em HST} \label{hstobs}}
\begin{tabular}{@{}ccccccccccccc}
Name&RA (2000)&Dec (2000)&$z$& Type$^1$ & $L_{60}$$^2$ & $L_{V}$$^3$ & $m_{606}$ &$M_{606}$&$M_{814}$$^4$ & $f_{25}/f_{60}$$^5$& Class.$^6$&$A^7$\\
 00275-2859 & 00 30 04.1 & -28 42 24.3 & 0.280 & Sy1    & 12.30    & 11.34      & 16.63     & -24.06& -24.76 &  $0.25$   & 6/C &0.398\\   
 02054+0835 & 02 08 06.9 & 08 50 04.6  & 0.345 & Sy1    & 12.48    & 11.11      & 17.69     & -23.48& -24.49 &  $<0.417$ & 1/C &0.242\\   
 02587-6336 & 02 59 43.2 & -63 25 03.4 & 0.255 & Sb     & 12.24    & 10.46      & 18.62     & -21.85& -23.10 &  $0.123$  & 4/A &0.277\\
 04384-4848 & 04 39 50.9 & -48 43 16.5 & 0.204 & Sb     & 12.21    & 10.56      & 17.83     & -22.13& -22.89 &  $0.07$   & 6/B &0.381\\
 06268+3509 & 06 30 13.3 & 35 07 49.1  & 0.170 & Sb     & 12.00    & 10.61      & 17.32     & -22.24& -23.18 &  $<0.27$  & 6/A &0.414\\
 06361-6217 & 06 36 35.9 & -62 20 32.4 & 0.160 & Sb     & 12.19    & 10.34      & 17.84     & -21.58& -22.45 &  $0.10$   & 6/B &0.335\\  
 06561+1902 & 06 59 05.7 & 18 58 20.3  & 0.188 & Sb     & 12.11    & 10.46      & 17.93     & -21.86& -23.04 &  $<0.27$  & 5/A &0.252\\  
 07381+3215 & 07 41 21.3 & 32 08 23.9  & 0.170 & Sb     & 11.90    & 10.10      & 18.57     & -20.98& -22.37 &  $<0.20$  & 6/B &0.333\\   
 10026+4347 & 10 05 42.0 & 43 32 40.8  & 0.178 & Sy1    & 11.85    & 11.07      & 16.28     & -23.39& -23.81 &  $0.33$   & 0/C &0.330\\
 10579+0438 & 11 00 33.8 & 04 22 08.2  & 0.173 & Sb     & 11.84    & 10.14      & 18.54     & -21.06& -22.06 &  $<0.44$  & 6/B &0.272\\
 13469+5833 & 13 48 40.3 & 58 18 52.2  & 0.158 & HII    & 12.08    & 10.51      & 17.39     & -22.00& -23.17 &  $<0.06$  & 6/B &0.438\\
 14337-4134 & 14 36 59.2 & -41 47 06.6 & 0.182 & Sb     & 11.87    & 10.42      & 17.95     & -21.76& -22.86 &  $<0.42$  & 6/B &0.399\\
 16159-0402 & 16 18 36.5 & -04 09 42.8 & 0.213 & Sb     & 12.22    & 10.50      & 18.09     & -21.97& -23.16 &  $0.31$   & 6/B &0.484\\
 17431-5157 & 17 47 09.9 & -51 58 46.3 & 0.175 & LINER  & 11.85    & 9.98       & 18.96     & -20.66& -22.15 &  $<0.38$  & 4/B &0.283\\ 
 18520-5048 & 18 55 59.1 & -50 44 53.2 & 0.152 & Sb     & 11.78    & 10.51      & 17.30     & -22.00&   --   &  $<0.23$  & 5/A &0.434\\  
 18580+6527 & 18 58 14.0 & 65 31 26.4  & 0.176 & Sb/Sy2 & 11.97    & 11.24      & 15.81     & -23.82& -24.52 &  $0.09$   & 6/B &0.587\\   
 20037-1547 & 20 06 31.9 & -15 39 06.7 & 0.192 & Sy1    & 12.36    & 11.25      & 16.00     & -23.82& -24.44 &  $<0.17$  & 5/C &0.483\\
 20109-3003 & 20 14 05.6 & -29 53 52.7 & 0.143 & Sb     & 11.83    & 10.48      & 17.26     & -21.91& -22.67 &  $<0.28$  & 6/B &0.295\\
 20176-4756 & 20 21 11.0 & -47 47 08.6 & 0.178 & Sb     & 11.98    & 10.44      & 17.84     & -21.81& -23.11 &  $<0.08$  & 4/A &0.252\\  
 20253-3757 & 20 28 37.5 & -37 47 10.2 & 0.180 & Sb     & 11.93    & 10.61      & 17.44     & -22.24& -23.28 &  $<0.40$  & 5/A &0.462\\  
 20507-5412 & 20 54 25.5 & -54 01 13.7 & 0.228 & Sb     & 12.05    & 10.20      & 18.98     & -21.23& -22.29 &  $<0.15$  & 6/B &0.313\\   
 23140+0348 & 23 16 35.2 & 04 05 18.8  & 0.220 & BLRG   & 12.02    & 10.85      & 17.29     & -22.85& -23.84 &  $<0.31$  & 0/B &0.215\\
 23220+2919 & 23 24 27.9 & 29 35 40.3  & 0.240 & HX     & 12.27    & 10.33      & 18.79     & -21.54& -22.51 &  $<0.15$  & 6/B &0.307\\
\end{tabular}

\medskip
Coordinates and {\em V} band magnitudes are taken from the {\em HST} Planetary Camera images and are in the 
{\em HST} flight filter system.
$^1$Spectral classification, taken from Lawrence et al 1999 and the NASA/IPAC Extragalactic Database:
`Sb' - Starburst, `Sy1/Sy2' - Seyfert 1/2, `BLRG' - Broad Line Radio Galaxy, `HX' - High Excitation but not Seyfert 2. 
$^2$60$\mu$m luminosity taken from Lawrence et al 1999 and converted to $H_{0}=65$, $\Omega_{0}=1$  
$^3$Logarithm of the F606W band luminosity, computed from the F606W band images and in units of bolometric solar 
luminosities ($3.823\times10^{33}$ ergs s$^{-1}$).
$^4$As measured from the WF3 F814W band images, but see also Borne et al 2001, in preparation.
$^5${\em IRAS} $25\mu$m to $60\mu$m flux ratio.
$^6$First number is from the 7 band system presented in Lawrence et al 1989 and described 
in \S 4.1., second letter is from the 3 band system described in \S 4.1. 
$^7$Asymmetry statistic, computed using Equation \ref{eqn:fasym} and described in \S4.1

\end{minipage}
\end{table*}

A subset of the {\em IRAS} galaxy population that has been extensively studied are the 
Ultraluminous Infrared Galaxies (ULIRGs), those with $L_{ir}>10^{12}h_{65}^{-2}L_{\odot}$
(or equivalently $L_{60}>10^{11.77}h_{65}^{-2}L_{\odot}$, where $L_{60}$ is the {\em IRAS} 
$60\mu$m flux). Recent deep submillimetre surveys \cite{hug,blai,lil,bcs,pea} suggest that 
ULIRGs are an important component of the Universe over $2 < z < 7$. The evolution of ULIRGs, 
their power source and the trigger mechanism behind the IR emission are still the subject 
of vigorous debate. Although it is now generally accepted that a mixture of star formation 
and AGN activity powers the IR emission, the {\em dominant} power source, and how ULIRGs 
evolve over time, are still unanswered questions. There are many similarities between ULIRGs 
and starburst galaxies \cite{jos,rrc,con}. Conversely, many ULIRGs display nuclear emission lines 
characteristic of Seyfert galaxies \cite{san2,kim}. A study of 15 ULIRGs \cite{gen} using the 
Infrared Space Observatory ({\em ISO}) found that the IR emission of 75\% were predominantly 
powered by star formation and 25\% by an AGN, although at least half the sources showed 
evidence for both types of emission. The spatial density, bolometric emission and luminosity 
function of ULIRGs are comparable to those of Quasars in the local Universe. It has been 
suggested \cite{san2} that ULIRGs represent the dust enshrouded precursors to optically 
selected QSOs and that all QSOs emerge from a luminous infrared phase. The majority of 
ULIRGs appear to lie in interacting or merging systems, although the exact fraction is still 
not certain \cite{san2,lch,cle1}. A merger would provide an ideal environment for triggering 
a nuclear starburst or AGN and thus would be a possible triggering mechanism for QSOs. A 
merger between two spirals is also likely to form an elliptical galaxy \cite{bar}, therefore 
ULIRGs may also have a pivotal role to play in elliptical galaxy evolution \cite{kor}. The 
relation of ULIRGs to their higher luminosity counterparts, Hyperluminous Infrared Galaxies 
(HLIRGs, those with $L_{ir}>10^{13}h_{65}^{-2}L_{\odot}$) is also not fully understood 
\cite{hin,row2}. Although HLIRGs as a class appear to be an extrapolation of ULIRGs to extreme 
luminosities, there exists the possibility that some fraction of HLIRGs are an entirely different 
population \cite{far}.

In this paper we present {\em Hubble Space Telescope} images, taken with the F606W filter 
using the Planetary Camera on WFPC2, for 23 ULIRGs and compare them to WFPC2 WF3 F814W band 
data for 22 of these sources. Sample selection and observations are described in $\S$2. Data reduction, 
photometry, colour map construction and point spread function subtraction are described in 
$\S$3. Results are presented in $\S$4. Discussion of the results can be found in $\S5$ and 
conclusions are summarised in $\S$6. 

We have taken $H_{0}=65$ km s$^{-1}$ Mpc$^{-1}$, $\Omega_{0}=1.0$ and $\Lambda=0$ unless otherwise stated.

\section{Sample}

\subsection{Sample Selection}
The 23 galaxies in our sample are randomly selected from those ULIRGs discovered in the QDOT 
all sky {\em IRAS} galaxy redshift survey \cite{law}. This survey consists of an all sky sample of 2387 
{\em IRAS} galaxies brighter than the {\em IRAS} PSC 60$\mu$m completeness limit of 0.6Jy, and is 
complete to greater than 98\%. All the galaxies in our sample have 60$\mu$m luminosities in the 
range $10^{11.77}-10^{12.50}h_{65}^{-2}L_{\odot}$. The redshift range is $0.14 < z < 0.35$ with a median 
redshift of $z = 0.18$ and a mean redshift of $z = 0.20$.

\begin{figure*}
\begin{minipage}{170mm}
\epsfig{figure=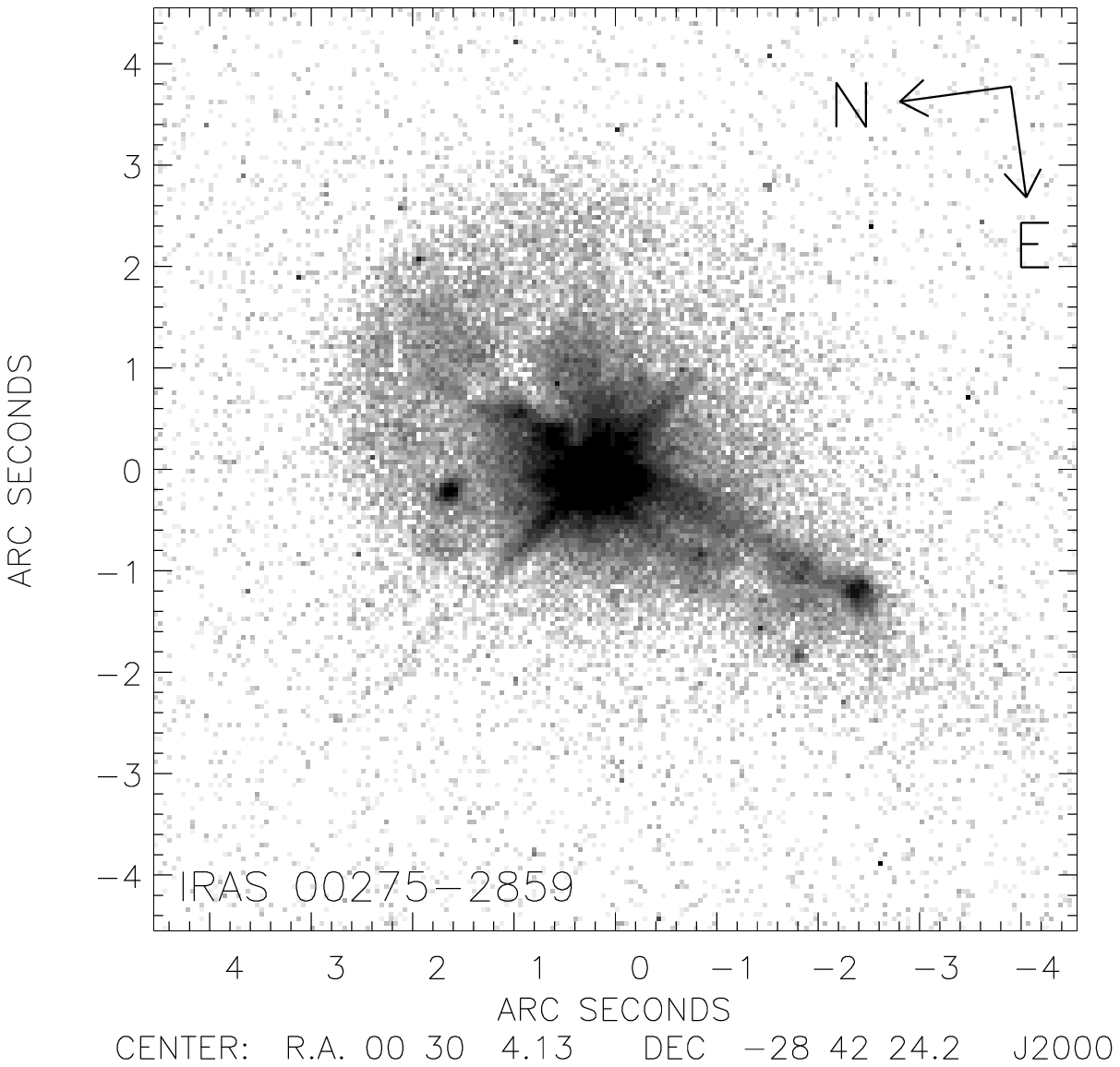,width=80mm}
\epsfig{figure=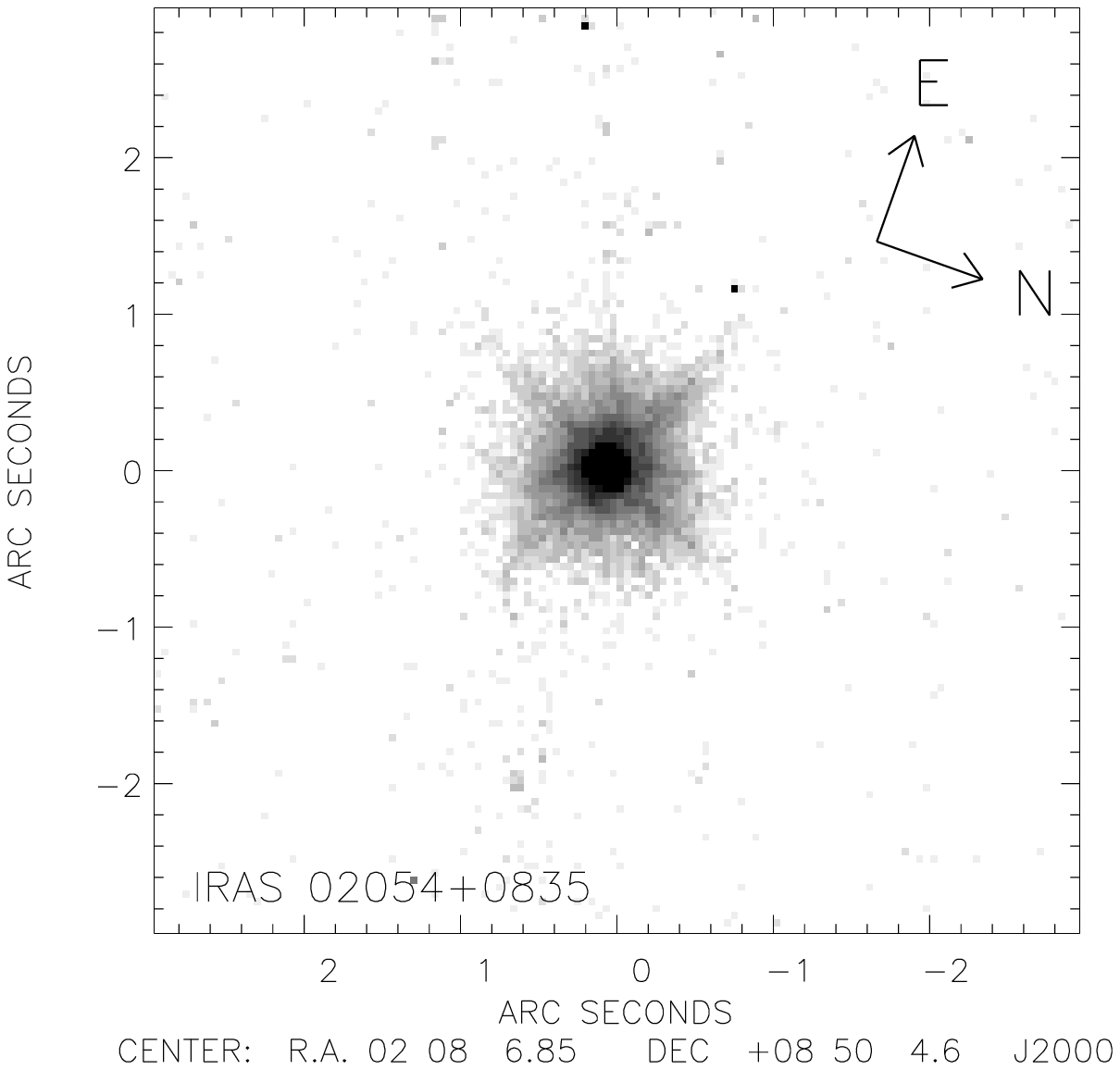,width=80mm}
\end{minipage}
\begin{minipage}{170mm}
\epsfig{figure=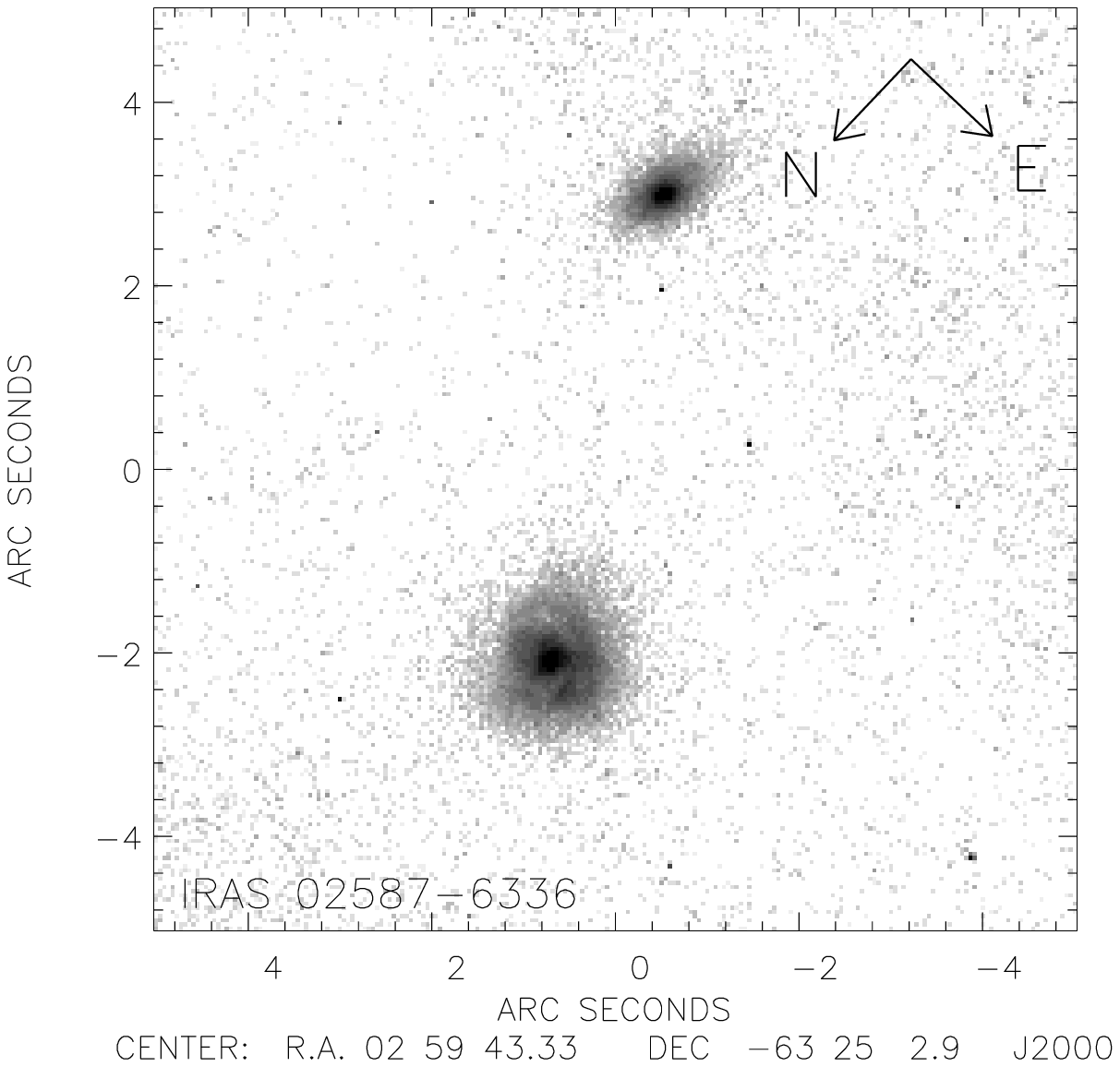,width=80mm}
\epsfig{figure=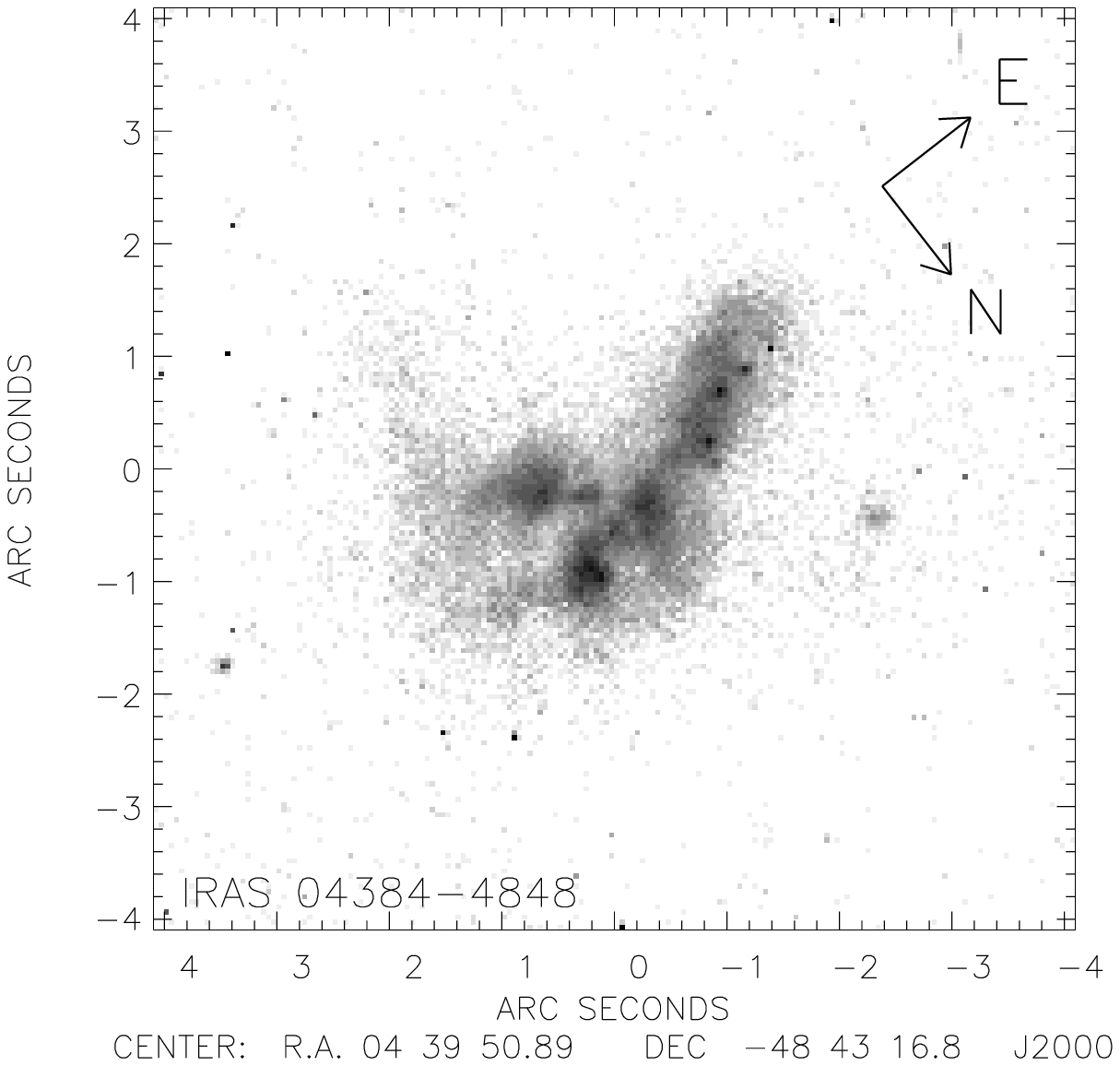,width=80mm}
\end{minipage}
\begin{minipage}{170mm}
\epsfig{figure=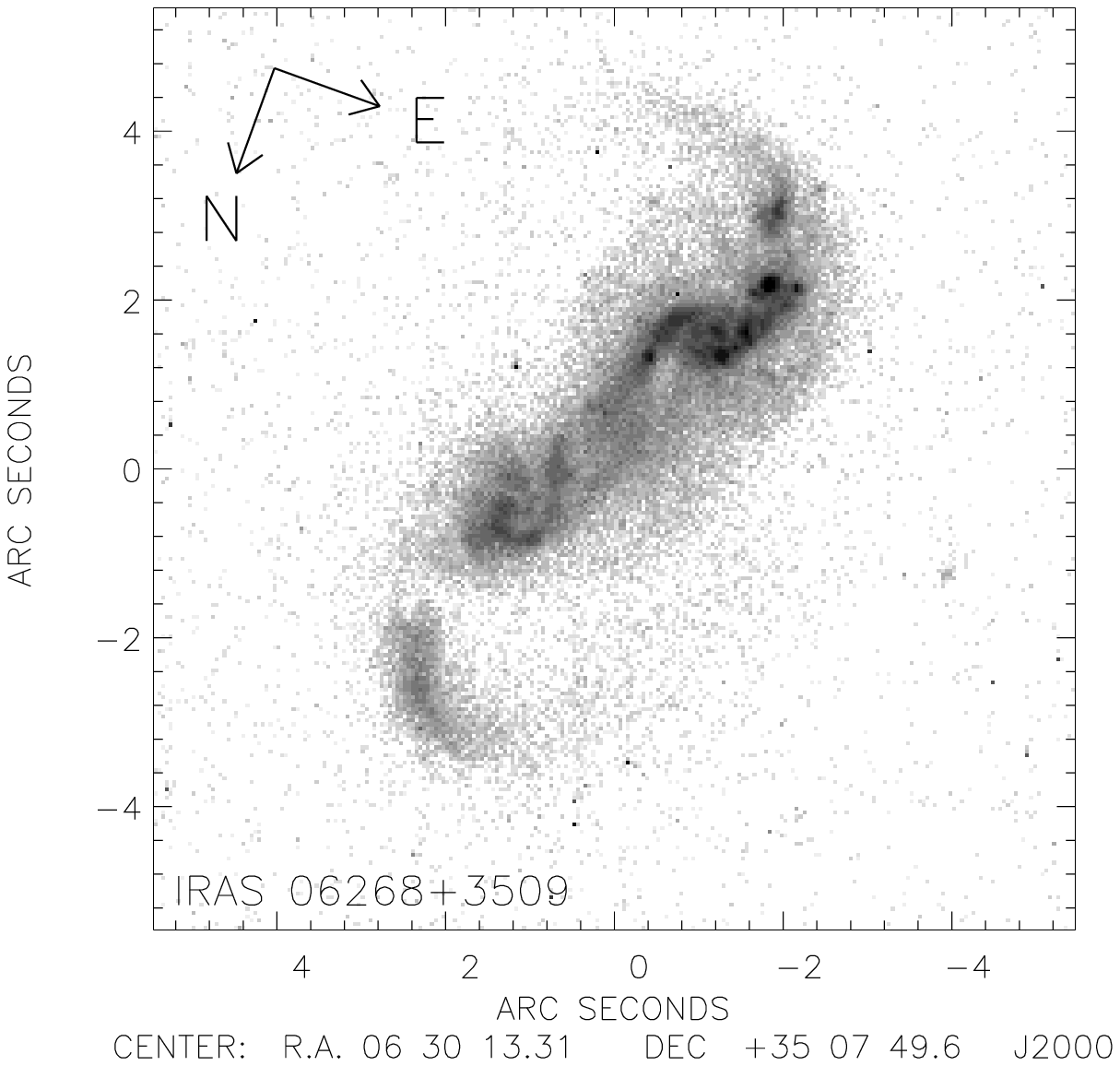,width=80mm}
\epsfig{figure=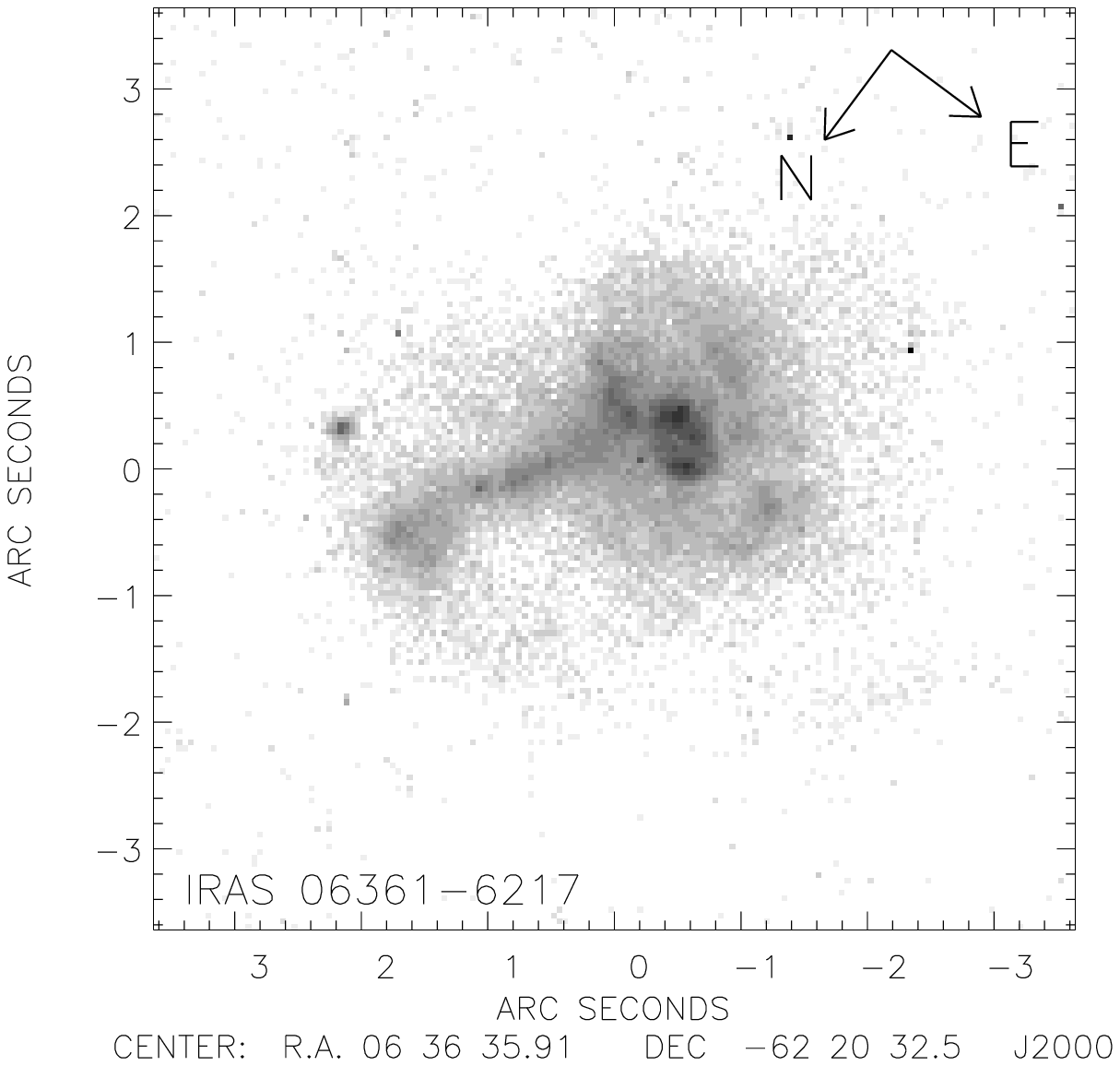,width=80mm}
\end{minipage}
\caption{{\em HST} F606W images of IRAS 00275-2859 to IRAS 06361-6217.
 \label{ulirgs_im_one}}
\end{figure*}

\begin{figure*}
\begin{minipage}{170mm}
\epsfig{figure=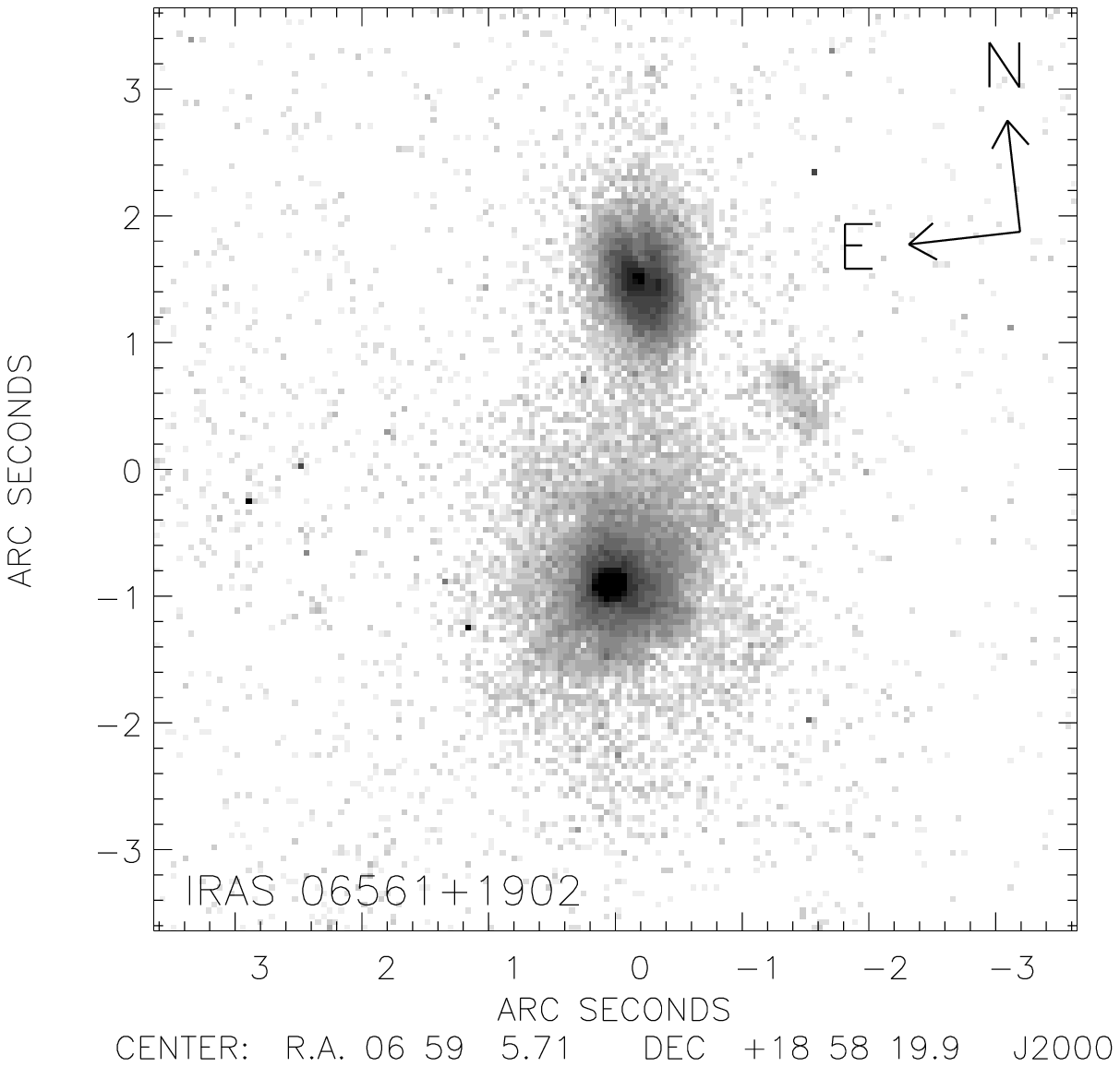,width=80mm}
\epsfig{figure=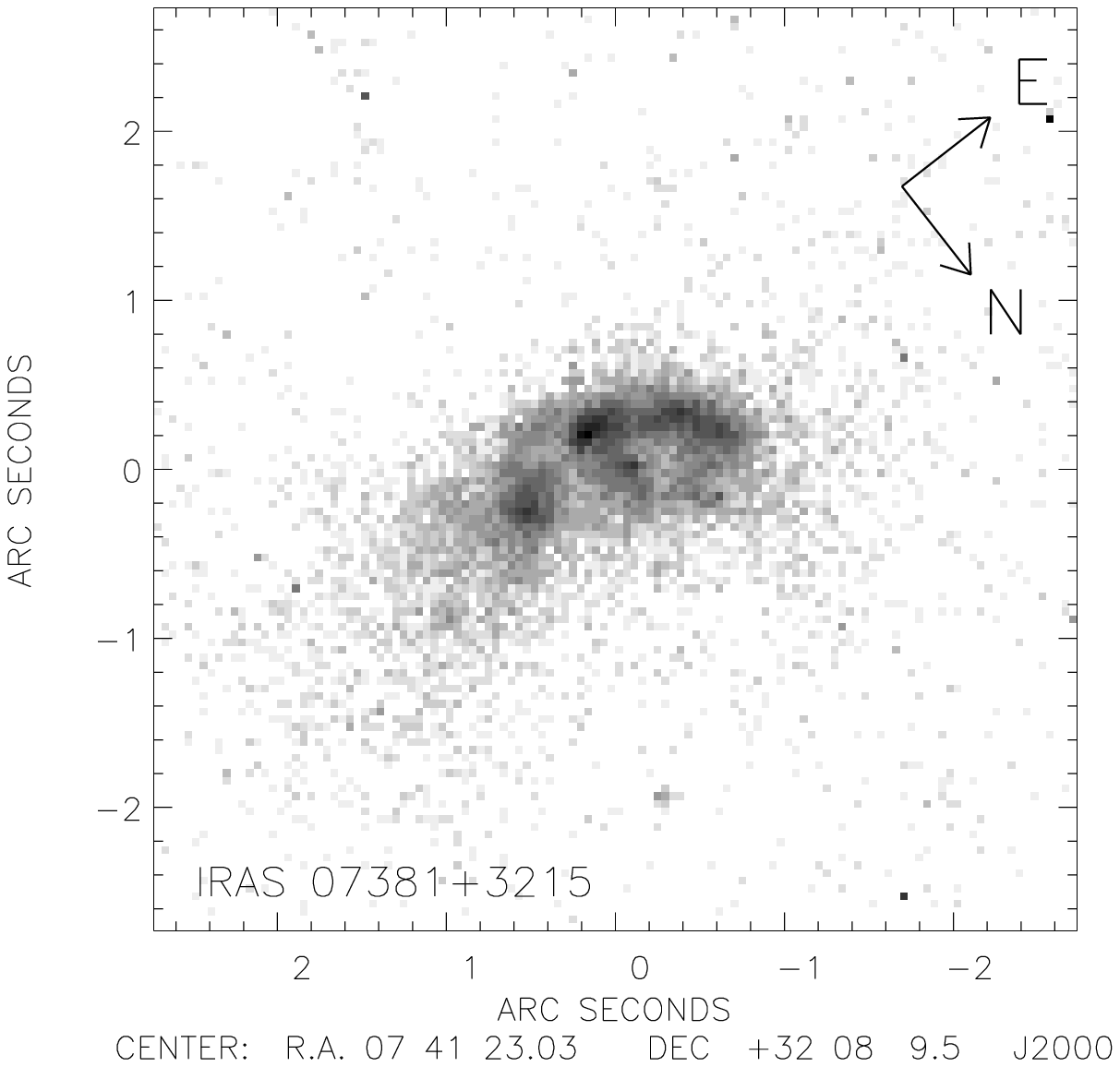,width=80mm}
\end{minipage}
\begin{minipage}{170mm}
\epsfig{figure=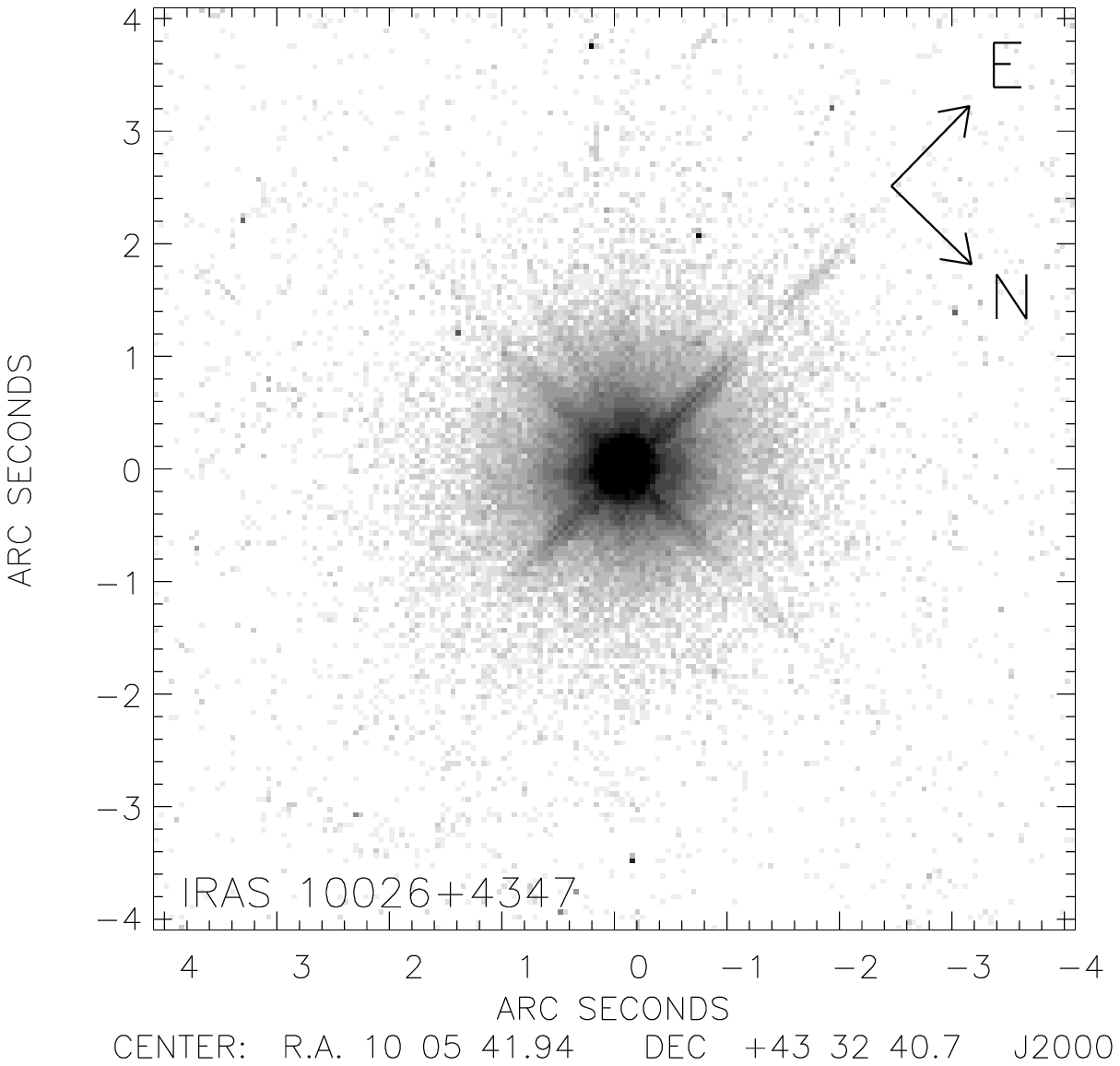,width=80mm}
\epsfig{figure=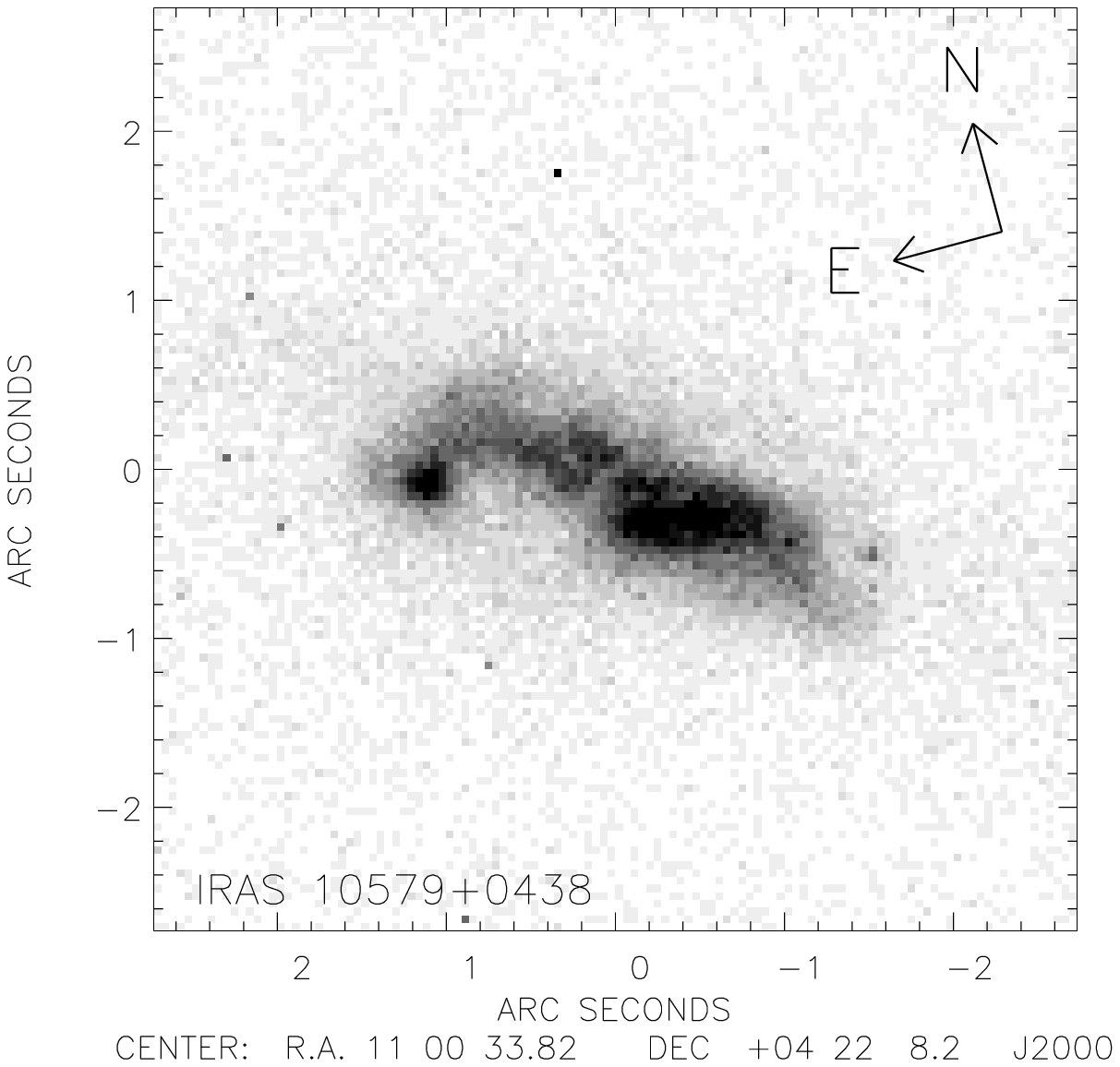,width=80mm}
\end{minipage}
\begin{minipage}{170mm}
\epsfig{figure=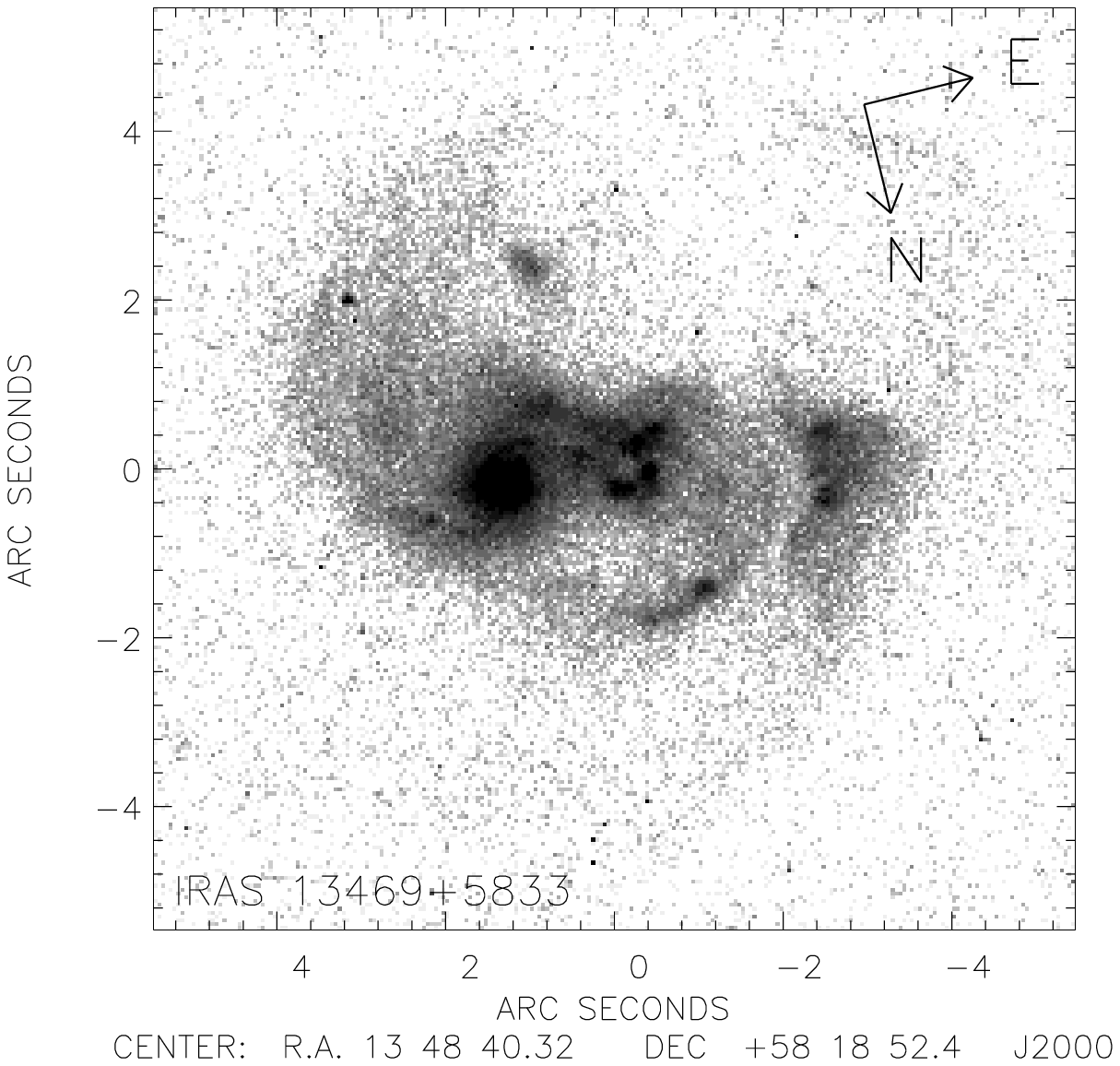,width=80mm}
\epsfig{figure=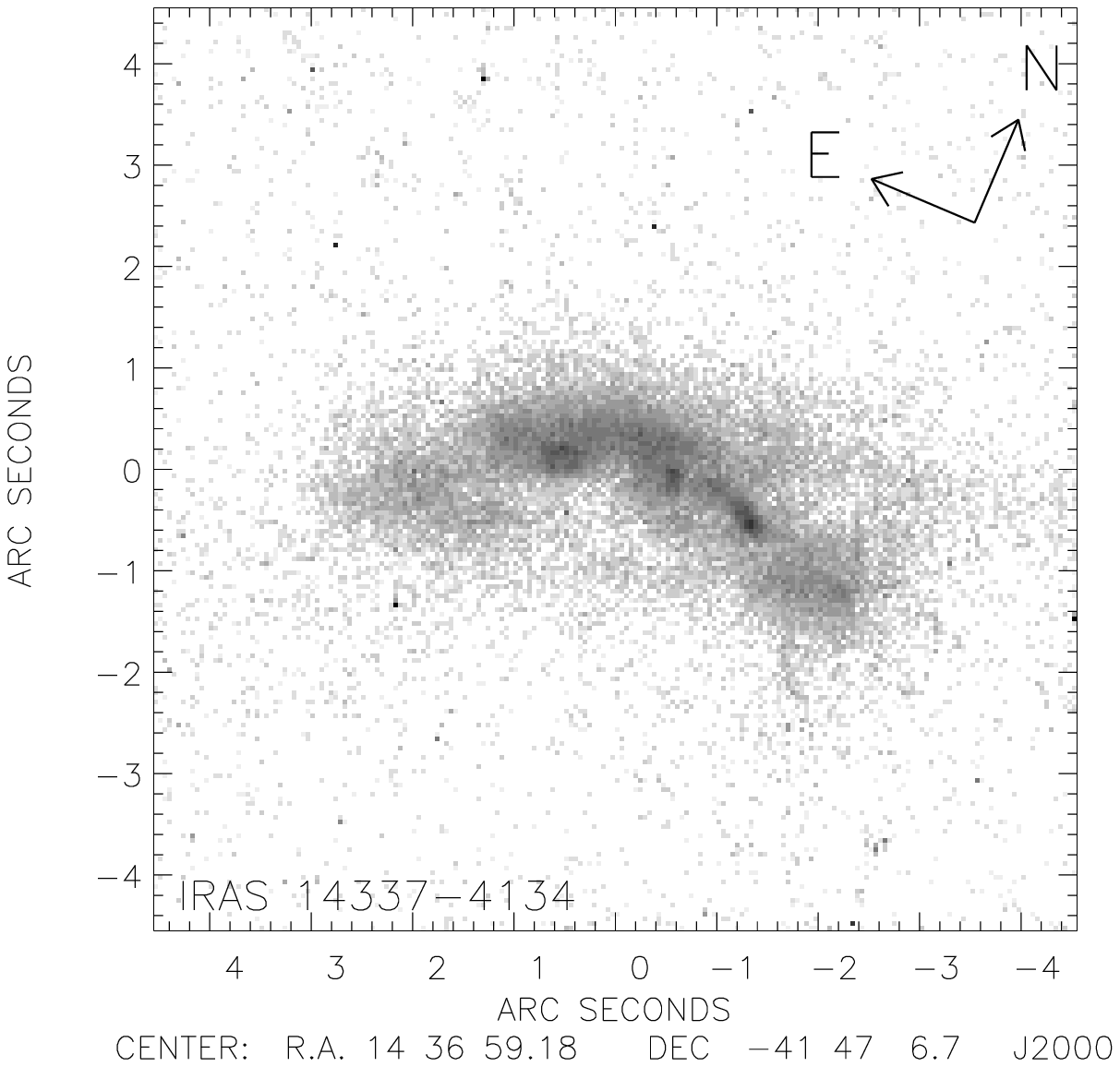,width=80mm}
\end{minipage}
\caption{{\em HST} F606W images of IRAS 06561+1902 to IRAS 14337-4134.
 \label{ulirgs_im_two}}
\end{figure*}

\begin{figure*}
\begin{minipage}{170mm}
\epsfig{figure=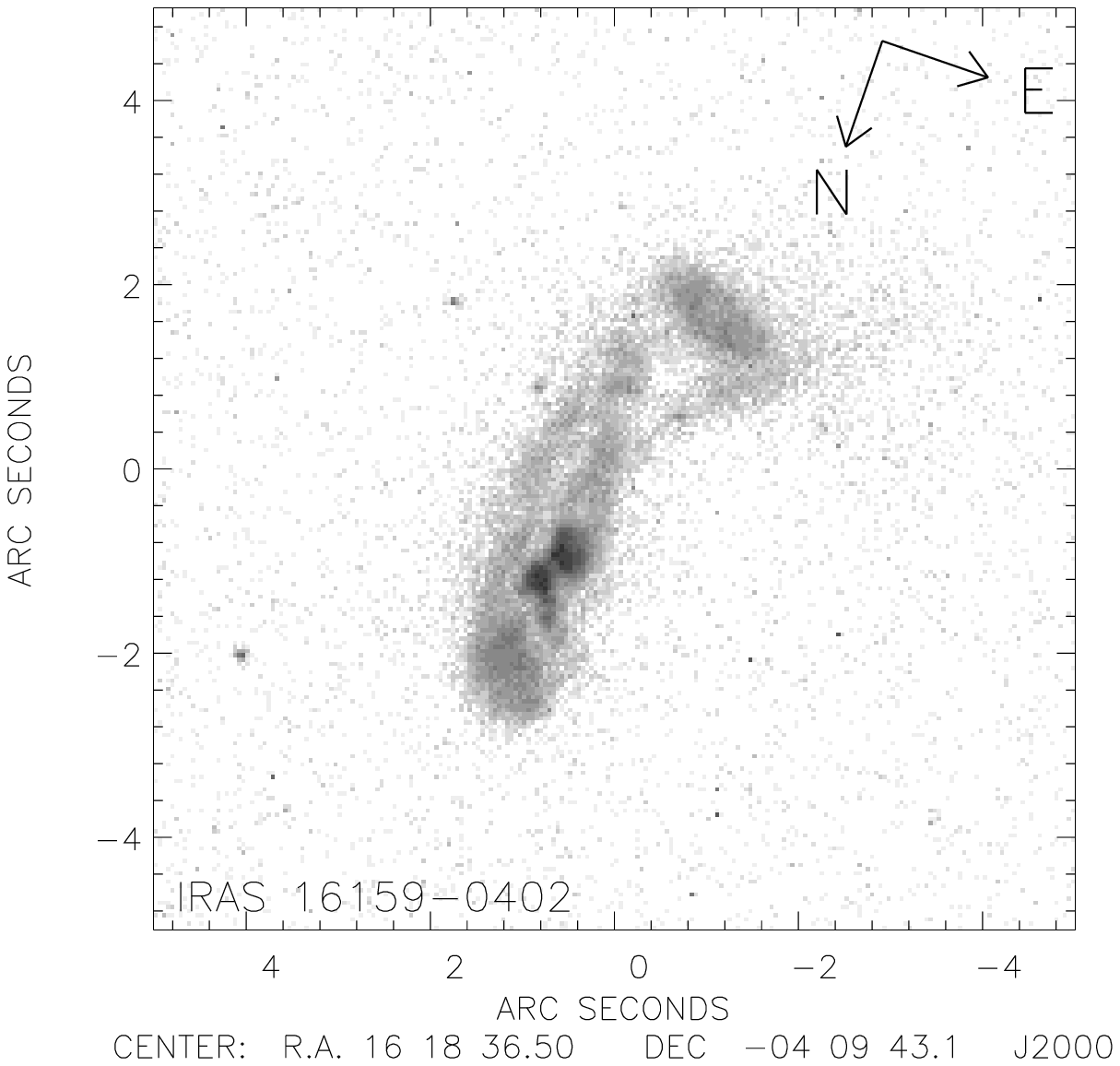,width=80mm}
\epsfig{figure=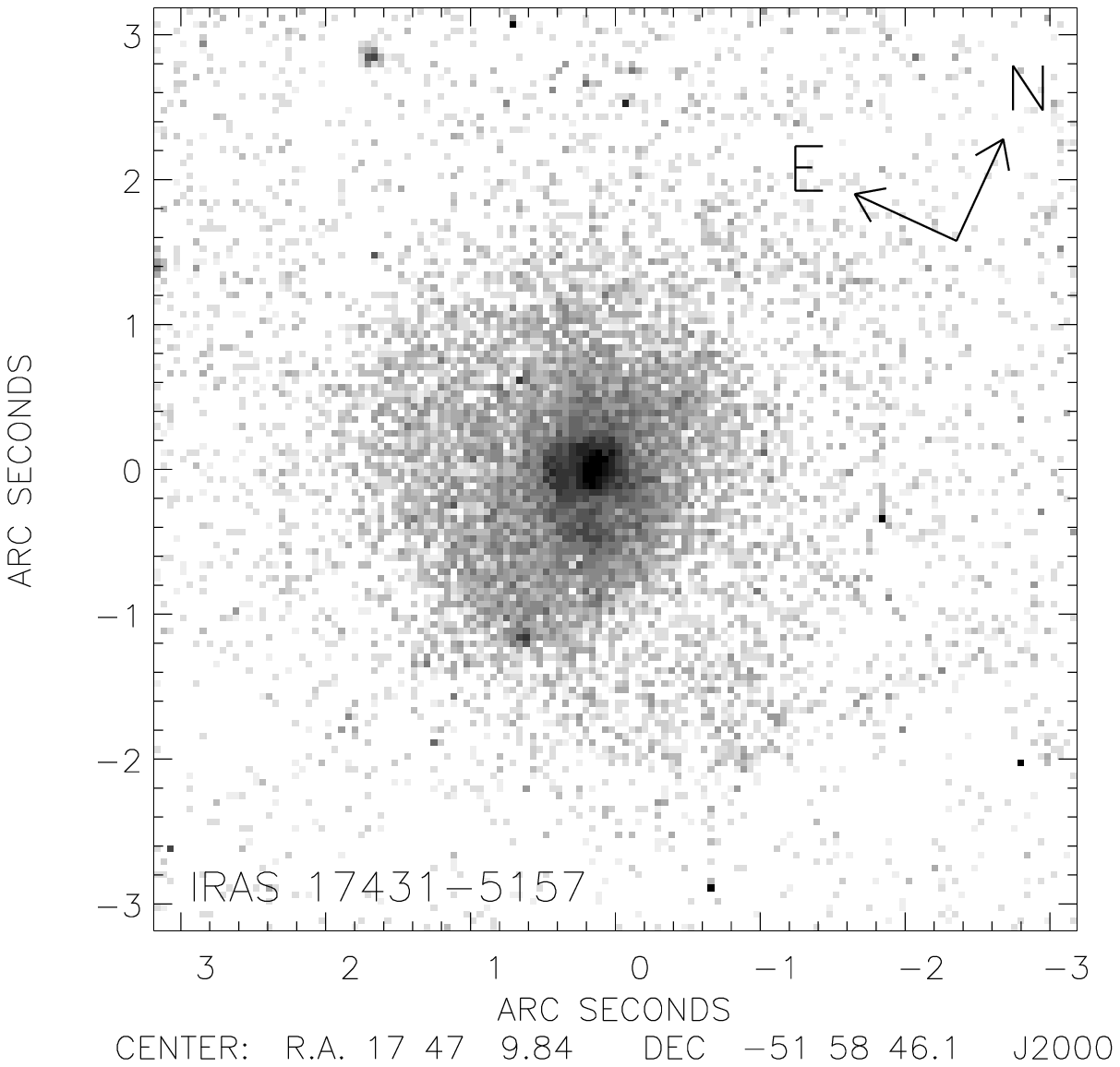,width=80mm}
\end{minipage}
\begin{minipage}{170mm}
\epsfig{figure=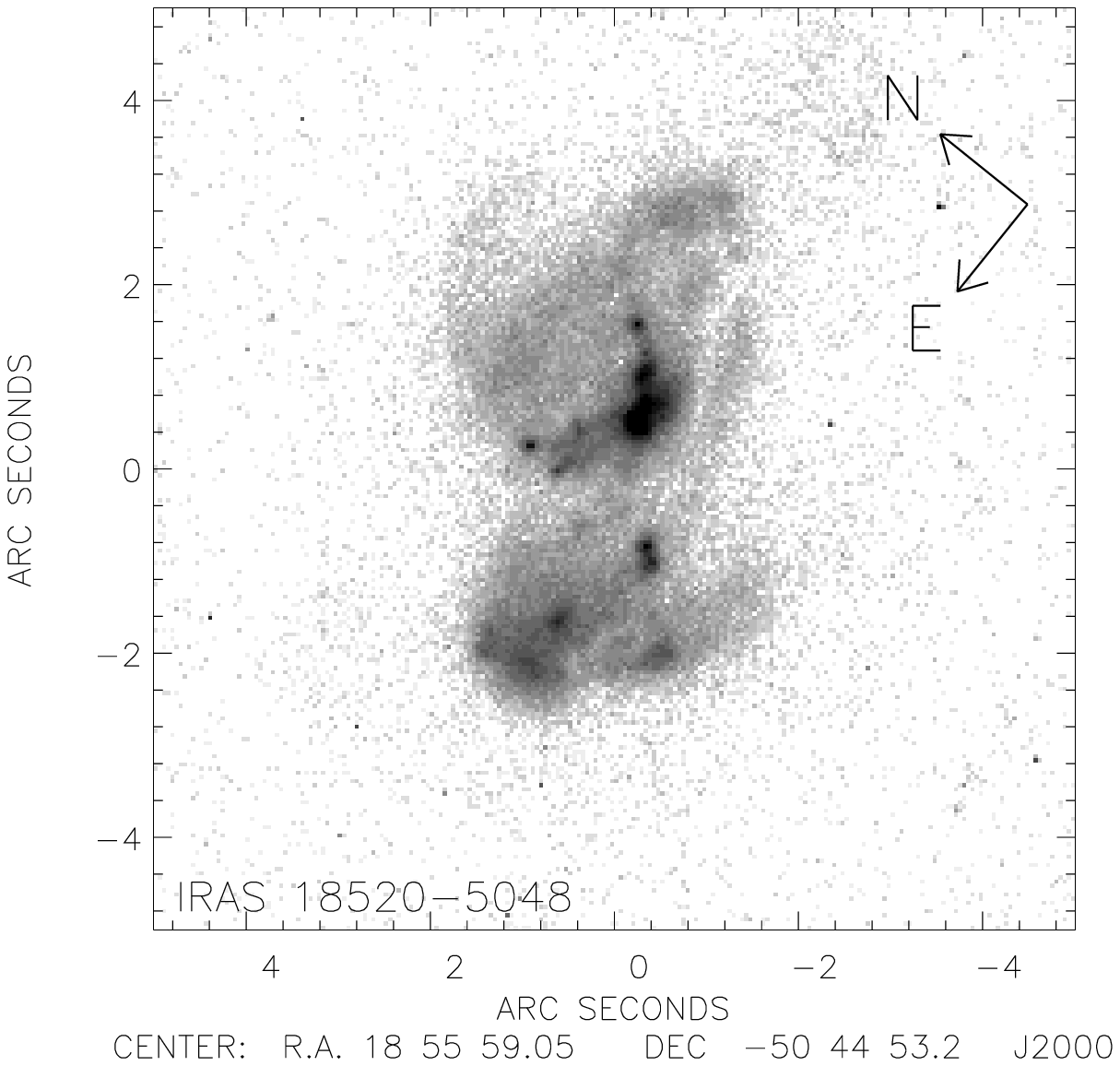,width=80mm}
\epsfig{figure=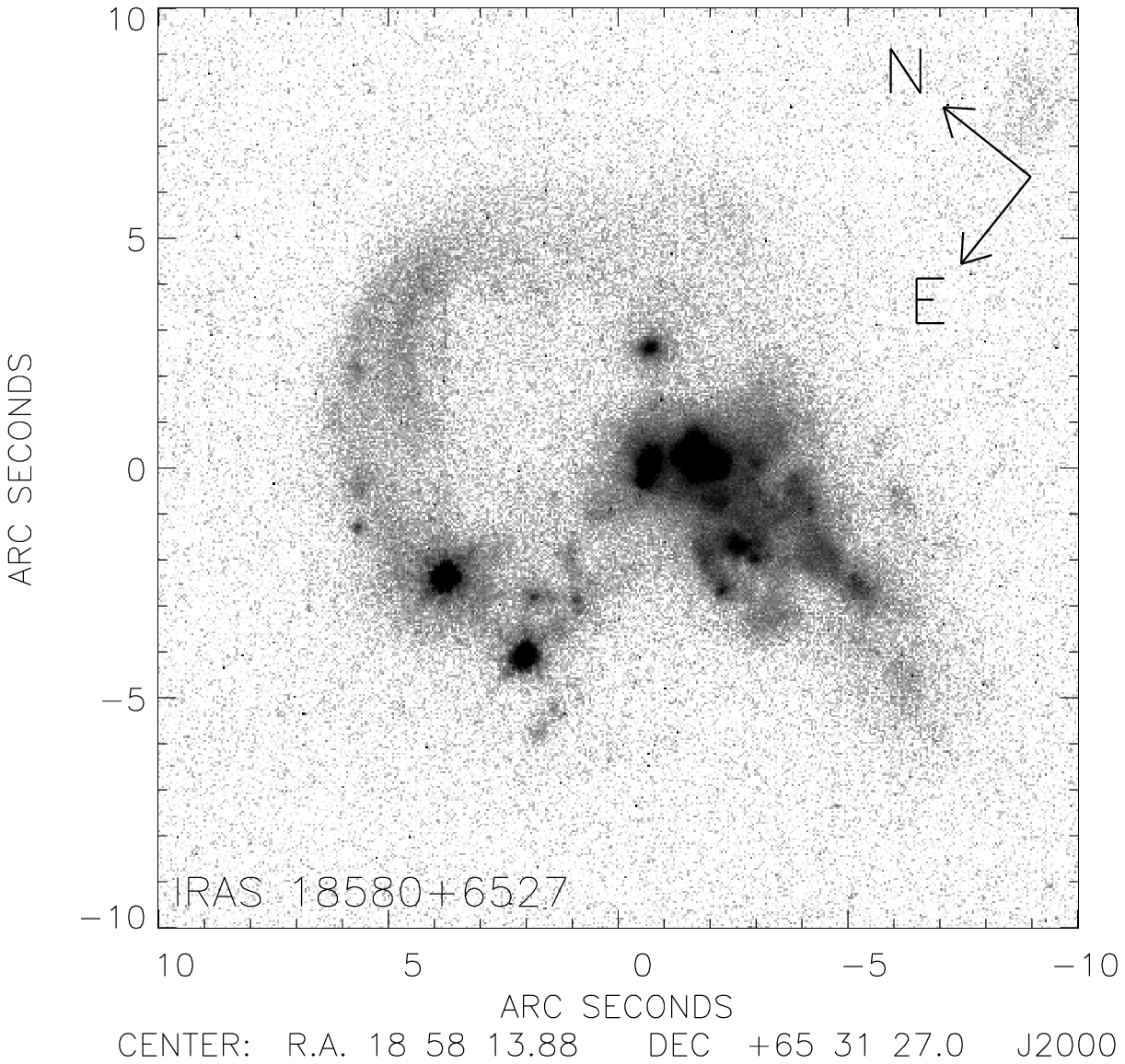,width=80mm}
\end{minipage}
\begin{minipage}{170mm}
\epsfig{figure=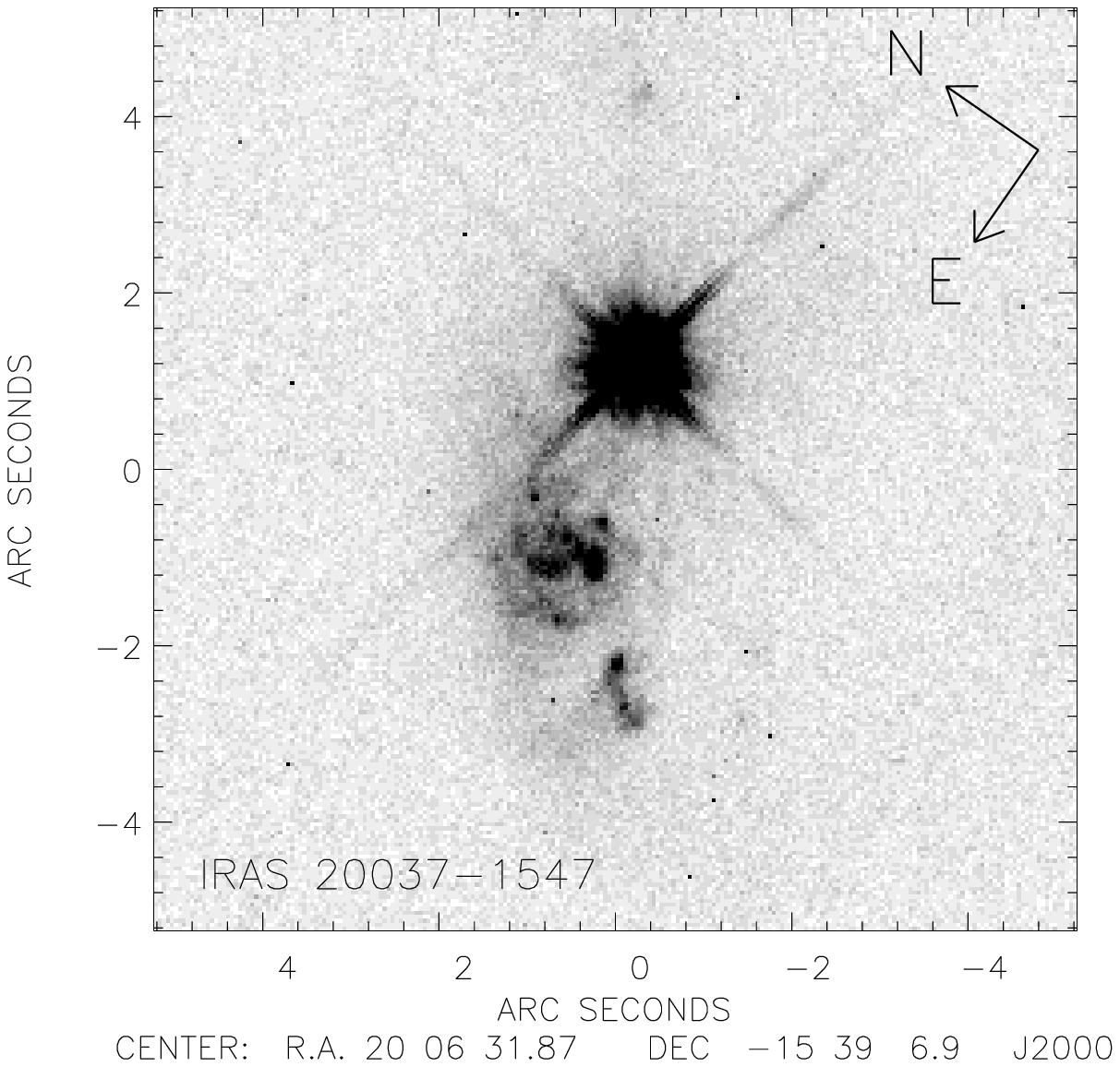,width=80mm}
\epsfig{figure=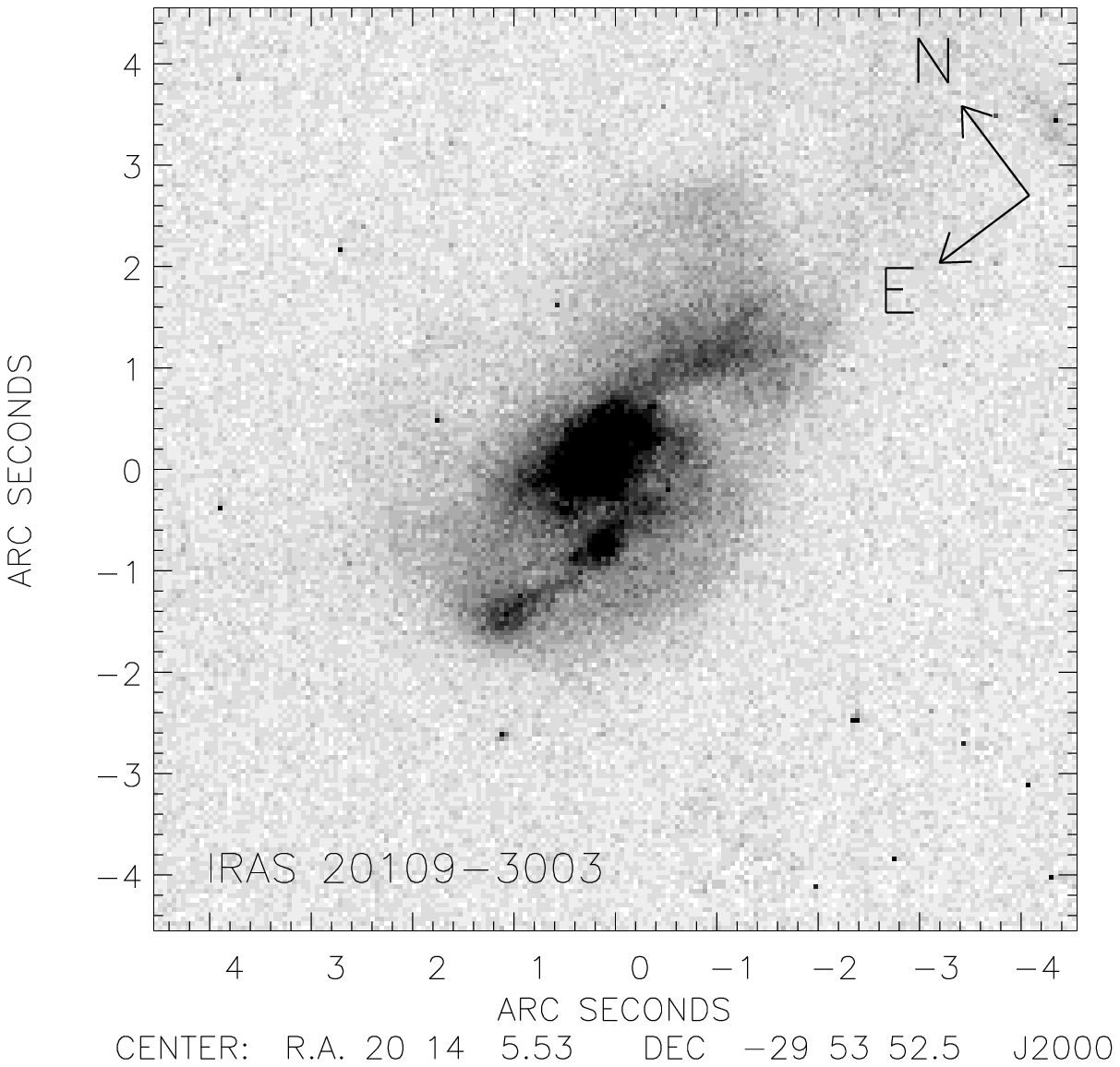,width=80mm}
\end{minipage}
\caption{{\em HST} F606W images of IRAS 16159-0402 to IRAS 20109-3003.
 \label{ulirgs_im_three}}
\end{figure*}

\begin{figure*}
\begin{minipage}{170mm}
\epsfig{figure=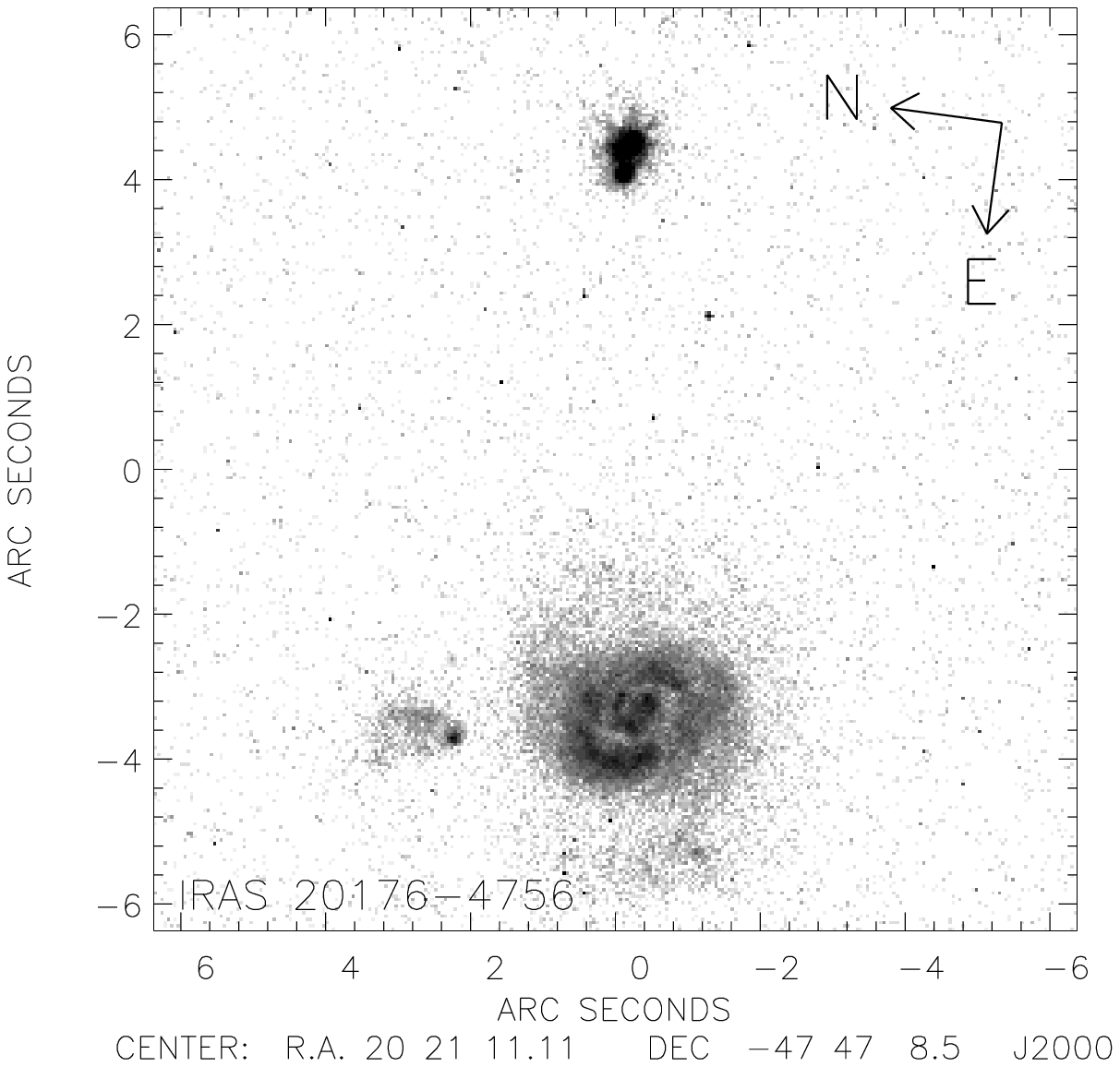,width=80mm}
\epsfig{figure=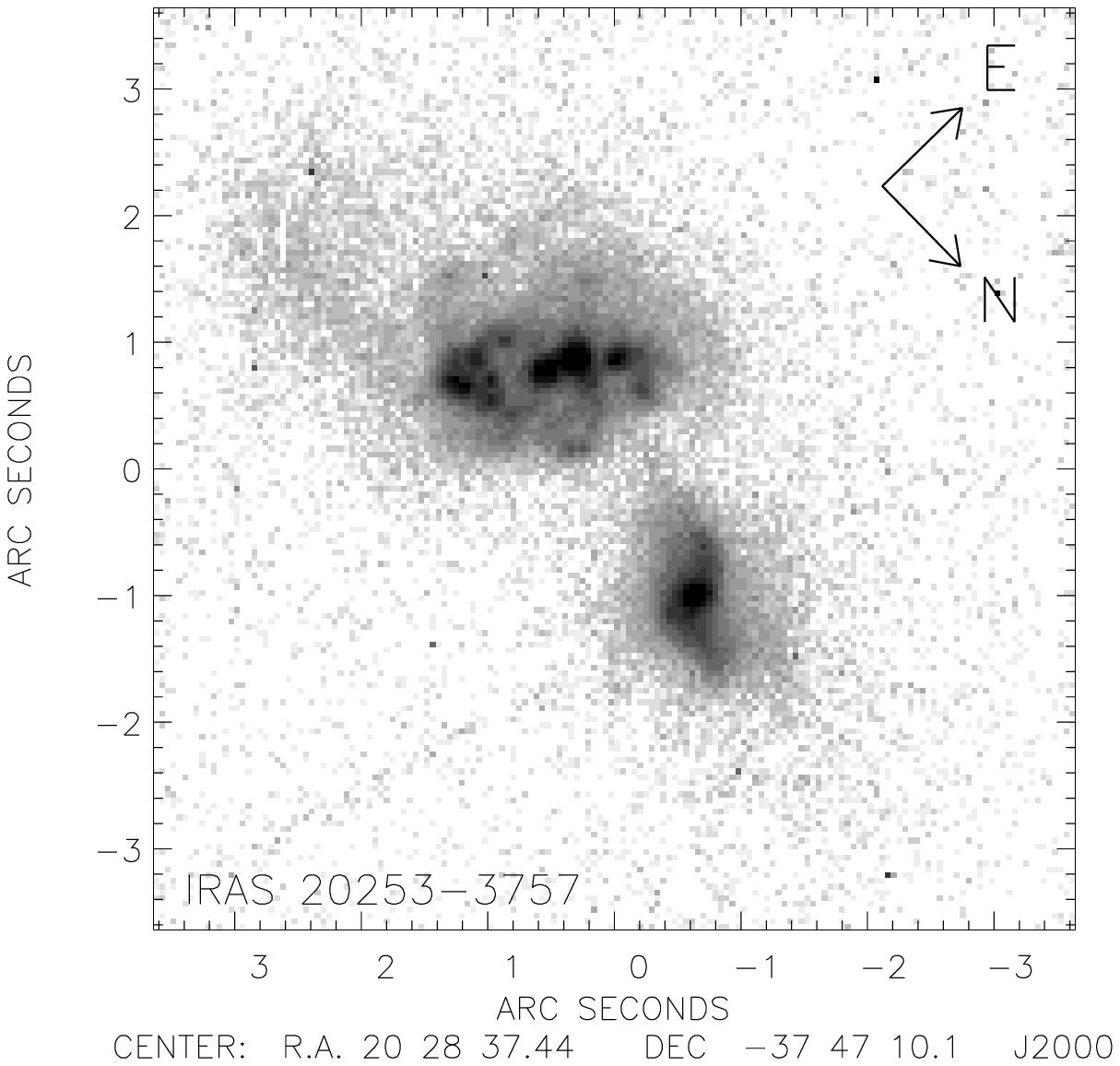,width=80mm}
\end{minipage}
\begin{minipage}{170mm}
\epsfig{figure=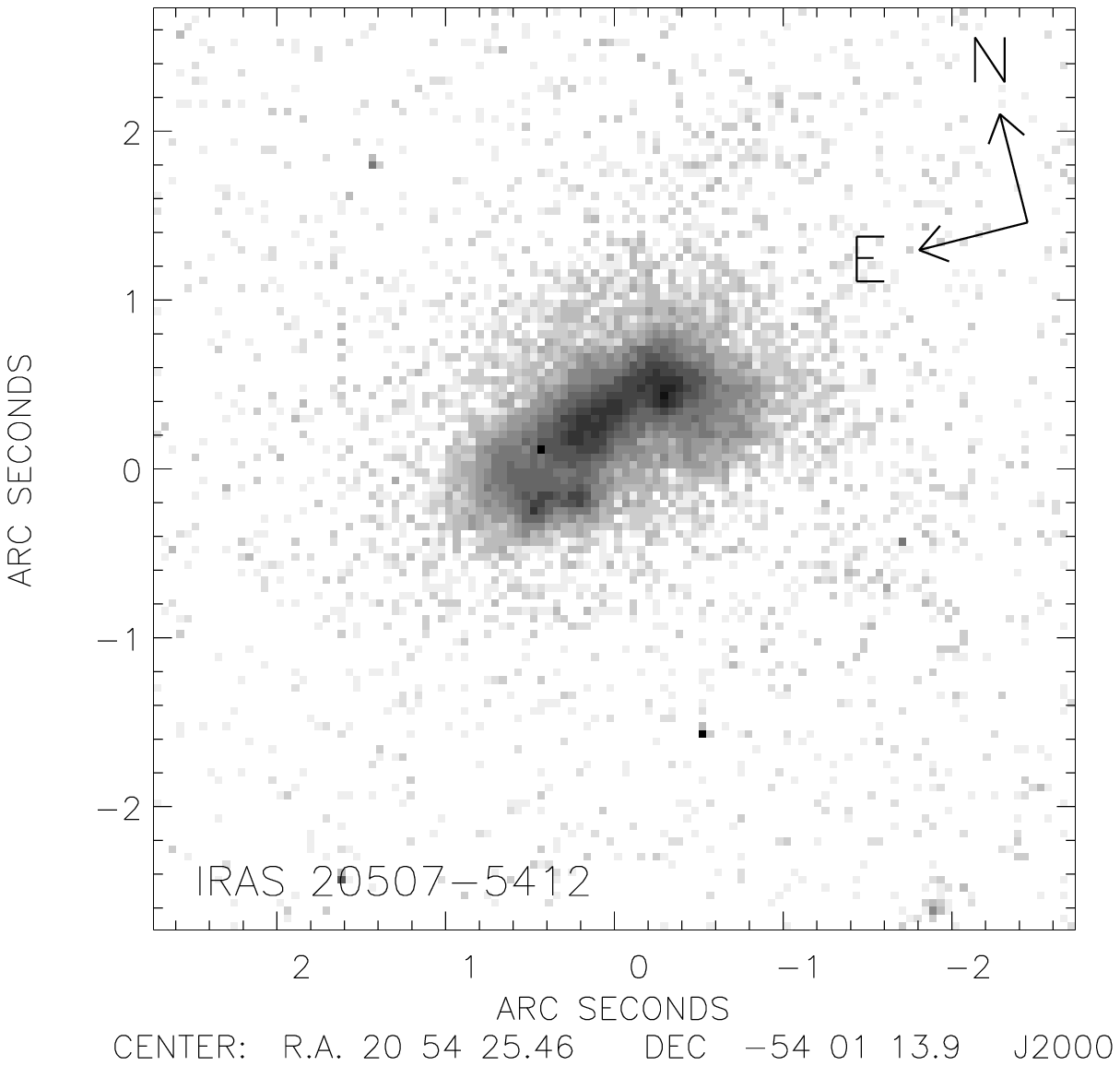,width=80mm}
\epsfig{figure=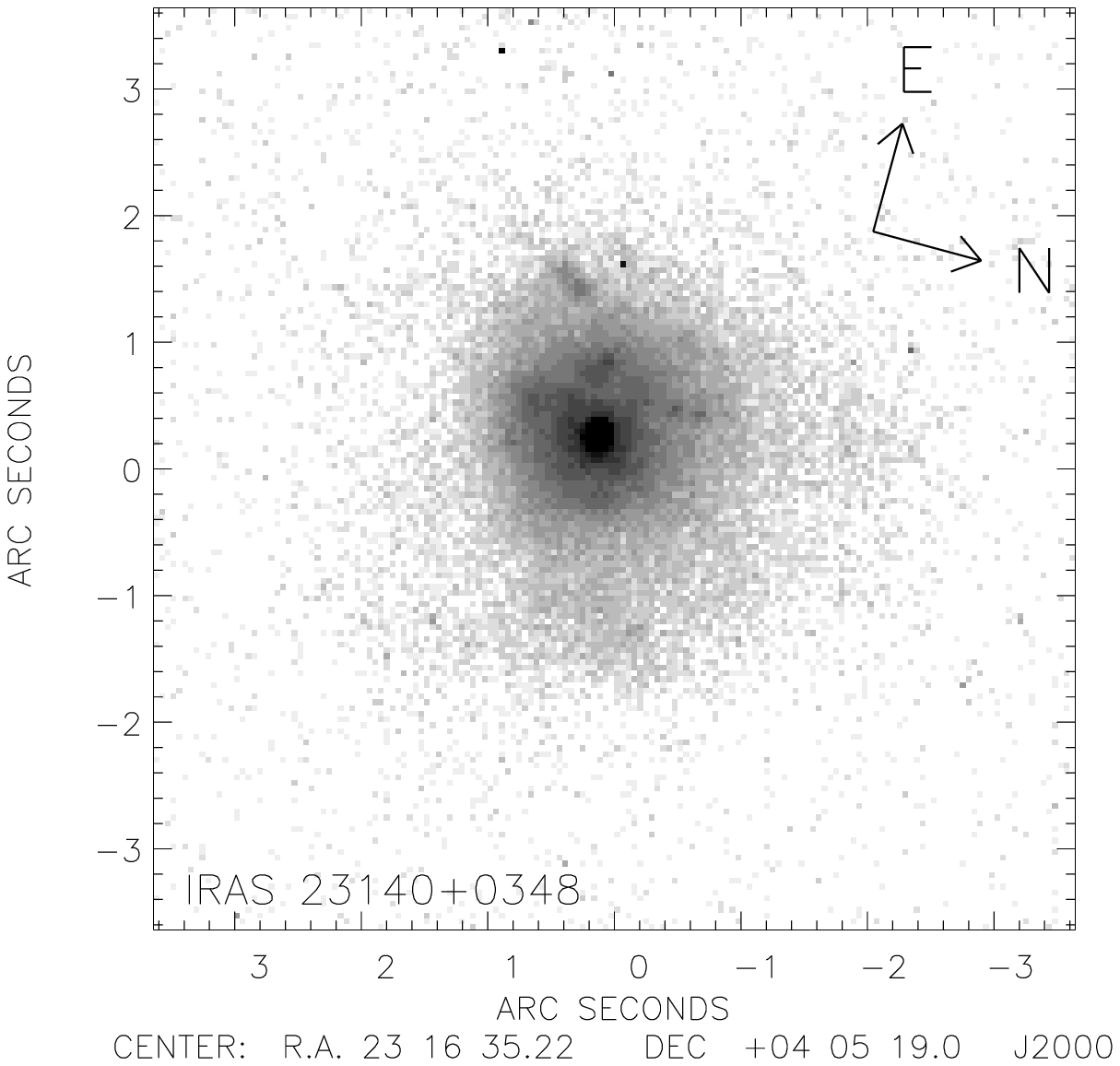,width=80mm}
\end{minipage}
\begin{minipage}{170mm}
\epsfig{figure=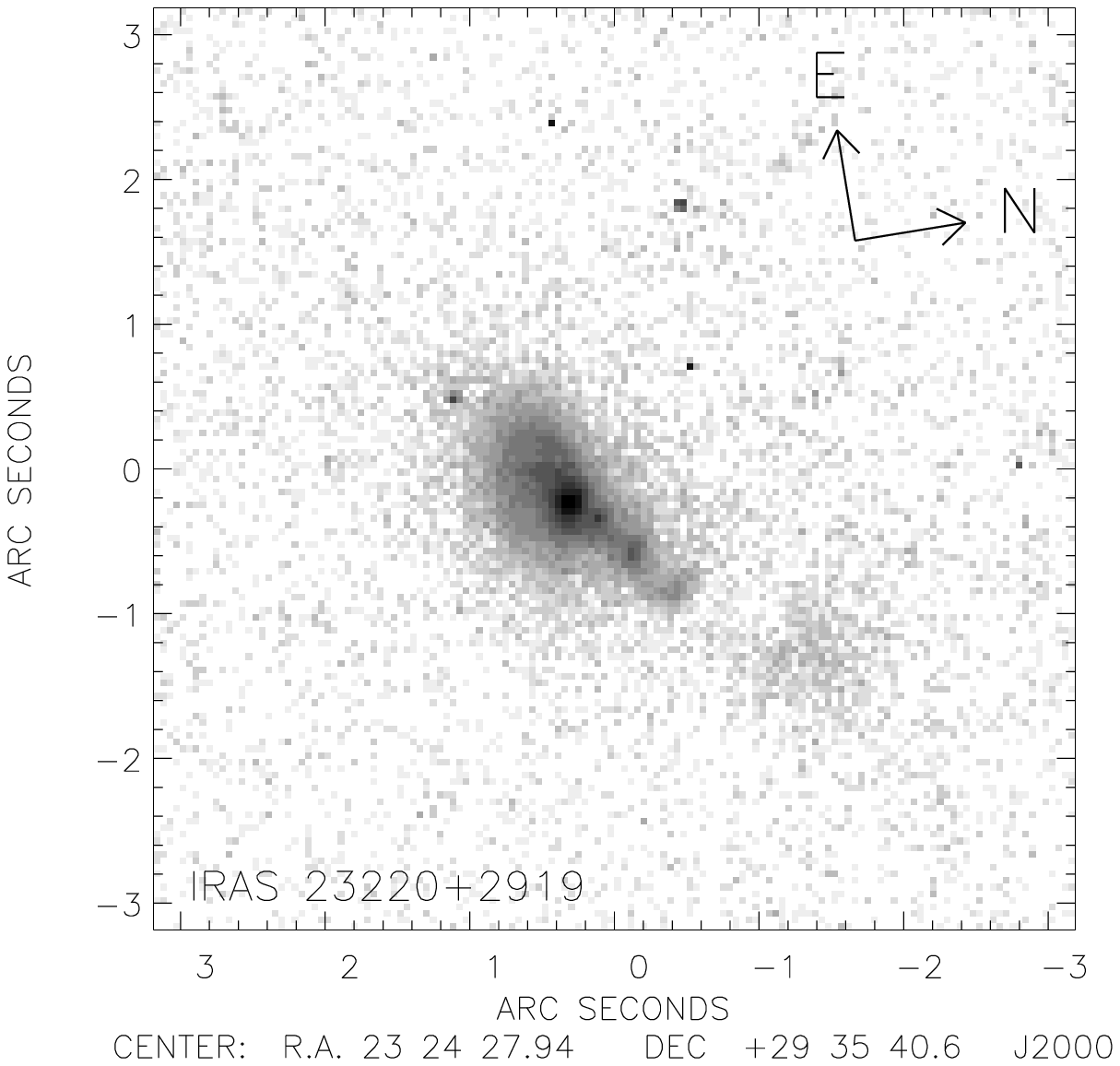,width=80mm}
\end{minipage}
\caption{{\em HST} F606W images of IRAS 20176-4756 to IRAS 23220+2919.
 \label{ulirgs_im_four}}
\end{figure*}

\subsection{Observations}
The data were taken in cycle 8 between October 1998 and July 1999 using the Wide Field Planetary Camera 2. 
The coordinates of each source were centred on the Planetary Camera CCD, selected for its superior pixel 
scale (0.046\arcsec\ pixel$^{-1}$ as opposed to $\sim$ 0.1\arcsec\ pixel$^{-1}$ for the WFC) and full sampling of the 
PSF over the Wide Field Camera CCD's. At the mean redshift of our sample the pixel scale of the Planetary 
Camera corresponds to a physical size of $0.2$kpc. Exposures were taken using the F606W filter, which 
corresponds approximately to the {\em V} band filter in the Cousins system. The observations were performed 
in `Snapshot' mode, each object being imaged for 600 seconds. Each observation was split into 2 exposures of 
300s to facilitate the subtraction of cosmic rays.

For 22/23 of the sample, data were taken using the Wide Field Camera 3 CCD and the wide band 
F814W filter, which corresponds closely to the {\em I} band filter in the Cousins system, as 
part of a separate program (prop. ID 6346, PI K. Borne). These observations were also taken in `Snapshot' 
mode, each object being imaged for 800s.

The conversion between {\em HST} filter system magnitudes and Cousins magnitudes is not 
straightforward. Depending on the filter used for the observations the applied conversion factor 
can be strongly related to the object spectrum. For the F606W filter the dependence of the 
conversion factor to Cousins {\em V} band on the object spectrum is quite large, and can introduce 
photometric errors of $\Delta$m = 0.2 or more. For the conversion between F814W magnitudes and 
Cousins {\em I} band the dependence is much smaller, generally $\Delta$m $< 0.05$. In order to 
avoid these associated uncertainties, magnitudes are given in the HST filter system using the 
V{\sevensize EGAMAG} synthetic zeropoints \cite{hol}, and have not been converted into the Cousins 
filter system. The differences 
between our F814W magnitudes and the equivalent Cousins {\em I} band magnitudes are likely to be very 
small, with $\Delta$m $< 0.1$. For the F606W filter, which is both wider and redder than the {\em V} 
band Cousins filter, the match is much poorer. Hence we do not make comparisons between our F606W band 
magnitudes and previously quoted {\em V} band magnitudes, but do make comparisons between our F814W band 
magnitudes and previously quoted Cousins {\em I} band magnitudes.

\begin{figure}
\rotatebox{90}{
\epsfig{figure=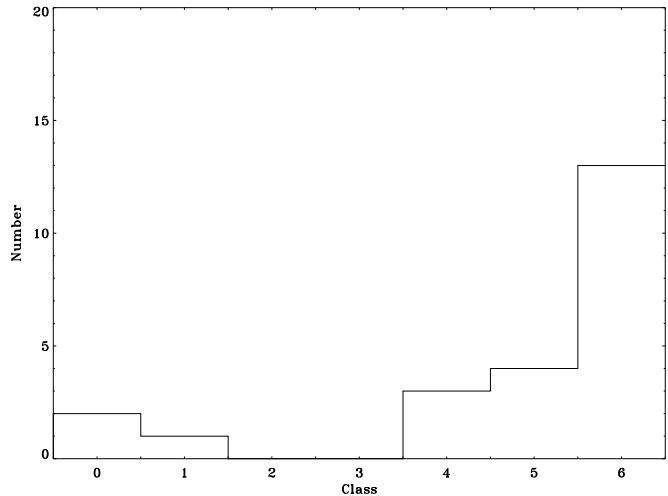,width=70mm}
}
\caption
{ 
Histogram showing the number of sources per band in the 7 band classification system of Lawrence et al 1989.
A summary of this classification scheme can be found in \S4.1.
\label{vi_plot}
}

\end{figure}

\section{Data Analysis}

\subsection{Data Reduction}
Upon arrival all datasets were recalibrated with the best available reference files, using the {\em IRAF} 
task C{\sevensize ALWP2}, and combined into a single image using the {\em IRAF} task C{\sevensize RREJ}. 
Following sky subtraction, warm pixels were removed by linear interpolation from their immediate neighbours. 
Photometric solutions were calculated using the S{\sevensize YNPHOT} package from the STScI, recently 
updated to match the solutions given by Holtzman et al \shortcite{hol}. For those QSOs in the sample that 
were saturated in the central regions magnitudes and fluxes were computed using PSF fitting photometry. 
Corrections were made for detector gain (for these observations the gain was $7e^{-}/DN$) and aperture size. 
Absolute magnitudes were derived using the expression:

\begin{equation}
M = m-25-5log[(2c/100h)(1+z)(1-(1+z)^{-1/2})]
\end{equation}

\noindent We estimate that our derived relative magnitudes have an associated photometric error of approximately 4\%. Fluxes 
in Jy were obtained by multiplying the flux in ergs cm$^{-2}$ s$^{-1}$ \AA$^{-1}$ by the central wavelength of the 
F814W filter in angstroms, and then dividing by the central wavelength in Hz. As such they do not include any bandpass 
correction. 

No {\em k} corrections have been applied to the calculated absolute magnitudes. Trial {\em k} corrections for both the F606W 
and F814W filters were computed 
using the spectral synthesis package P{\sevensize EGASE} \cite{fio}. At the mean redshift of the sample, $z = 0.2$, the {\em k} 
correction in the F606W band was found to be of the order $\Delta$m $= 0.15$ for a $10^{9}$ year old spiral galaxy, with the 
correction in the F814W filter being slightly lower. Even at the highest redshift in the sample, $z=0.345$, the {\em k} correction 
was not much greater than this ($\Delta$m $=0.18$). It was found that, if the age of the system was varied across $10^{7} - 10^{10}$ 
years and/or a different source spectrum was assumed, then the computed k correction varied across the range $-0.02 < \Delta$m$ < 0.3$.
The errors introduced in applying these synthetic {\em k} corrections therefore probably significantly outweigh any corrective 
effect. There are also no applied corrections for reddening due to dust extinction.

\subsection{Colour Maps}
Colour maps are an invaluable tool for highlighting morphological features not apparent in single band images. Constructing colour 
maps for the F606W and F814W band data presented here is not straightforward as the data were taken on different CCD's. The 
Point Spread Function (PSF) for any of the WFPC2 CCD's is strongly dependent on position, wavelength, and CCD characteristics. These 
differences must be carefully accounted for when matching the two images.

The F606W and F814W band images were first scaled to a common exposure time and gain. Although the readout noise does not scale 
linearly for the PC and WF3 chips, this effect is not significant when the exposure time of one image is at least $10\%$ of the other.
Following the advice of the STScI, the images were then deconvolved with appropriate PSFs using the Richardson-Lucy algorithm \cite{luc} 
as implemented in the IRAF task L{\sevensize UCY}. The TinyTIM v5.0 software 
\cite{kri} was used to generate PSFs for the appropriate filter and chip positions. The four frames in the F606W and F814W observations 
were then combined into single images using the IRAF task W{\sevensize MOSAIC}, which corrects for geometric distortion, offsets and scale 
differences between the Planetary Camera and the Wide Field Cameras. This task also degrades the Planetary Camera V band images 
to the resolution of the Wide Field Cameras, effectively convolving the PC image. Any residual PSF mismatch is thus due to 
differences in the effective PSF's of the images achieved by the initial deconvolutions. Since this PSF mismatch could in principle be 
significant we performed simulations to assess the accuracy of this procedure by constructing colour maps of stars visible in both 
the PC and WFC CCD's. It was found that the maps of these stars had a featureless colour structure. Although some level of PSF 
mismatch is undoubtedly still present it appears to be negligible compared to noise in the original images or large angle 
scattering within the WFPC2 optics.

A further potential source of extraneous colour structure with this method is that of deconvolution artifacts. These can include sharp 
colour peaks caused by amplification of poisson noise in the original images, or very bright point sources surrounded by dark rings which are 
caused by large numbers of iterations on a bright point source within an envelope of extended emission. To minimise the occurrence 
of these artifacts we examined the images after each iteration. It was found that deconvolution artifacts only became visible 
several iterations after convergence between the original image and the deconvolved image had been achieved (measured via a reduced 
$\chi^{2}$ statistic). As we halted the deconvolution immediately upon attaining convergence, the effect of artifacts on the achieved 
colour structure is insignificant.

\subsection{QSO Host Galaxies}
Host galaxies can only be resolved in QSOs if light from the central bright source, and from the host galaxy, can be separated. 
The most convenient way to accomplish this is by fitting a point source template to the QSO; the template being established from 
a set of observations of stellar PSFs that are a good match in colour, filter and detector position to the targets. For the optical 
QSOs in this sample no additional observations were made of nearby stars, and no suitable observed PSFs were available in the 
{\em HST} PSF Archive. Synthetic PSFs were therefore generated using TinyTIM.

Ten times oversampled PSFs were generated. These PSFs were then shifted by non-integer pixel distances, rebinned to Planetary Camera 
resolution and convolved with the PC pixel scattering function. The agreement between these synthetic PSFs and observed (stellar) 
PSFs is generally excellent within radii of about 2\arcsec. Beyond this radius scattered light in the WFPC2 optics can have a 
significant effect, an effect that is not modelled by the TinyTIM software. The effect of any systematic uncertainties in the 
PSF have also not been examined, such effects are however likely to be negligible.

The resulting PSFs were then centroided  and normalized to the image via a reduced $\chi^{2}$ fit. It was found that all the QSO 
images were saturated in the central regions to varying extents, mostly confined to the central 3x3 pixels and never beyond the 
central 5x5 pixels. In all cases the centralmost possible pixels were used, avoiding those that were saturated or that lay in the 
diffraction spikes. This method allows the image and PSF to be registered to within $\sim$0.2 pixels (estimated by visual 
inspection), and also gives a starting value for the PSF normalisation.

\subsection{Profile Fitting}
In order to determine the morphology of the host galaxies of the QSOs in the sample, both de Vaucoleurs and exponential disk profiles were 
fitted to the surface brightness profiles of the host galaxies. The de Vaucoleurs profile is of the form:

\begin{equation}
I(r) = I_{0}e^{-7.67[(r/r_{e})^{1/4} - 1]}
\end{equation}

\noindent and the exponential disk profile is of the form:

\begin{equation}
I(r) = I_{0}e^{-(r/h)}
\end{equation}

One-dimensional surface brightness profiles of the sources were extracted by fitting annuli spaced at 1-pixel intervals, fixing 
the ellipticity (at $\epsilon = 0.05$), semimajor axis and position angle for each fit. The PSF profile together with either a radial de 
Vaucoleur or disk profile were then fitted to the source surface brightness profile, treating the PSF normalisation as a third 
free parameter. This helps to minimise as far as possible the systematic errors introduced by subtracting off a `best guess' PSF 
and fitting a galaxy profile to the remaining light distribution. Fitting was carried out using Levenberg-Marquardt least-squares 
minimization in order to robustly derive the best fit parameters. Two dimensional isophote fitting, using the IRAF task 
E{\sevensize LLIPSE}, was also used to investigate the ellipticities and isophotal twisting (large changes in azimuthal angle as a 
function of radial position) for each host by iteratively fitting isophotes of constant surface brightness whilst allowing the 
ellipticity and position angle to vary.

In order to quantify whether a host galaxy was detected or not we imposed the condition that a PSF + galaxy to QSO fit must be 
better than a pure PSF fit to at least 99\% confidence, computed via an {\em F} test. We excluded pixels within the central 
0.2\arcsec\ as the PSF within this radius is highly undersampled. We also excluded all pixels 
beyond 2\arcsec\ because of light scattered within the WFPC2 optics.

\section{Results}

\subsection{Morphology and Colours}

Positions, redshifts, magnitudes, spectral types, luminosities and merger classification for the sample can be found in Table \ref{hstobs}. The 
{\em HST} Planetary Camera images of the sample can be found in Figures \ref{ulirgs_im_one}, \ref{ulirgs_im_two}, \ref{ulirgs_im_three} 
and \ref{ulirgs_im_four}. For the two binary systems where the galaxies are clearly separated the magnitude of both sources together 
is given. Magnitudes of individual galaxies in binary systems can be found in Table \ref{ulirg_binmag}.

Contributions to the measured magnitudes from emission lines could in principle be significant, as many ULIRGs are known to be strong 
emission line galaxies \cite{vei2}. At the mean redshift of our sample the $H\alpha$ lies near the centre of the F814W bandpass, and 
the $OIII$ line lies nearly in the centre of the F606W bandpass. We evaluated uncertainties in our measured magnitudes by comparison 
with $H\alpha$ and $OIII$ luminosities taken from Veilleux, Kim \& Sanders \shortcite{vei2}. The F606W band luminosities of our sample 
span the range $10.0 < L_{V} < 11.4$, with all but three having $L_{V} > 10.3$ (where $L_{V}$ is the logarithm of the F606W band luminosity 
in units of bolometric solar luminosities). The F814W band luminosities, although not quoted in Table \ref{hstobs}, span a similar range. 
In contrast, the dereddened $H\alpha$ luminosities from Veilleux, Kim \& Sanders 
\shortcite{vei2} lie in the range $7.0 < L_{H\alpha} < 10.5$, with most lying in the range $8.0 < L_{H\alpha} < 9.3$. Similarly, the 
$OIII$ luminosities span the range $7.0 < L_{OIII} < 10.0$, with most falling between $7.0 < L_{OIII} < 8.0$. The uncertainty in our 
quoted magnitudes and colours due to these emission lines is therefore almost certainly much less than $\Delta$m$=0.1$ for all the objects in 
the sample.

\begin{figure}
\rotatebox{90}{
\centering{
\scalebox{0.36}{
\includegraphics*[12,40][514,724]{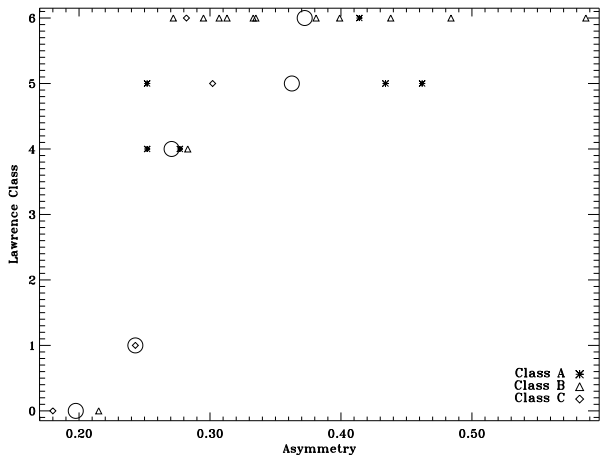}
}}}
\caption
{
Comparison between the Lawrence et al classification scheme, the qualitative classification scheme described in \S4.1 and the 
asymmetry statistic $A$ described in \S4.1. The three different systems, although based on different criteria, overall give 
similar descriptions. The three systems are therefore measuring related physical quantities, as expected. The circles denote 
the mean asymmetry for each Lawrence et al band.
\label{check_class}
}
\end{figure}

In order to provide a quantitative description of the morphologies in the sample we have classified the sources 
according to the seven band prescription given by Lawrence et al (1989):

\noindent 0: Source has no companions within 200Kpc and shows no signs of interaction or merger.

\noindent 1: Source has a faint companion (between 4 and 2 magnitudes fainter than the source) between 40Kpc and 200Kpc away.

\noindent 2: Source has a bright companion (less than 2 magnitudes fainter than the source) between 40Kpc and 200Kpc away.

\noindent 3: Source has a faint companion less than 40Kpc away but shows no signs of interaction. 

\noindent 4: Source has a bright companion less than 40Kpc away but shows no signs of interaction.

\noindent 5: Source is interacting with companion, and there are signs of tails, loops, or bridges.

\noindent 6: Source is merging; either there is obvious disturbance and/or there are two or more nuclei in a common envelope.

No source in the sample is an isolated undisturbed spiral or elliptical galaxy. Using the above classification scheme, 3/23 of 
the sample are classified as not interacting (groups 0 -- 2), 3/23 of the sample are classified as having a nearby 
companion but showing no signs of interaction (groups 3 -- 4), and 17/23 are classified as interacting or peculiar 
(groups 5 -- 6). Figure \ref{vi_plot} shows a histogram of the number of sources in each band. If it is assumed that the 
three sources with a nearby companion are interacting then this gives a fraction of ULIRGs observed to be interacting of 
87$\%$. If it is assumed that the sources with a nearby companion are not interacting then this gives a 
minimum observed interacting fraction of 74$\%$.

For some purposes it is informative to classify the ULIRGs into broader categories based solely on qualitative descriptions of 
their optical morphology. We have therefore classified our sample as follows:

{\em Class A}: There are two separate sources, but close enough to reasonably assume that they are interacting, or they are interacting 
but still physically distinguishable as separate objects.

{\em Class B}: The source is not distinguishable into separate objects and there is no QSO activity, but there are clear signs of 
interaction such as tails, bridges and and multiple nuclei.

{\em Class C}: The source contains an optical QSO with or without signs of ongoing interaction.

These classes are chosen such that sources can be classified unambiguously. They are not intended to imply an 
evolutionary sequence. 
By these criteria, 5 sources are identified as class A, 14 as class B and 4 as class C. The classification of 
individual objects is given in Table \ref{hstobs}. Analysis using this qualitative classification scheme is given in \S 5.

A quantified estimate of the degree of disturbance in an object can be obtained by computing some form of asymmetry 
statistic. Numerous different methods exist, a popular approach is to compute the $180\degr$ 
statistic $A$. Here we use a modified form of the equation presented in Brinchmann et al \shortcite{brin} and used by e.g. 
Serjeant et al \shortcite{serj}:

\begin{equation}
A=\frac{\sum_{ij}|I_{ij} - I^{T}_{ij}|}
{2\sum_{ij}|I_{ij}|} - k
\label{eqn:fasym}
\end{equation} 

\noindent where $I_{ij}$ is the original image, $I^{T}_{ij}$ is the image following a $180\degr$ rotation, and $k$ is 
the sky asymmetry statistic:

\begin{equation}
k=\frac{\sum_{ij}|G_{ij} - G^{T}_{ij}|}
{2\sum_{ij}|G_{ij}|}
\label{fasym2}
\end{equation} 

\noindent where $G_{ij}$ and $G^{T}_{ij}$ are areas of sky equal in area to the object image. There are two differences between 
the formula used by Brinchmann et al and our formula. The extra factor of 2 in both equations ensures that the asymmetry statistic 
is normalised to between 0 (perfect $180\degr$ symmetry) and 1 (perfect $180\degr$ antisymmetry). The modulus on the denominator 
ensures that sky values are not overestimated. For the QSOs in the sample the asymmetries are computed for the host galaxies rather 
than for the QSOs themselves as the QSOs will otherwise dominate and lead to unrealistically low values. For each object the 
asymmetry was calculated for all possible positions within the object and the lowest value was then selected. The computed asymmetry 
values can be found in Table \ref{hstobs}.

A comparison between the Lawrence et al classification scheme, the qualitative classification scheme described above, and the 
computed asymmetries can be found in Figure \ref{check_class}. It is apparent that, despite the different criteria used in these three 
systems to classify the objects in the sample, the asymmetry statistic $A$ gives similar results to both of the more qualitative 
classification schemes. The three different statistical measures are therefore measuring related physical quantities, as expected.

With the enhanced resolution of {\em HST} it is informative to examine the morphology of the sample in more detail. Two of the 
sample (IRAS 02054+0835 and IRAS 10026+4347) are QSOs that show no apparent signs of interaction. Two objects 
(IRAS 00275-2859 and IRAS 20037-1547) are mergers containing QSO activity. Three sources (IRAS 06268+3509, IRAS 
18520-5048 \& IRAS 20253-3757) are merging galaxy pairs. One source (IRAS 06361-6217) resembles a collisional ring galaxy. 
13/23 sources show evidence for tidal tails, plumes,  and other signs of ongoing interaction. 
In addition these sources all show at least one, and in most cases several, compact bright regions or `knots'. Two sources 
(IRAS 02587-6336 \& IRAS 20176-4756) are binary systems. Discussion of the morphology of individual sources can be found in 
\S 4.4.

Three sources from our sample can be identified as having `warm' infrared colours ($f_{25}/f_{60} > 0.2$), whereas ten sources 
have `cool' infrared colours ($f_{25}/f_{60} < 0.2$). The remaining ten sources have {\em IRAS} upper limits that do not allow us 
to distinguish between these two possibilities.
Of the three `warm' sources, two are QSOs, one of which shows signs of interaction. Of the ten `cool' sources, one is an 
interacting system containing QSO activity, one is a close pair, and the rest are interacting systems. Two of these 
interacting systems (IRAS 13469+5833 \& IRAS 18580+6527) show evidence for a violent merger, with large numbers of bright knots 
and large tidal tails.

\begin{table}
\caption{Binary System magnitudes \label{ulirg_binmag}}
\begin{tabular}{@{}cccccc}
Source             & RA       & Dec      & $M_{V}$ & $M_{I}$   \\ 
02587-4336 {\em A} & 59 43.07 & 25 05.43 & -20.50  & -21.50    \\
02587-4336 {\em B} & 59 43.48 & 25 00.88 & -21.62  & -22.81    \\
20176-4756 {\em A} & 21 10.67 & 47 08.12 & -20.97  & -22.16    \\
20176-4756 {\em B} & 21 11.44 & 47 09.16 & -21.45  & -22.52    \\
\end{tabular}

\medskip
Coordinates are J2000 and are given as mm ss.ss 
\end{table}

\subsection{Colour Maps}

\begin{figure*}
\begin{minipage}{180mm}
\epsfig{figure=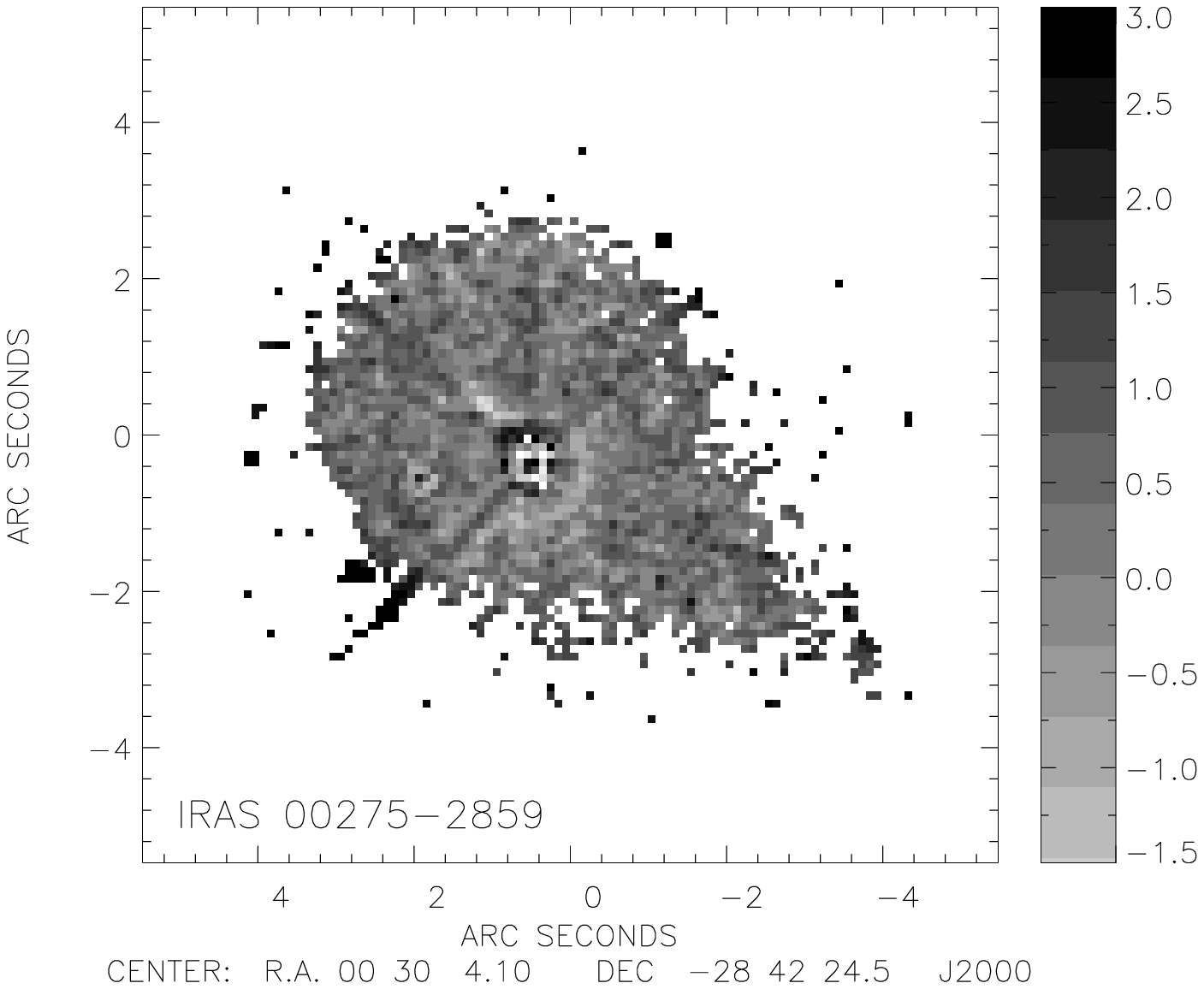,width=55mm}
\epsfig{figure=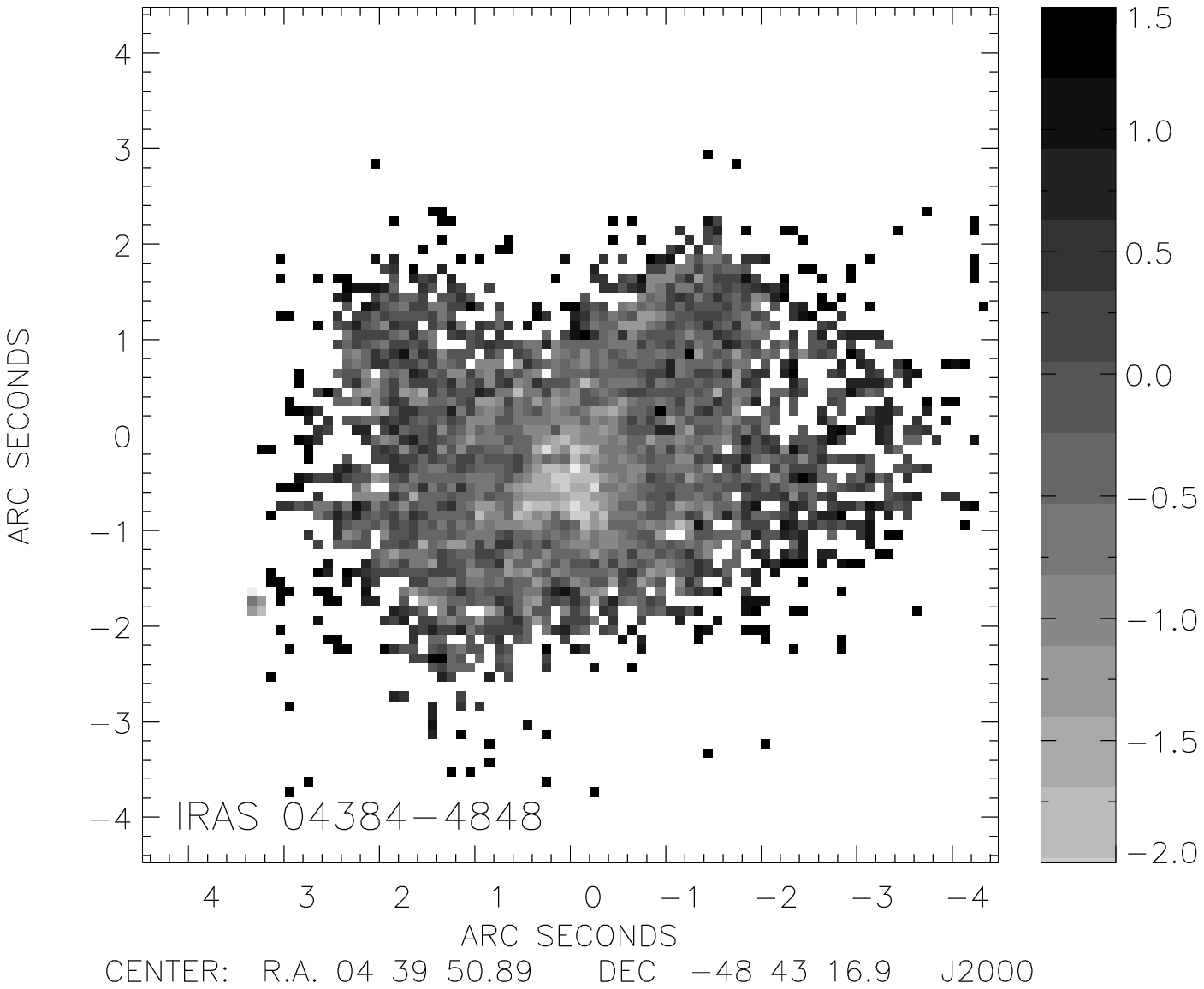,width=55mm}
\epsfig{figure=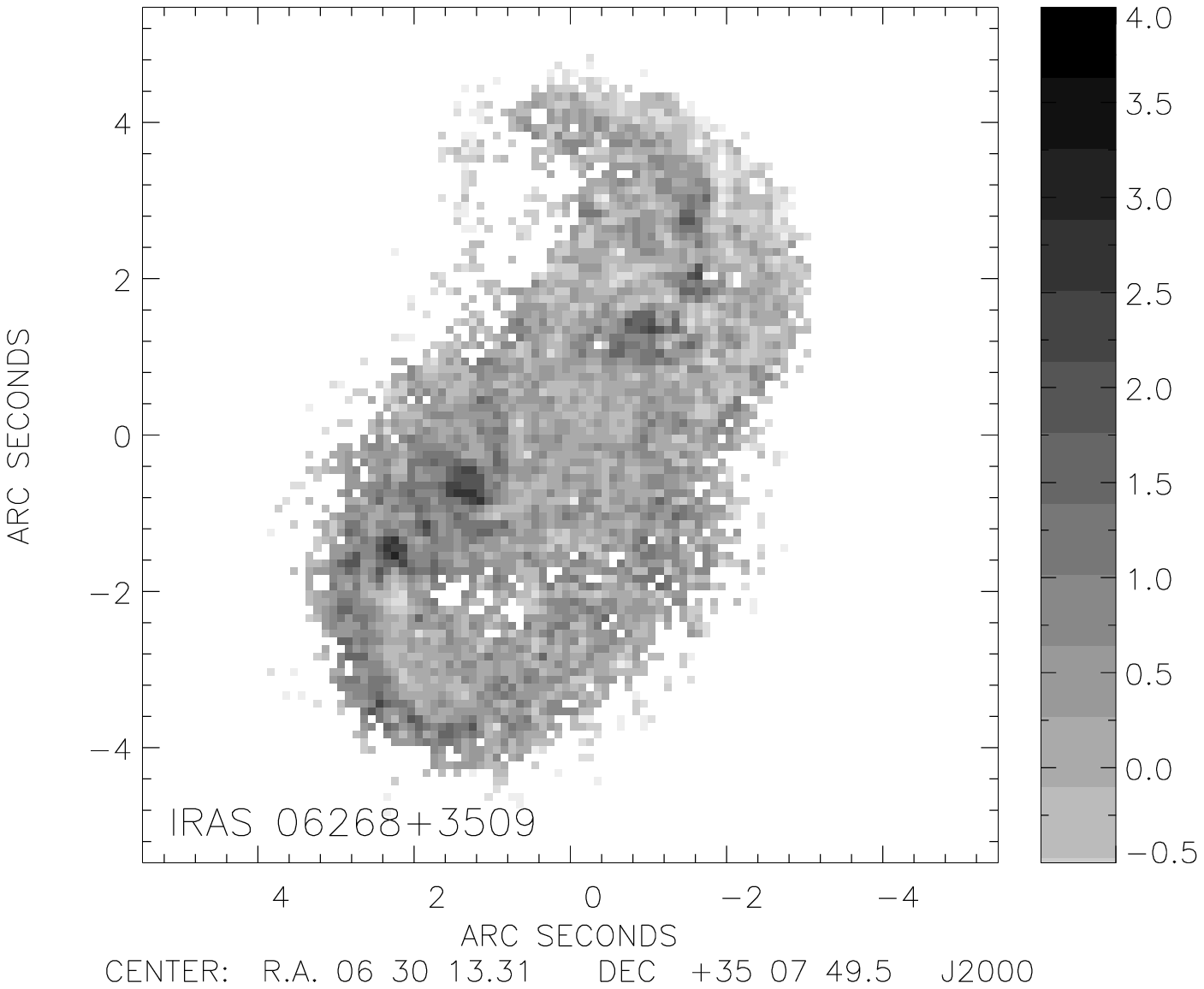,width=55mm}
\end{minipage}
\begin{minipage}{180mm}
\epsfig{figure=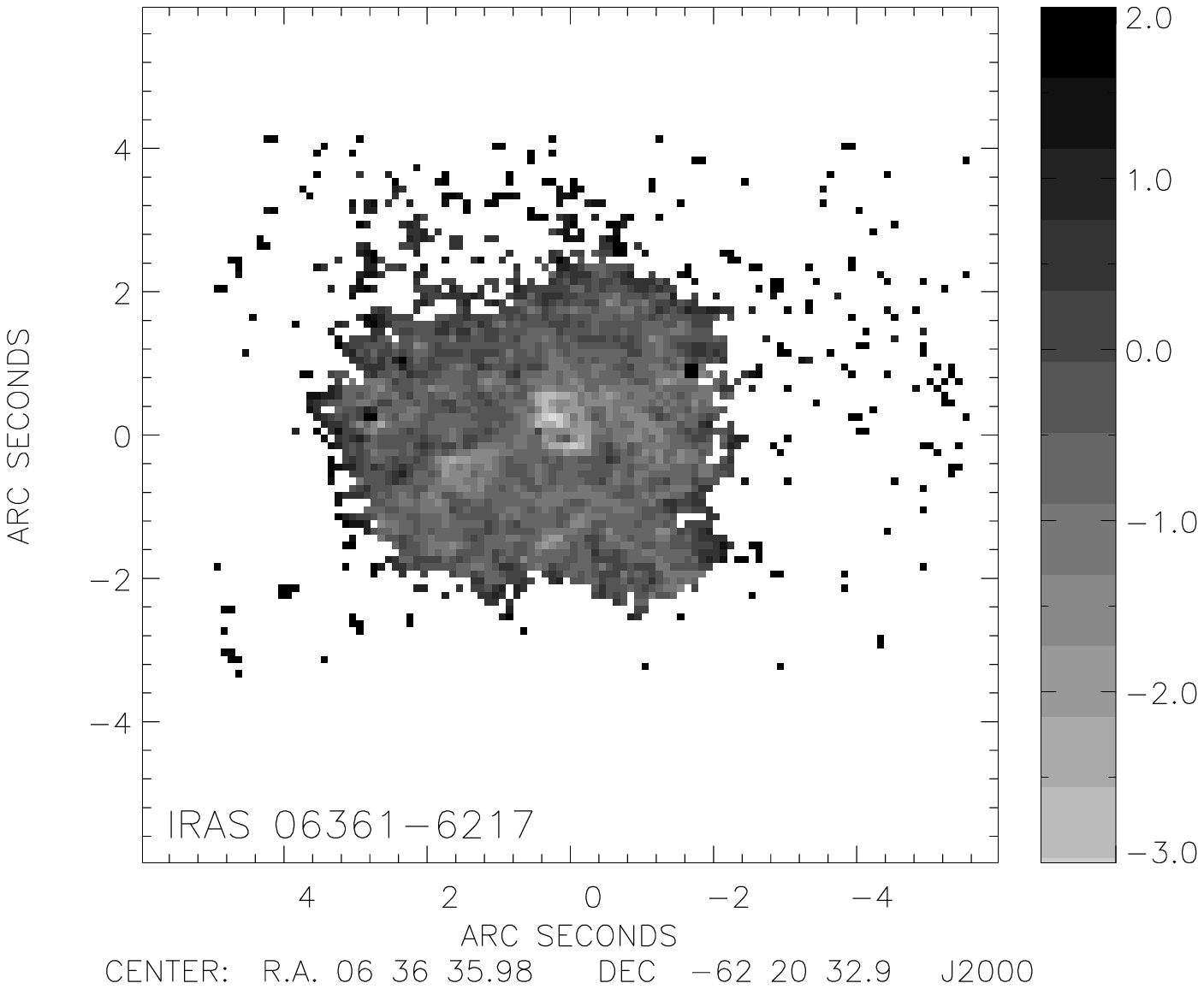,width=55mm}
\epsfig{figure=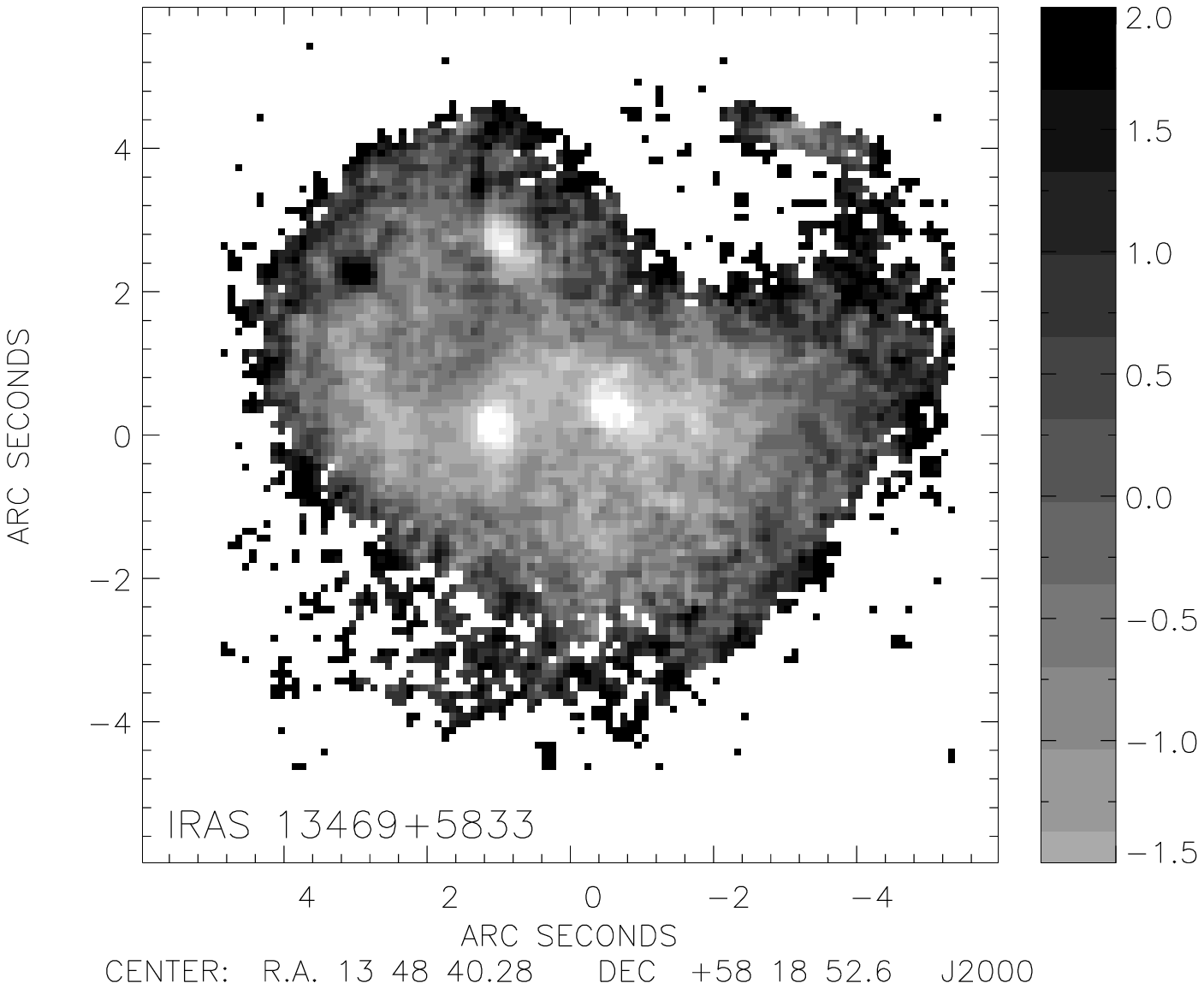,width=55mm}
\epsfig{figure=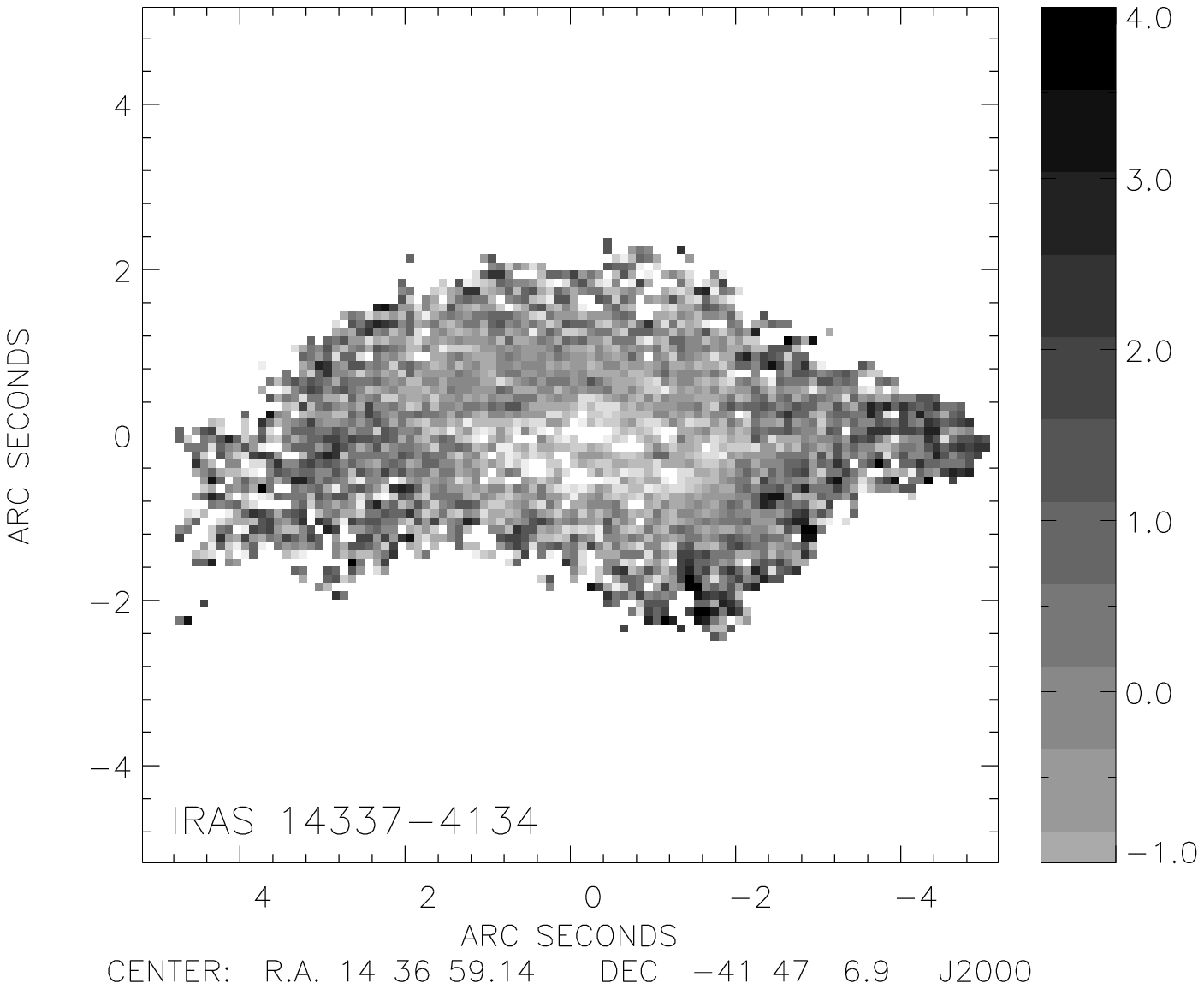,width=55mm}
\end{minipage}
\begin{minipage}{180mm}
\epsfig{figure=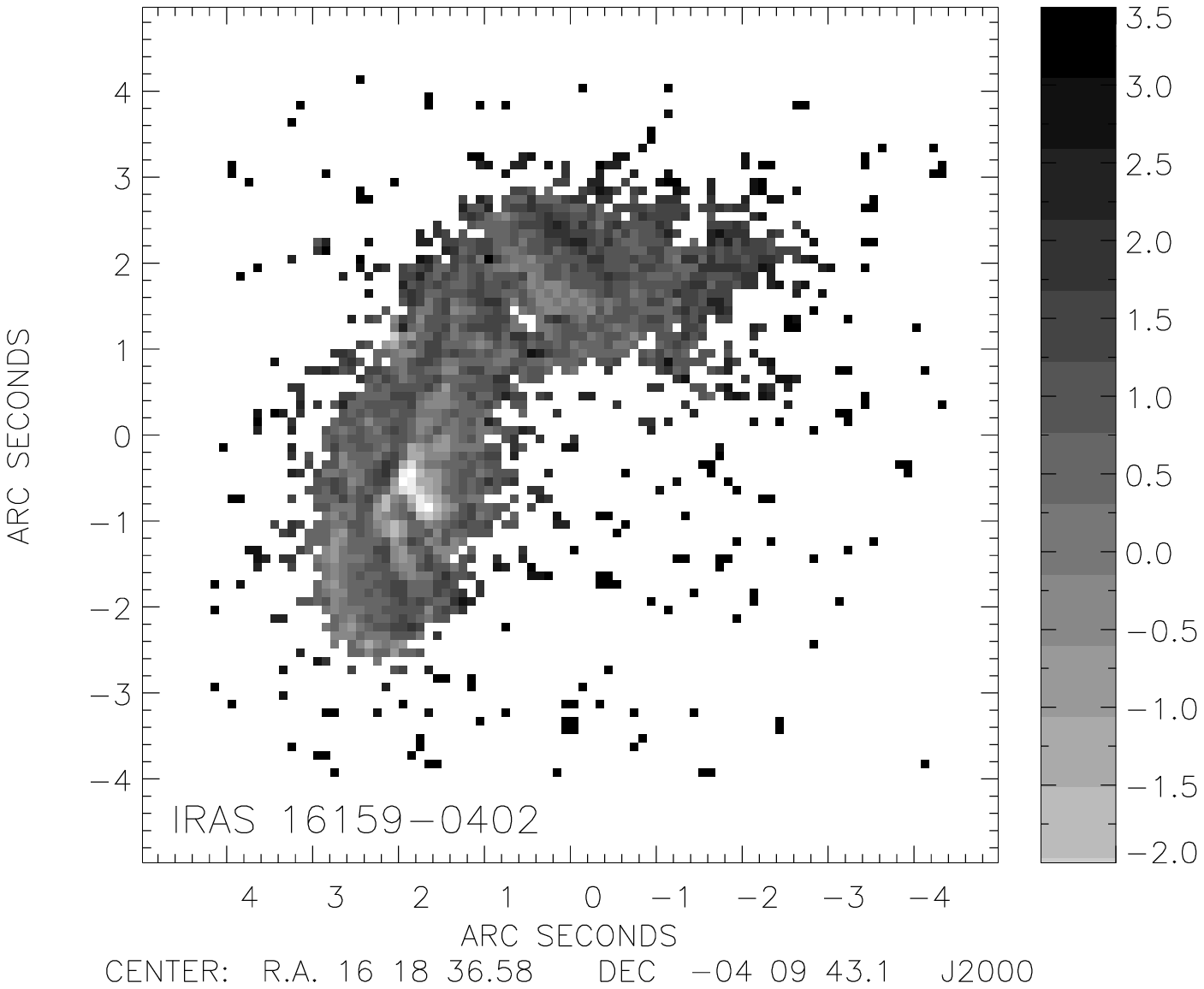,width=55mm}
\epsfig{figure=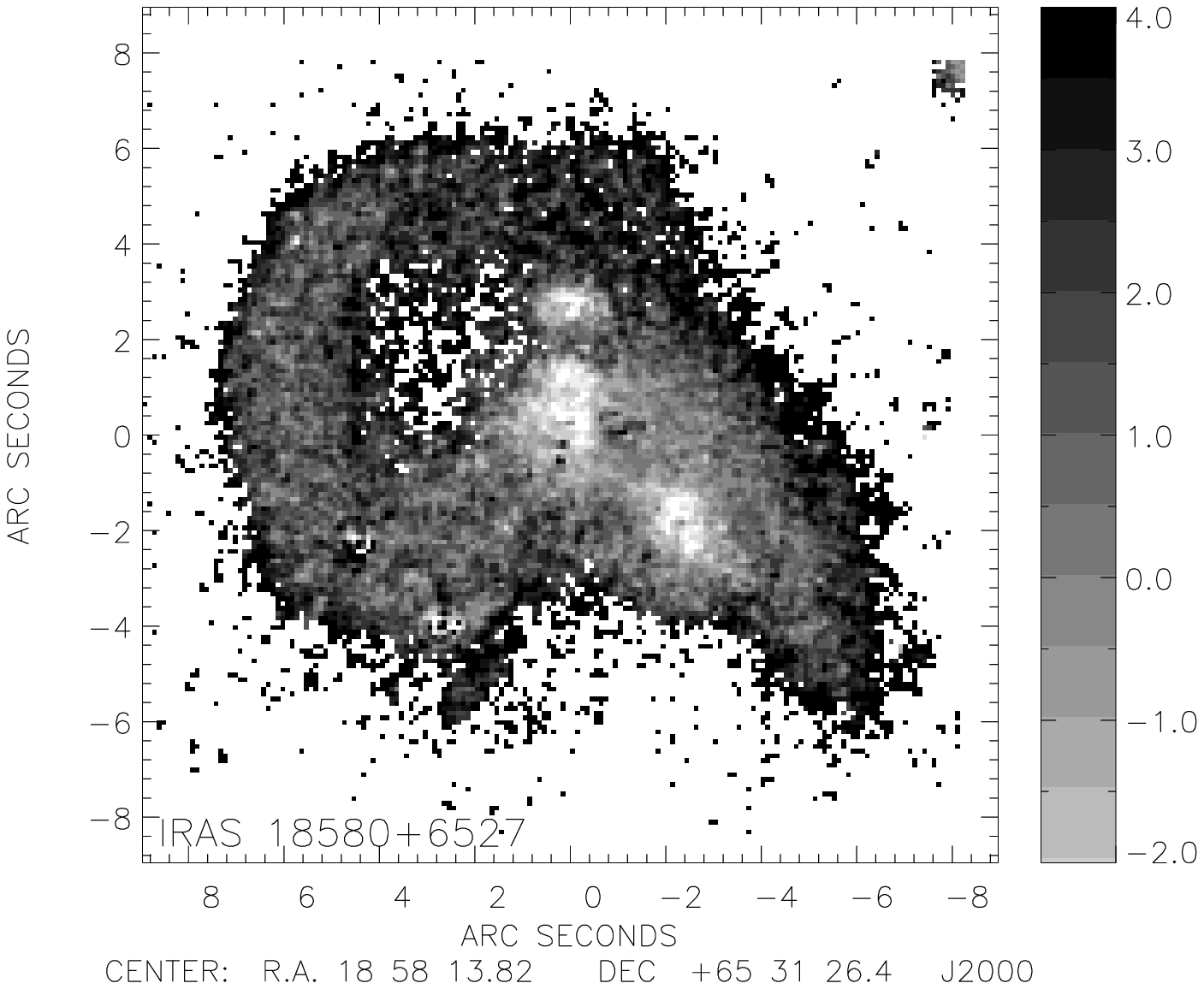,width=55mm}
\epsfig{figure=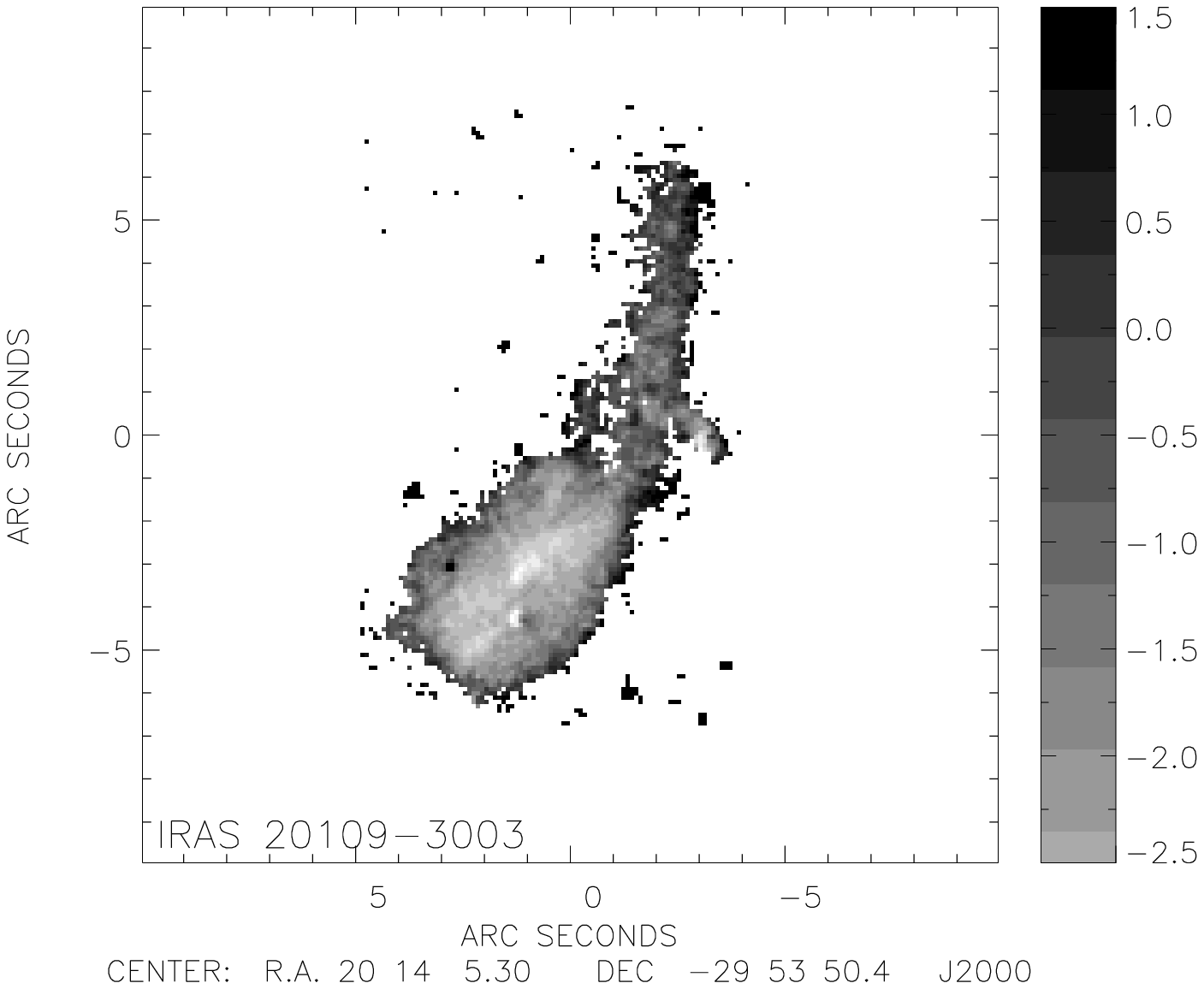,width=55mm}
\end{minipage}
\caption{$F606W-F814W$ colour maps for nine of the 22 sources. The colourbar gives the range in 
colour for each source. The orientation of each map is the same as the source image. The F606W 
band data has been degraded to F814W band resolution. The central region of IRAS 00275-2859 is 
aberrated due to the very bright central point source. \label{ulirgs_col_one}}
\end{figure*}

Colour maps have been constructed for those nine sources that show sufficient spatial extension in both the F606W and F814W 
band images, and are presented in Figure \ref{ulirgs_col_one}. Sky background regions, and those regions with low signal to noise, 
have been masked out. The slightly lower S/N in the F606W band means that excessively reddened (dark) regions with low S/N are 
masked off. This effect is most apparent in IRAS 18580+6527 and IRAS 20109-3003, where the maps become increasingly dark towards 
the edges of the objects before being masked off.

One object (IRAS 00275-2859) displays a smooth colour structure (discounting the slightly aberrated centre), perhaps indicating 
the presence 
of a uniform dust shroud or screen obscuring the central illuminating source. The other colour maps all display evidence for a large scale 
non-uniform colour structure which cannot be attributed to artifacts of amplified noise spikes in the deconvolved images. Two 
objects (IRAS 04384-4848 \& IRAS 14337-4134) show a central, extended, diffuse blue region. One object (IRAS 06268+3509) shows 
several red, compact regions coinciding with the centres of the two merging galaxies. We interpret these as heavily obscured 
star formation regions. The remaining sources show a number of blue or red compact `knots', in most cases located towards the 
central regions. The blue regions are most plausibly interpreted as areas of enhanced, unobscured star formation, and the red 
regions as either dust shrouded star formation or remnant nuclei from the merger progenitors. Care must be taken in interpreting 
such regions from imaging data alone. An alternative interpretation is that these regions are reflected light from an obscured AGN, 
as observed in Mrk 463 \cite{uom}. Although most ULIRGs ($\geq80\%$) are starburst dominated, AGN reflected light could still be 
a potential source of uncertainty in classifying knots as most starburst dominated ULIRGs have a significant AGN 
component, especially at higher luminosities. Another possibility could be shock fronts from an expanding superwind \cite{hek0}. 
Although a superwind might be expected to produce a systematic colour gradient as opposed to a point-like structure comparisons 
between locally observed superwinds \cite{hek0,wan} show that, at the mean redshift of our sample, a superwind could create a small 
structure only a few pixels ($\sim5$) across. A final possibility is that of small holes in locally obscured regions, caused 
by photodestruction of dust grains by hot young stars. These alternate possibilities are impossible to completely discount without 
spectroscopic data.

\subsection{Host Galaxies}
\begin{figure*}
\begin{minipage}{170mm}
\epsfig{figure=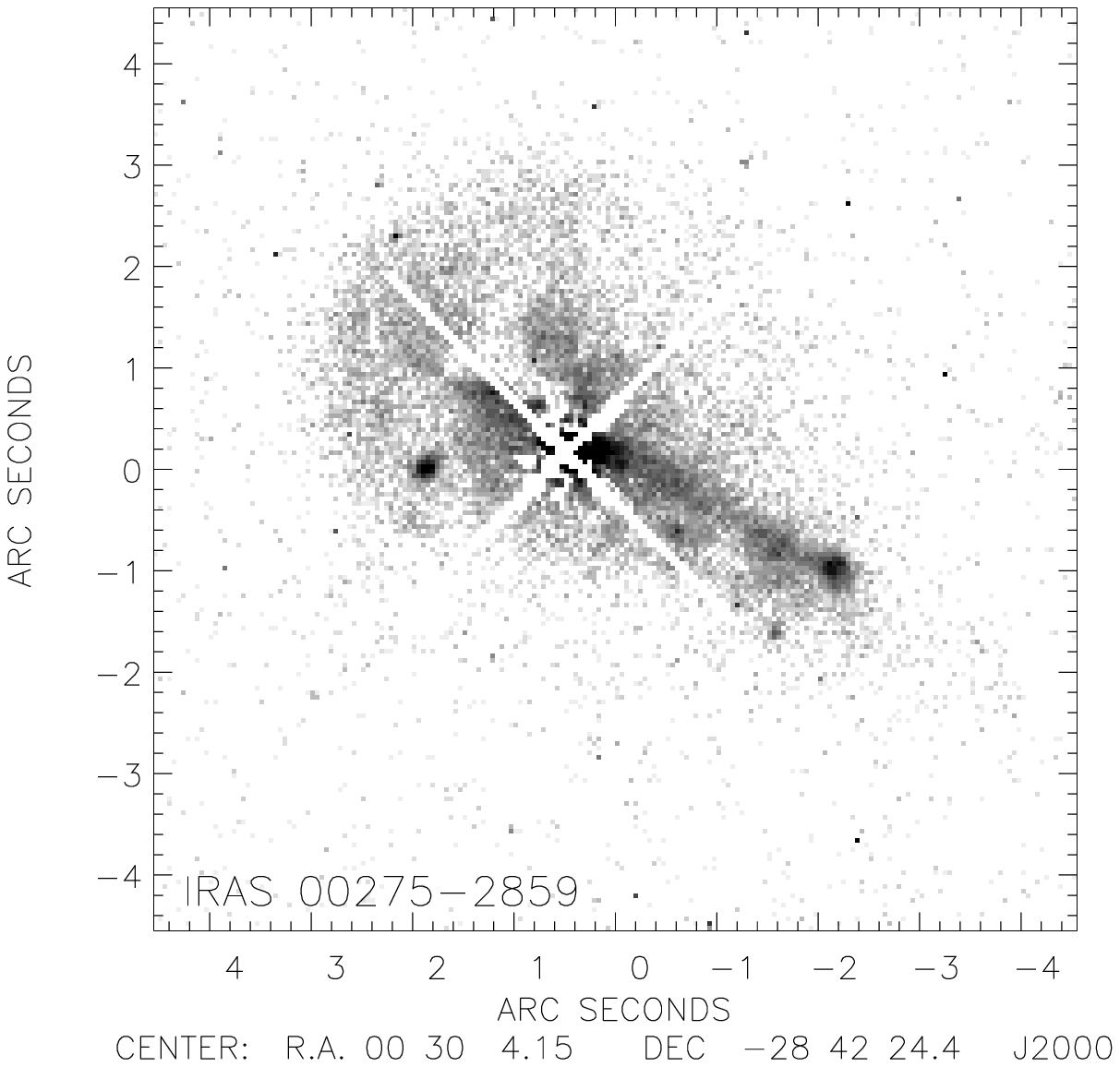,width=80mm}
\epsfig{figure=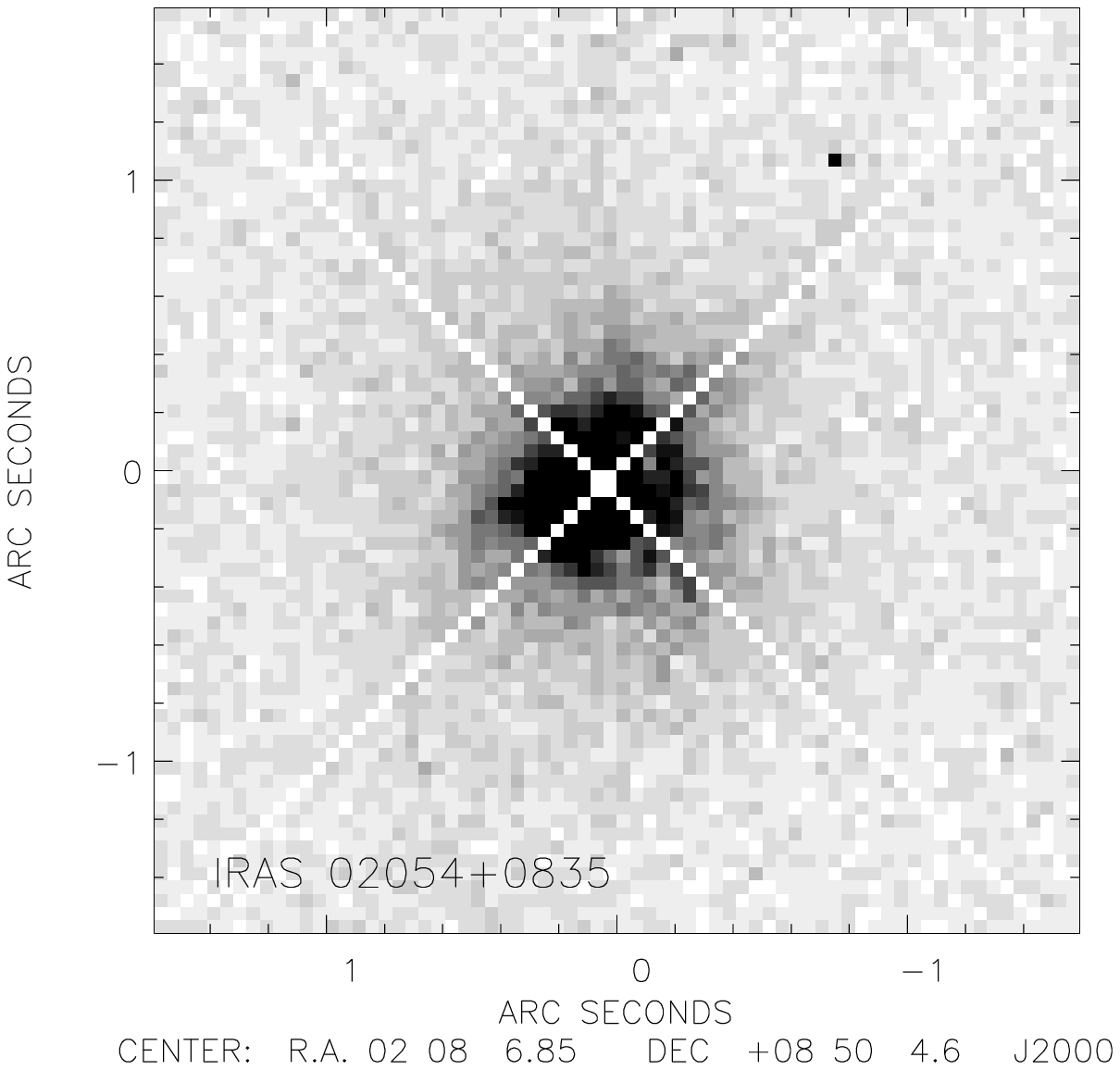,width=80mm}
\end{minipage}
\begin{minipage}{170mm}
\epsfig{figure=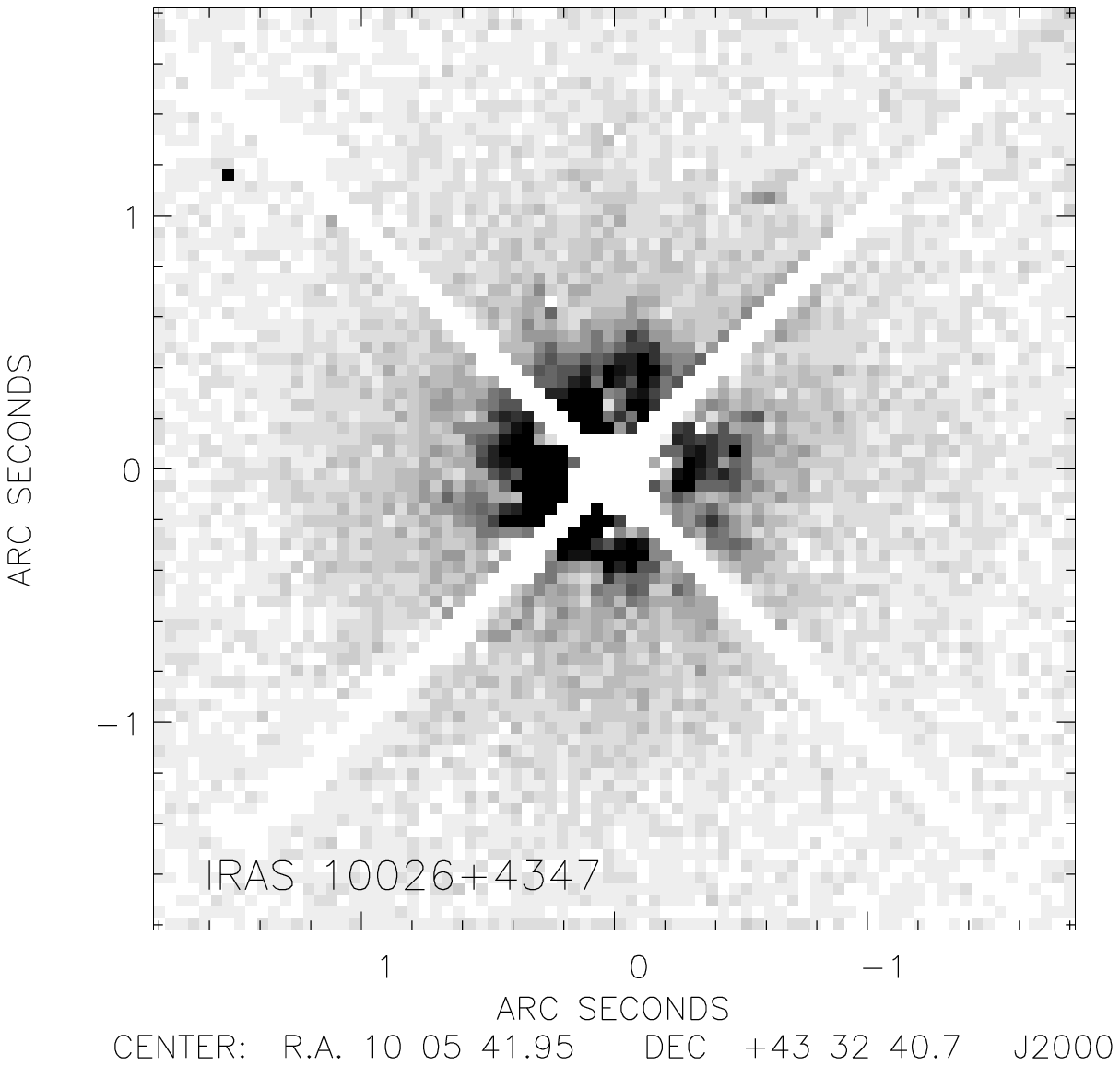,width=80mm}
\epsfig{figure=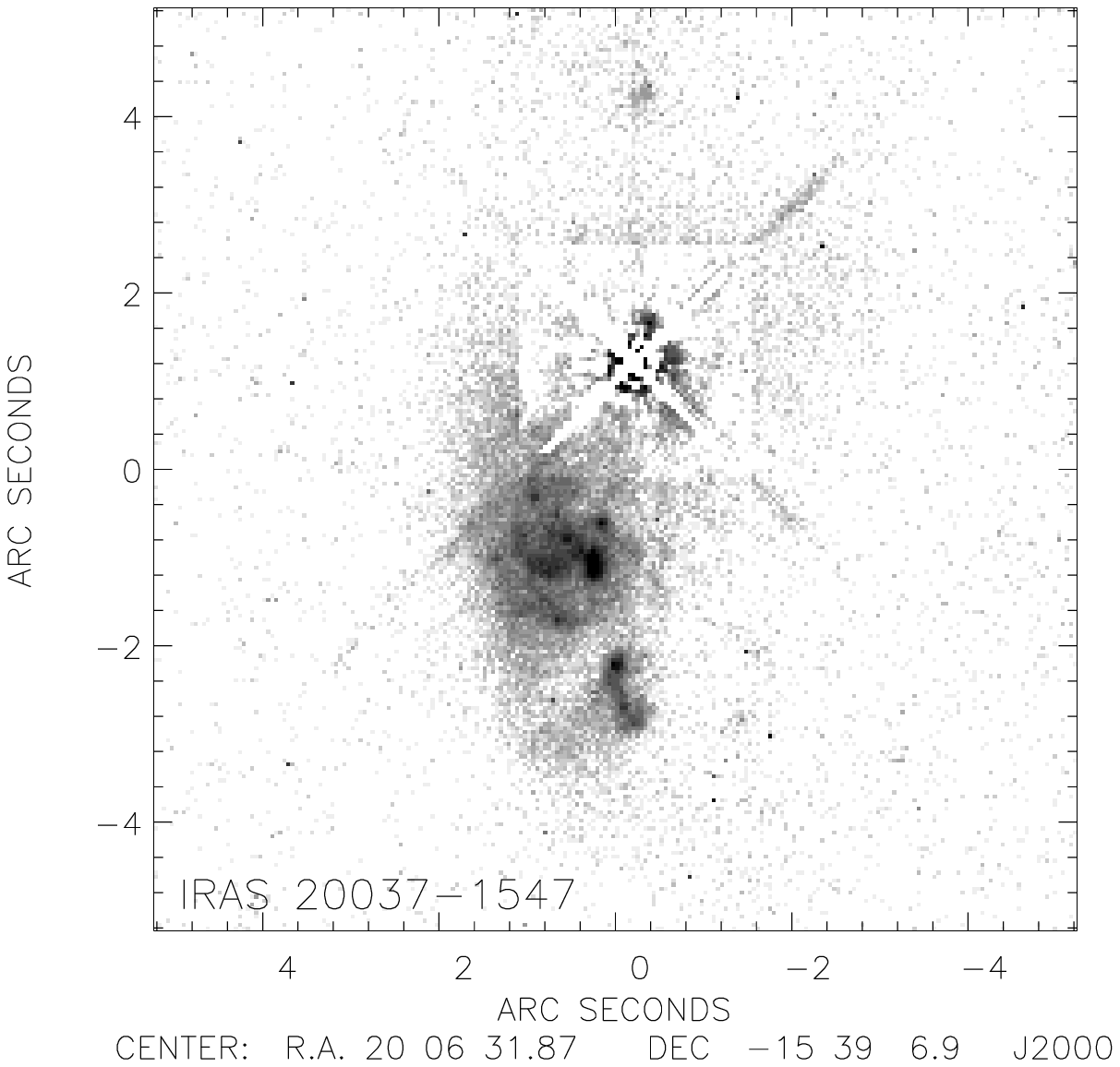,width=80mm}
\end{minipage}
\caption{{\em HST} F606W images of the ULIRG QSO host galaxies. Residual PSF features 
have been masked out. \label{ulirgs_im_host}}
\end{figure*}

Four of the objects in the sample contain some form of optical QSO. Host galaxy images for these four QSOs are presented in Figure 
\ref{ulirgs_im_host}. Data for the host galaxies are presented in Table \ref{ulirg_hosts}. 

Host galaxies were clearly detected in all four QSOs. For IRAS 20037-1547 \& IRAS 00275-2859 both spiral and elliptical light profiles 
were very poor fits to the underlying light distribution and the host galaxies are clearly interacting. For IRAS 02054+0835 \& IRAS 
10026+4347, the host galaxies were well fitted by an elliptical profile ($\chi^{2}\sim1.4$), with a spiral profile being a much poorer 
fit to the data ($\chi^{2}\sim4.0$). We investigated the possibility that we were merely fitting the bulges of spiral host galaxies by 
fitting simultaneously elliptical and spiral components in the central and outer regions respectively, the radius of each component 
being allowed to vary. In both cases it was found that a `pure' elliptical profile was strongly preferred. 

IRAS 02054+0835 has a host galaxy which displays no evidence for ongoing interactions. The ellipticity is low ($\epsilon\sim0.05$) and 
there is no isophotal twisting. The other resolved host galaxy (IRAS 10026+4347), although well fitted by an elliptical profile, also 
shows moderate disturbance. Although having only two resolved hosts means comparisons with other samples are necessarily limited, our 
host galaxy properties are similar to Early type galaxies in the Fornax and Virgo clusters \cite{cao}, and to Elliptical galaxies at 
higher redshifts \cite{kel}. 

Of the remaining 19 objects in the sample without an optical QSO, 17 are too disturbed to attempt fitting any form of galaxy profile. 
The remaining two objects (IRAS 17431-5157 \& IRAS 23140+0348) have a sufficiently regular morphology such that it might be expected that 
fitting galaxy profiles would yield informative results. Although one profile was clearly preferred in each case it was not possible to 
obtain a `good' fit for any profile. Reduced $\chi^{2}$ values were of the order 3 for the best fit in each case. We therefore conclude 
that these galaxies, despite being regular in appearance, have alight profile not consistent with a dynamically relaxed spiral or elliptical 
galaxy and are therefore disturbed.

\subsection{Individual Sources}

In the following descriptions, comments regarding {\em HST} images refer to the F606W Planetary Camera images.

{\em IRAS 00275-2859:} This IR luminous QSO was discovered by Vader and Simon (1987).
Clements et al \shortcite{cle2} calculate a dust temperature of 54K and a dust mass of $7.3\times10^{11}M_{\odot}$ 
for this object, assuming a linear emissivity - frequency relation and a monolithic dust population whose 
temperature is solely defined by the 60/100$\mu$m flux ratio. Zhenlong et al \shortcite{zou} classify this object 
as having a near ($<10\arcsec$) faint companion but not interacting. Clements et al \shortcite{cle1} classify this 
object as 'disturbed'. The {\em HST} image shows strong interactions and a double nucleus. The northeastern nucleus 
contains a QSO. There is also a spatially (but not luminally) symmetric halo extending about 2$\arcsec$. There is 
one very bright `knot' near the QSO, and a number of dimmer knots near the southwestern nucleus. Canalizo \& Stockton 
\shortcite{can} classify this second nucleus as a giant HII region rather than an AGN. The faint companion described 
by Zhenlong et al is revealed to be an undisturbed spiral.  

{\em IRAS 02054+0835:} A QSO with no signs of ongoing interaction. There are, however, five 
small, faint sources within 10$\arcsec$.

{\em IRAS 02587-6336:} Previous optical imaging \cite{zou} classified this system as a close pair with interaction. The 
{\em HST} image shows a binary system. Both sources show signs of disturbance, although there are no tails or loops and the 
sources appear still physically separate. The brightest source is a disturbed spiral, with a smaller source to the southwest.

{\em IRAS 04384-4848:} A small, disturbed source with no QSO activity. There are a number of compact bright regions and two 
faint tails.

{\em IRAS 06268+3509:} A pair of merging spirals. The northern spiral contains several compact bright `knots'. There are four 
further small sources within 12$\arcsec$.

{\em IRAS 06361-6217:} Zhenlong et al \shortcite{zou} classify this object as being non-interacting, but with a nearby 
companion. The {\em HST} image shows a source with a collisional ring galaxy morphology. The 
south eastern nucleus is very bright, with several nearby knots. A faint, elliptical halo surrounds these regions, with a 
semimajor axis length of about 8$\arcsec$. The western region is smaller and dimmer, with no bright knots. There is a 
small, faint companion 10$\arcsec$ to the north, and an object 15$\arcsec$ to the south which is either a QSO or 
foreground star.

{\em IRAS 06561+1902:} An interacting pair with a close companion. Although still physically distinct, the two interacting 
galaxies are connected by a faint bridge of nebulosity. This source is striking in that there is a similar group 
of three objects 10$\arcsec$ to the southwest, one of which is a QSO.

{\em IRAS 07381+3215:} A small, peculiar galaxy with some bright starforming regions. Notable are the two foreground stars 
15$\arcsec$ south east of the source.

{\em IRAS 10026+4347:} This object shows less than 1\% polarisation in the optical/UV, rapid X-ray variability and no  
optical absorption, indicative of a low quantity of dusty gas in the object and zero intrinsic polarisation, and probably 
indicative of a direct, unobscured line of sight to an AGN disk or nucleus \cite{gru}. The optical and X-ray properties of 
this object are discussed extensively by Xia et al \shortcite{xia}, who describe the source as a post-merger system with 
strong FeII emission. It shows a steep X-ray continuum slope, an X-ray luminosity comparable to that of Sy1's or QSO's, 
large rapid X-ray variability, and optical continuum variability. The intense FeII emission is cited as being 
the result of gas enrichment by supernovae. The {\em HST} image shows a QSO with no apparent signs of interaction. There are 
three very faint sources within 12$\arcsec$. 

{\em IRAS 10579+0438:} The HST image reveals a small source with two bright regions separated by about 1.5$\arcsec$, linked by 
a small tail. The western region is brighter, larger and more disturbed than the eastern region. The only object within 
30$\arcsec$ is a foreground star.

{\em IRAS 13469+5833:} Ground based optical imaging \cite{lch} showed an interacting 
system with giant tails. The {\em HST} image shows a large and complex morphology. There is a large, bright, circular 
region to the west with a radius of approximately 1$\arcsec$. To the east of this there is a 
cluster of smaller `knotlike' regions. There are two large tails on the east and west sides of the 
source. Large numbers of bright `knots' are also apparent throughout the source. There are two faint sources 
15$\arcsec$ to the west.

\begin{table}
\caption{QSO Host Galaxy Properties \label{ulirg_hosts}}
\begin{tabular}{@{}ccccc}
Source     &  $M_{V}$ & $r_{e}$ (Kpc) & $I_{e}$ & Notes \\ 
00275-2859 &    -     &      -        &    -    & Interacting          \\
02054+0835 &   -22.63 &   3.7         & 20.7    & Elliptical, low $\epsilon$ \\
10026+4347 &   -21.14 &   6.6         & 22.2    & Elliptical, disturbed      \\
20037-1547 &    -     &       -       &   -     & Interacting         \\
\end{tabular}

\medskip
Properties of the host galaxies of the 4 objects in the sample containing an optical QSO. 
Data for the two interacting hosts are not given.
\end{table}

{\em IRAS 14337-4134:} Zhenlong et al \shortcite{zou} identify this source as a close pair with a tidal tail, and 
classify it as interacting. The {\em HST} image shows a merging system with several bright regions and two small tidal tails. 
There are four bright sources within 25$\arcsec$.

{\em IRAS 16159-0402:} This interacting source shows a bright, peanut shaped region with
some small knots immediately to the north. Further north there is a larger, dimmer region. A tail or 
plume stretches south to an inclined disklike structure.

{\em IRAS 17431-5157:} A bright, disturbed core surrounded by a dim asymetric halo. There are 
a large number of other sources within the PC field of view.
   
{\em IRAS 18520-5048:} Zhenlong et al \shortcite{zou} describe this object as being non-interacting but having 
a close faint companion. Clements \& Baker \shortcite{cle3} however argue that this system in fact possesses 
two nuclei separated by 2.5$\arcsec$. The {\em HST} image reveals an interacting 
pair of spirals. The two spirals are similar in size and brightness, both having a number of small, bright 
'knots'.
 
{\em IRAS 18580+6527:} This source is cited as a possible multiple nucleus ULIRG \cite{aur}. Leech
et al \shortcite{lch} tentatively identify this source as a merger system interacting with a companion.
The western source has a Sy2 spectrum and the eastern source has an HII spectrum. The {\em HST} image shows a 
source that is strongly interacting. There is a 
large, bright region to the west of the source, with several smaller bright regions surrounding 
it. To the east there are two smaller bright regions,  together with a very large tail extending
approximately 10$\arcsec$ to the northwest. The western region appears brighter and more 
active than the eastern region.

{\em IRAS 20037-1547:} An interacting system containing a QSO. Remnants of one of 
the merger progenitors lie to the east. Physical merging appears to have only just started in this system.
   
{\em IRAS 20109-3003:} A pair of merging spirals. There are two compact, bright regions separated 
by about 1$\arcsec$, which can plausibly be identified as the galaxy nuclei. The northwestern region is both larger 
and brighter than its southwestern sounterpart. A long `tail' extends about 10$\arcsec$ to the west, and there are no 
companion sources.

{\em IRAS 20176-4756:} This source is difficult to interpret. The most plausible picture is a binary system consisting of a 
very compact double nucleus source and a slightly disturbed spiral with a small companion, separated by 30kpc. Under this 
interpretation, this source would be in the earliest stages of interaction. It is however not possible to draw solid conclusions 
without redshifts for both sources. The quoted redshift is for the spiral.

{\em IRAS 20253-3757:} An interacting pair with no nearby companions. The northern galaxy is a 
disturbed spiral, with arms extending to the north and south. A number of luminous 'knots' are apparent in the 
southern galaxy, extending east-west.

{\em IRAS 20507-5412:} This source is small, luminous and displays a number of compact bright regions. There are two bright 
sources within 15$\arcsec$, to the south east is a spiral and to the north east is a small QSO.

{\em IRAS 23140+0348:} Yee and Oke (1978) reported broad wings on both H$\alpha$ 
and H$\beta$ for this radio-loud source. An optical polarisation study \cite{dra} reveals two polarisation 
components, one intrinsic to the nucleus and the other restricted to 
elongated or `fan-like' areas. This implies that significant amounts of radiation are escaping 
at large angles to the radio axis. The core is described as having a compact steep radio spectrum 
with extensions to the east and west. The optical spectrum of the associated galaxy is characteristic 
of young A type stars. An intensity contour map of the source shows circular isophotes, whereas 
polarised intensity isophotes are very anisotropic, showing a 'fan' to the southwest and a slight 
enhancement to the northeast. The radio axis of the source does not coincide with any of the 
polarisation axes of the source. The flux from the radio core of this object is uncertain. Early 
studies \cite{ulv} do not agree with later VLA studies \cite{har}. Heckman et al \shortcite{hek} identify 
this source as a narrow line radio galaxy. Interestingly, this source is a member of one of the `compact 
groups' identified by Hickson (1982), these groups being characterised by possessing at least three 
members, having a high density and a low velocity dispersion. Members of such groups would therefore 
undergo almost continuous gravitational perturbation and could be candidate sites for triggering ULIRG 
activity \cite{bor}. The {\em HST} image shows a bright source with small, dim knots to the east and north. 
There is a dim asymmetric halo, extending towards the northwest. 

{\em IRAS 23220+2919:} A small source with a bright nucleus to the southeast and an extension to the southwest which 
probably contains a second smaller, dimmer nucleus. There are no companion sources within a 15$\arcsec$ radius.

\section{Discussion}

\subsection{Sample Morphology}

The large scale morphology of ULIRGs has been studied by many authors, often with conflicting 
results. Sanders et al \shortcite{san2} found that 100\% of a sample of 10 ULIRGs showed signs 
of interaction. Lawrence et al \shortcite{lawr} found that, although 19/41 {\em IRAS} galaxies 
with $L_{60} > 10^{11}L_{\odot}$ were interacting (where $L_{60}$ is the $60\mu$m luminosity 
computed from the {\em IRAS} $60\mu$m flux), only 2/6 {\em IRAS} galaxies with 
$L_{60} > 10^{12}L_{\odot}$ were interacting. Near IR imaging of a sample of nine ULIRGs \cite{car} 
showed that four of the sample possessed double nuclei with high activity levels, and they estimated 
the mean starburst lifetime to be $4 \times 10^{8}$ years. {\em VLA} observations of two complete 
samples of ULIRGs \cite{sop} found several binary starbursts, and that starburst activity can be 
triggered in both, either or neither galaxy in an interacting pair. Zhenlong et al \shortcite{zou} 
derived an interacting fraction of 61\% for their sample of 41 ULIRGs. Leech et al \shortcite{lch} 
imaged 35 ULIRGs and found that 23/35 showed signs of interaction. Their results also revised the 
interaction fraction quoted by Lawrence et al \shortcite{lawr} from 2/6 to 4/6. Clements et al 
\shortcite{cle1} derive an interacting fraction of 90\% for their sample of 60 ULIRGs. They explain 
the discrepancy between their results and previous work by saying that deep optical imaging is 
needed to detect faint signs of interactions and that previous studies may have missed such signs. 
Observations of the {\em IRAS} 1Jy sample of ULIRGs \cite{kim2} found a luminosity function of the 
form $\Phi(L)\propto L^{-2.35}$, strong density evolution with $\alpha = 7.6\pm3.2$, and no evidence 
for clustering. Murphy et al \shortcite{mur}, using optical and near-IR imaging of a sample of 56 
ULIRGs, quote 95\% as involved in current or recent interactions. Surace et al \shortcite{sur} 
performed deep {\em B} and {\em I} band imaging using {\em HST} for a sample of nine `warm'  ULIRGs 
and found that all showed signs of ongoing interactions. They also found that most of their sample 
contained a number of compact blue `knots' of star formation that do not contribute significantly 
to the bolometric emission, and that a small number of these knots were plausibly active nuclei. 
Overall they found that their results were consistent with `warm' ULIRGs representing a critical 
transition phase between a galaxy merger and a QSO. A complementary study of a sample of `cool' 
ULIRGs \cite{sur2} found that all of the sample were either advanced mergers, or early stage mergers 
with evidence for separate ($>600$pc) nuclei. Emission from the central regions was, with one exception, 
consistent with high rates of star formation. Murphy et al \shortcite{mur2}, using integral field 
spectroscopy of a sample of 4 ULIRGs, propose that the ULIRG phase may be bimodal in time with some 
ULIRGs being early stage mergers and some being late stage mergers. They further propose that multiple 
starburst events may take place during the ULIRG phase and that some LIRGs may be `resting' ULIRGs.

The number of sources observed to be interacting in our sample of ULIRGs is 87\%, with a lower bound 
of 74\%. These results are in excellent agreement with the more recent ULIRG morphology studies using 
deep, high resolution ground and space based imaging, although care must be taken in comparing this 
fraction to lower resolution samples. Five of the sources are binary systems where the progenitors have 
yet to coalesce. The merging systems show a wide range of morphological features and asymmetries, most 
containing at least one compact luminous `knot'. Of the four QSOs two (IRAS 02054+0835 \& IRAS 10026+4347) 
show no signs of recent interaction. The other two QSOs are still interacting. In one case (IRAS 00275-2859) 
the merger is advanced, whereas in the other (IRAS 20037-1547) the QSO and galaxy have only just started 
to physically merge.

The Sanders et al picture of ULIRG evolution \cite{san2} postulates that ULIRGs are the dust shrouded 
precursors to optically selected QSOs. According to this picture (hereafter referred to as the S88 picture), 
interactions and mergers between gas rich spirals transport gas to the central regions of the galaxies. 
This large central gas concentration triggers starburst activity, and in the latter stages of the merger 
creates or commences the fuelling of a central active supermassive black hole. This central black hole rapidly 
comes to dominate the IR luminosity of the ULIRG. In the last stages the dust screen shrouding the black hole 
is blown away and the ULIRG evolves into an optical QSO.

The sample presented here provides excellent evidence that ULIRGs are strongly linked with both galaxy 
interactions and (to a lesser extent) QSO activity. That QSO activity is physically connected with ULIRGs 
is clearly demonstrated by the number of optical QSOs in the sample. If the two phenomena were unconnected 
then the number of objects in the comoving volume out to $z=0.4$ which are by chance both ULIRGs and QSOs 
is 0.1 \cite{can}. The sample includes still physically separate but 
interacting spirals, violent mergers with large amounts of associated star formation, QSOs with signs of 
ongoing interaction, and QSOs with no apparent signs of interaction. Those sources that are classified 
as isolated and non-interacting according to the system of Lawrence et al are all QSOs. 

Looking solely at the optical morphologies of the sample it is, however, far from clear that the S88 picture 
describes the sample well. It is not possible to rule out 
from this data alone that, for the two QSOs with no signs of ongoing interaction, the IR emission was 
not triggered by interactions, a counterpart of which exists for the HLIRG population as well \cite{far}. 
In two sources, a QSO is coeval with galaxy interactions. If it is assumed that the e-folding timescale of 
an Eddington luminosity black hole with accretion efficiency $\epsilon=0.1$ is a good estimate for the 
lifetime of a QSO \cite{ree} then it is entirely plausible that for these two sources the QSO will burn 
out before the merger is completed and the system is dynamically relaxed. A straightforward interpretation 
of all ULIRGs being a transition stage between galaxy mergers and QSOs is not supported by this data.

\begin{figure}
\rotatebox{90}{
\centering{
\scalebox{0.35}{
\includegraphics*[12,40][514,724]{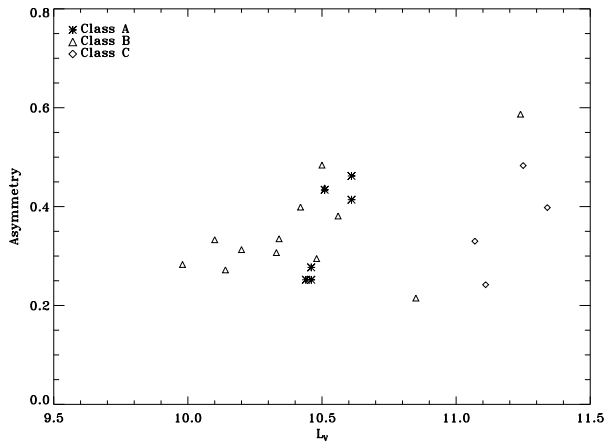}
}}}
\caption
{
Luminosity in the V band, $L_{V}$, versus asymmetry factor (after subtraction
of QSO in case of sources shown in Fig \ref{ulirgs_im_host}).  The different symbols
correspond to the classification of section 4.1 (class A - two
separate sources; class B - interacting galaxies; class C  - QSO).
 \label{mag_asym}
}
\end{figure}

Previous studies have also highlighted potential problems with the S88 picture. Mid-IR spectroscopy of  
62 ULIRGs \cite{rig} found that the majority were powered by starbursts, and that most did not contain a buried QSO. 
There was no evidence that advanced mergers are more AGN-like, based on the ULIRG nuclear separations. 
That AGN activity becomes more prevalent with merger advancement is a natural prediction of the S88 picture. A mid-IR 
spectroscopic survey of 15 ULIRGs \cite{gen} found that $\sim80\%$ were powered predominantly by starbursts and that $\sim20\%$ 
were powered by AGN, although at least half the sample showed clear evidence for both types of activity. There was no detected 
trend for the AGN-like systems to reside in the more compact (and hence more advanced merger) systems.  
Canalizo \& Stockton \shortcite{can}, in a study of a sample of 9 QSOs with FIR colours intermediate between ULIRGs and QSOs, 
proposed a model in which some ULIRGs evolve to become classical QSOs in a short time ($\leq300$Myr) and that some QSOs are 
born under the same conditions as ULIRGs but with a very short ($\leq300$Myr) lifetime. They further propose that the QSO 
nuclear regions are initially shrouded in a dust cocoon, an idea qualitatively similar to the S88 picture.

\begin{figure}
\rotatebox{90}{
\centering{
\scalebox{0.36}{
\includegraphics*[12,40][514,724]{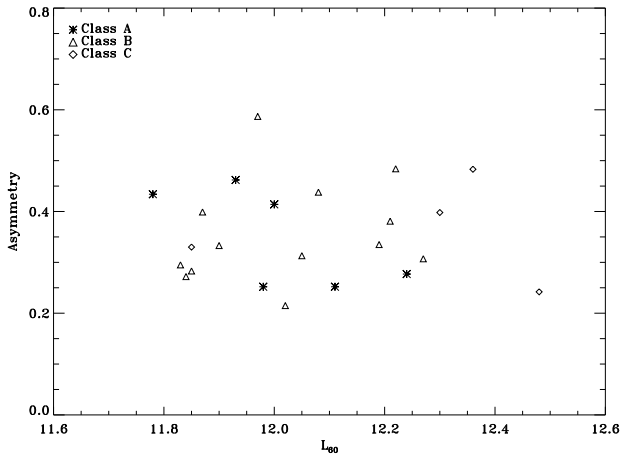}
}}}
\caption
{
60 micron luminosity plotted against Asymmetry for all 23 ULIRGs. Description of the classes A, B and C can be found in \S4.1
 \label{l60_asym}
}
\end{figure}

Our data does not allow us to discriminate between IR emission due to starbursts and that due to AGN activity. We can, however, 
examine this topic in alternative ways. If the IR luminosities of the ULIRGs with a QSO are compared to the IR luminosities of 
the ULIRGs without a QSO via a Kolmogorov-Smirnov (K-S) test, the the probability $P$ that the IR luminosities of these two 
populations are drawn from the same distribution is found to be $P = 0.0411$. Thus it seems very likely that there is a difference 
in the distributions of the IR emission between those sources with a QSO and those without a QSO (as also found by numerous previous 
studies), consistent with some form of evolution between a starburst and AGN phase. A plot of $L_{V}$ vs. asymmetry statistic $A$ 
(Equation \ref{eqn:fasym}) can be found in Figure \ref{mag_asym}. The sources that contain a QSO have 
comparable asymmetries to the non-QSOs in the sample, with $\langle A \rangle = 0.36$ for the QSOs as opposed to 
$\langle A \rangle = 0.35$ for the non-QSOs. This suggests that QSOs (and hence AGN) are not found in more advanced mergers, 
a result in direct contradiction of the S88 picture and indirectly supportive of Rigopoulou et al \shortcite{rig}. A plot of 
$L_{60}$ vs. $A$ can be found in Figure \ref{l60_asym}. Amongst those sources with no QSO there is no correlation between 
asymmetry and IR luminosity. Both class A and class B objects are found across a wide range of IR luminosities with no 
apparent trends. There is some support for the idea that ULIRGs become more AGN-like at the highest IR luminosities 
\cite{vei,shi} as three of the four most IR luminous objects in the sample are QSOs.

We performed simulations to assess the impact of both large scale features such as tidal tails, and small scale features such as 
starforming knots, on our derived asymmetry values. It was found that, although both large and small scale features contributed 
to the derived asymmetry values, the asymmetries were determined predominantly by the large scale morphologies. The 
asymmetry values are therefore an accurate measure of the degree of disturbance in our sample, rather than solely a measure of the 
number of starforming knots or other small scale structures.

Unfortunately there are no other galaxy samples for which direct comparisons between asymmetry values can be made. The closest 
samples available are the optical starburst sample of Brinchmann et al \shortcite{brin} and the sub-ultraluminous radio selected 
starbursts of Serjeant et al \shortcite{serj}, both of which lie in a similar redshift range, 
and were imaged with the Planetary Camera on {\em HST} but using the F814W filter. Although the difference in filters means that 
different stellar populations are being sampled, the closeness in wavelength between the F606W and F814W filters 
means that this difference should not significantly affect large scale morphologies. After accounting for 
the extra factor of two in our version of the asymmetry statistic, our computed values are very similar to those computed by 
Serjeant et al \shortcite{serj}, with a slightly higher mean value. They are much higher than those values presented by Brinchmann 
et al \shortcite{brin}. This implies increasing asymmetry with increasing star formation rate. 

The images presented here have demonstrated the efficacy of {\em HST} for detecting signs of interaction where lower resolution 
ground based images have either shown no signs of interaction or have shown ambiguous results. Three sources (IRAS 00275-2859, 
IRAS 06361-6217 \& IRAS 18520-5048) have, using ground based imaging, been classified as non-interacting or ambiguous. The very 
high resolution {\em HST} images of these sources have however shown clear signs of interaction, implying that ground based 
surveys of ULIRGs may underestimate the interacting fraction by 15\% or more.

Studies of samples of `warm' and `cool' ULIRGs have, quite recently, become active areas of research. {\em HST} imaging of a sample of 
warm ULIRGs \cite{sur} revealed that all the sources lay in advanced merger systems. Imaging of a complementary sample of cool 
ULIRGs \cite{sur2} revealed that all of the sample were either advanced mergers or `premergers' where the galaxies were still separate.
It has been suggested that warm ULIRGs may represent an important transition stage between cool ULIRGs and QSOs. Two of the three 
sources in our sample with warm colours are QSOs, consistent with warm ULIRGs containing a higher fraction of QSOs than cool ULIRGs. 
The cool sample however contains one QSO with ongoing interactions. The mean asymmetry for the warm ULIRGs in our sample is 0.404, 
whereas the mean asymmetry for the cool ULIRGs is lower, at 0.376. Although our sample supports QSO activity being more prevalent in warm 
ULIRGs than in cool ones, our data does not support the hypothesis that warm ULIRGs are more advanced mergers than cool ULIRGs.

Comparison between the colour maps and the F606W band images shows that, in most cases, bright regions in the F606W band images 
correspond with very blue or very red regions in the colour maps. For three objects however (IRAS 13469+5833, IRAS 04384-4848 \& IRAS 14337-4134) 
this is not the case. These objects show blue regions that do not coincide with bright regions in the F606W band 
images or the F814W band images. As these regions are located towards the centres of the sources, this result can be interpreted 
as a generally slightly higher star formation rate, a local underdensity of dust, or a local IMF skewed towards high mass stars in 
what are probably the most turbulent and gas rich regions of ULIRGs. These optically bluer regions may also therefore correspond to sites of 
buried starbursts and/or AGN.

\subsection{Host Galaxies}

Previous surveys of QSOs using {\em HST} have generally been extremely successful in detecting host galaxies. 
Urry et al \shortcite{ur} give results for a sample of BL Lacertae objects, detecting host galaxies to a redshift of 
$z = 0.7$. They find that the hosts are all extremely luminous ellipticals with $\langle M_{I} \rangle = -24.6$. The derived 
ellipticities are generally low and morphologies are smooth. Imaging of a sample of 14 quasars (6 radio loud, 5 
radio quiet, 3 radio quiet ultraluminous infrared quasars) by Boyce et al \shortcite{bo} showed that all the radio loud 
quasars had elliptical hosts, the radio quiet quasars possessed both elliptical and spiral hosts, and that the 
radio quiet ultraluminous infrared quasars lay in violently interacting systems. Host galaxies were on average 
0.8 mag brighter than $L^{*}$. Results from imaging of two luminous radio 
quiet quasars \cite{ba1} show that both quasars reside in apparently normal host galaxies, one being an elliptical and 
one a spiral. Disney et al \shortcite{di} present {\em HST} imaging for a sample of four QSOs, finding that all four have elliptical 
hosts. Although they find that the morphologies of the hosts are all featureless, they argue that the presence of 
multiple very close companions is evidence for interactions as the trigger for the quasar activity.

We resolved host galaxy morphologies in only two of the four QSOs in the sample, finding one to be a smooth elliptical and one 
to be an elliptical with signs of disturbance. The hosts of the other two sources were detected at $>99$\% confidence  
but it proved impossible to fit an elliptical or spiral profile to the underlying light distributions. The PSF subtracted images 
show these hosts to be interacting. Although discussion based on only four sources can only be very limited, these results do not 
contradict the S88 picture, and are also supportive of a link between merging spirals and emerging ellipticals.

\subsection{Optical Colours}

\begin{figure*}
\rotatebox{90}{
\centering{
\scalebox{0.75}{
\includegraphics*[7,7][514,724]{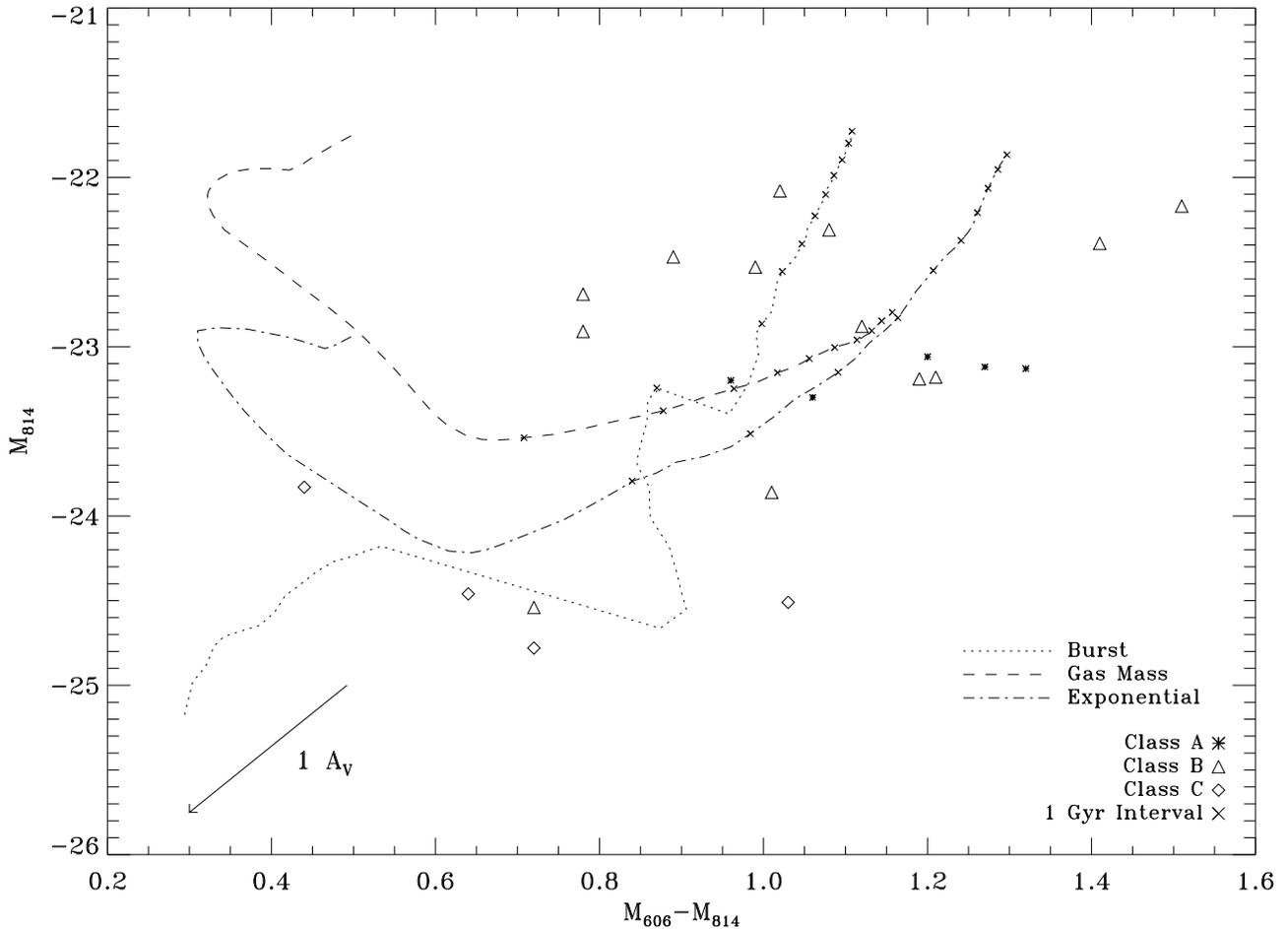}
}}}
\caption{Colour-Magnitude diagram for the QDOT ULIRGs. Evolutionary tracks are for a Salpeter IMF and run from 10Myr to 10Gyr. The 
dereddening vector is shown in the bottom left. Time direction for the tracks is L-R. `Burst' - instantaneous burst of star formation 
at $t=0$. `Exponential' - star formation law of the form $SFR=e^{(-t/1000)}/1000$. `Gas Mass' - $SFR= M_{gas}/3000$. The class C 
points are not QSO subtracted, and are plotted for comparison only. \label{colmag}}
\end{figure*}

A colour magnitude diagram ($M_{606} - M_{814}$ vs. $M_{814}$) is presented in Figure \ref{colmag}. The average F814W band 
magnitude of those sources which do not contain a QSO is $M_{814} = -22.9$. This is 0.3 magnitudes less bright than the 
field elliptical galaxy magnitude in the Cousins {\em I} band, $M^{*}_{I} = -23.2$ \cite{poz,ef}. Still physically 
separate but interacting systems (class A) have a similar optical magnitude to those systems that are physically merging 
(class B), with the class A objects having an average magnitude of $M_{814} = -22.93$ and the class B objects having 
$M_{814} = -22.86$. The QSOs (class C) are much brighter than the field galaxy magnitude, with $M_{814} = -24.4$, and 
are plotted for comparison only. Similarly, the two objects without an optical QSO but with broad lines (18580+6527 \& 
23140+0348) are plotted only for comparison.

Evolutionary tracks for three star formation scenarios have been plotted to estimate the contribution of stellar populations 
of various ages to the optical emission from the class A/B objects. These tracks were computed using the P{\sevensize EGASE} 
package \cite{fio}, suitably modified to produce output in the {\em HST} flight filter system. An initial mass normalization 
was selected by assuming that the Milky Way is a suitable template. The total mass of the Milky Way is currently thought to 
be about $10^{12}M_{\odot}$ \cite{wil}. A rough estimate for the mass of stars for two merger progenitors (excluding any 
dark matter component) is therefore about $2\times10^{11}M_{\odot}$. Dust extinction effects are included in the models for 
a spherically symmetric system and are computed individually for each timestep. The calculated optical depth varies between 
$0 < \tau < 35$. 

The $A_{V}=1$ dereddening vector, plotted on the lower left of Figure \ref{colmag}, lies almost parallel to the three 
evolutionary tracks. Coupled with the uncertainty in the mass normalisation, this means that only upper limits can be 
set on the ages of the stellar populations.Similarly it is not possible to draw conclusions about which star formation 
scenario is applicable, or to construct more precise constraints on the ages of the stellar populations, as data is only 
available in two bands which are quite close in wavelength. These limitations considered, the optical emission from nearly 
all the class A/B ULIRGs is best described by stellar populations of several Gyr or less in age. The most likely 
interpretation is that the optical emission is dominated by old stellar populations from the merger progenitors, a result 
consistent with previous studies \cite{sur}, although a heavily dust shrouded starburst cannot be ruled out.

\subsection{Pair Mergers vs. Multiple Mergers}

The number of objects involved in a typical ultraluminous galaxy merger can have a fundamental impact on both the star formation 
rate and (to a lesser extent) the merger morphology. Although the physical mechanisms behind starburst and AGN activity are likely 
to be no different in pair mergers and in multiple mergers, the fraction of ULIRGs in multiple merger systems can give a good estimate 
of the fraction of ULIRGs in compact groups of galaxies \cite{bek}, as mergers between more than two galaxies are expected to occur 
naturally in compact groups \cite{bar,hic2}. Numerical simulations of a merger between five disk galaxies \cite{bek} showed that 
repetitive massive starbursts are characteristic of multiple galaxy mergers, in contrast to pair mergers where only one or two 
starbursts occur. Multiple galaxy merging has been previously linked to Arp 220 \cite{dia}. Models for multiple galaxy mergers 
\cite{ta0} propose that the existence of more than two AGN or starburst nuclei is evidence for a multiple merger history. 
It has been suggested \cite{bor} that Hickson Compact Groups, where multiple mergers may be expected to take place, may be the 
progenitors of ULIRGs. It has also been suggested \cite{mur2} that multiple starbursts, as may be expected to occur in mergers 
between more than two galaxies, may be common in ULIRGs.

For those objects in our sample where the progenitor galaxies are no longer distinct (class B/C) it is difficult to draw reliable 
conclusions about the number of merger progenitors, as the complex geometries of gas and dust in these systems can easily give rise 
to multiple bright regions. Even using optical spectroscopy to identify separate AGN regions as a method for counting the progenitors 
could be ambiguous. AGN regions could be heavily obscured by dust making them invisible in the optical. Alternatively some models 
suggest \cite{ta} that an AGN could form in a galaxy merger via the collapse of a star cluster containing compact supernova remnants, 
without the need for a `seed' black hole from the merger progenitors. Although we cannot use our data to establish accurate fractions 
of ULIRGs in pair mergers and multiple mergers, we can examine whether pair merging and multiple merging does in fact take place. For 
all the ULIRGs in class A the merger appears to be taking place between only two galaxies. This is most apparent in IRAS 06268+3509 and 
IRAS 18520-5048. Conversely, some ULIRGs in the sample (e.g. IRAS 13469+5833 and IRAS 18580+6527) show more than two large, well separated 
bright regions which cannot plausibly be interpreted as having arisen due to dust obscuration patterns. According to multiple galaxy 
merger models \cite{ta0} this can only arise in mergers between more than two progenitor galaxies. From this, we conclude that ULIRGs 
can be both pair mergers and multiple mergers, suggesting that ULIRGs 
can be formed both in compact groups and in isolated pairs.

\subsection{ULIRG Evolution}

\subsubsection{Merger, Starburst, \& ULIRG lifetimes}

A {\em prima facie} picture of how long each stage in a merger is ultraluminous in the IR can be gained from looking at the 
number of sources in the sample in each merger stage. In five of the sources the galaxies are still physically separate, or 
they are interacting but still physically distinguishable as separate objects (class A). In 14 sources (class B) separate 
galaxies are not distinguishable and there are clear signs of ongoing interaction, such as tidal tails, loops, and multiple 
nuclei. In the remaining 4 sources an optical QSO is visible. It is therefore plausiblethat physical mergers are the most 
prevalent ULIRG phase, although the optical QSO phase and the IR luminous stage before the galaxies are physically merging 
in ULIRGs are both significant.

Using N-body simulations to assess galaxy-galaxy interactions and mergers has proven to be an invaluable tool \cite{bar2}. 
The course of a merger event is affected by many variables, including (but not limited to), angle of approach, relative 
velocity, relative disk inclinations, and gas masses \cite{bar3}. Hence a complete discussion will not be presented here. 
We instead refer the reader to Barnes \& Hernquist \shortcite{bar2}. As an example we describe the N-body simulations 
of Dubinski, Mihos \& Hernquist \shortcite{dub}, specifically their models A and D. Model A considers a direct coplanar 
collision between two equal mass galaxies with very low mass dark matter haloes. After the first close aproach at $t = 0$ 
the galaxies reach a maximum separation of $\sim 50$kpc at $t = 2.5\times10^{8}$ years. After the second close approach 
at $t \sim 6\times10^{8}$ years the merger proceeds rapidly and is completed by $t \sim 9\times10^{8}$ years. The same 
model but with high mass dark matter haloes (model D) yields a similar pattern of evolution but with a longer merger time 
of $t \sim 2\times10^{9}$ years.
 
Overall, N-body simulations of galaxy mergers predict merger times (from first close approach to coalescence) of approximately 
$10^{9}$ years. The lifetime of a starburst event has been estimated by several authors. Thornley et al \shortcite{thor} 
find that starburst events are relatively short lived, with a lifetime range of $10^{6} - 10^{7}$ years, and that galactic 
superwinds produced by supernovae may be responsible for the short duration. Genzel et al \shortcite{gen} derive a starburst 
age range of $10^{7} - 10^{8}$ years. Evolutionary models applied to interacting and merging galaxies \cite{bern} give a 
starburst lifetime of $2\times10^{7}$ years, but with gaps between starburst events of up to $2\times10^{8}$ years. The 
lifetime of a ULIRG can be estimated indirectly in two ways, both independent of the power source behind the IR emission. 
The space density of ULIRGs is approximately the same as that of optically selected QSOs at the same bolometric luminosity 
in the local Universe \cite{san3}. If it is assumed that both ULIRGs and QSOs are an event in a host galaxy largely 
independent of environment then the range in ULIRG lifetimes should be approximately the same as that of QSO lifetimes. 
Estimating the lifetimes of QSOs is difficult. Recent results \cite{mar} give a lifetime range of $10^{7} - 10^{8}$ years. 
Alternatively the maximum ULIRG lifetime can be estimated by comparing the morphologies in our sample to the N-body 
simulations discussed above. The maximum observed separation in the sample is 30kpc. Assuming that these galaxies are 
receding from each other after their first close approach gives a time remaining to merger of  $1.6\times10^{9}$ years 
(model D) or $8.1\times10^{8}$ years (model A). Overall therefore we derive a lifetime range for a ULIRG of approximately 
$1\times10^{8} - 1\times10^{9}$years. This range is in broad agreement with previous estimates \cite{mur}, which quote a 
range of $2\times10^{8} - 2\times10^{9}$ years but argue that the real lifetime probably lies towards the lower end of 
this range.

Clearly there is some discrepancy between these lifetimes. It seems unlikely that ultraluminous IR activity can persist 
for the entire duration of a merger event, and yet our sample show ULIRGs at all stages of a galaxy merger. Similarly 
it seems implausible that a single starburst event can supply the necessary IR luminosity for the entire lifetime of all 
ULIRGs. On the other hand an AGN cannot be responsible for the IR emission for the majority of the ULIRG phase as most 
ULIRGs are powered predominantly by starbursts \cite{gen,rig}. To achieve consistency multiple starburst events, similar 
to those predicted by numerical simulations of multiple galaxy mergers \cite{bek} or by previous observations \cite{mur2}, 
must therefore be invoked before the final merger. Alternatively the evolution of ULIRGs may be more complex than that 
proposed by Sanders et al, or follow a different pattern.

\subsubsection{Starburst \& QSO Triggering}

Starburst triggering in galaxy mergers has previously been examined using numerical simulations where star formation is 
assumed to follow a density dependent Schmidt law \cite{mih,mih2}. In a merger between two bulgeless disk/halo galaxies 
it was found that at first perigalacticon bars were formed in each galaxy due to tidal forces, causing a rapid inflow of 
gas into the central regions. This central gas concentration triggered powerful starbursts as the galaxies reached their 
maximum separation. This starburst lasts for approximately $10^{8}$ years, dying out well before the galaxies finally 
merge at $\sim10^{9}$ years. The ensuing starburst during the merger is very weak. In a merger between two disk galaxies 
with a significant bulge component the bulges stabilize the galaxies against gas inflow, hence no bars are formed. The 
resulting starbursts triggered whilst the galaxies are still separate are therefore much less intense than those in 
bulgeless disks. When the galaxies merge however very large quantities of gas are driven to the central regions of the 
merger, triggering an extremely intense starburst lasting $\sim5\times10^{7}$ years. A few Gyr after this starburst has 
ended, the merger remnant strongly resembles a peculiar elliptical galaxy. 

Remarkably, these results depend less on orbital dynamics, such as approach angle and relative disk orientation, than 
progenitor galaxy morphology. The role of orbital dynamics in triggering starbursts is unclear, although the relative 
orientations of the galaxy disks and the pericentric separation after the first close approach appear important. Direct 
encounters produce rapid inflows whereas highly inclined passages do not, implying that strong axisymmetric forces are 
needed for prompt inflows \cite{bar3}. The predicted star formation rates also imply that an AGN is not needed to explain 
the IR emission in ULIRGs. Based on numerical simulations and the then available optical imaging of ULIRGs, Mihos \& Hernquist 
\shortcite{mih2} concluded that most ULIRGs are  mergers between disk/bulge/halo systems, with the few ULIRGs composed of 
widely separated galaxies being mergers between gas rich disk/halo galaxies.

Observations suggest that QSOs contain black holes with mass $>10^{8}M_{\odot}$ \cite{ree}. Spiral galaxies are however 
thought to contain black holes with mass $\leq10^{7}M_{\odot}$ \cite{sal}. Although the masses of black holes in ellipticals 
cover the range $10^{7}M_{\odot} < M_{BH} < 10^{10}M_{\odot}$ \cite{mag}, the quantities of gas and dust in such systems 
are too small to trigger ultraluminous IR activity in a merger. Hence a $\geq10^{8}M_{\odot}$ black hole must be formed in 
a galaxy merger if ULIRGs are the transition stage between mergers and QSOs. The formation of quasar nuclei in ULIRGs has 
been investigated by Taniguchi Ikeuchi \& Shioya \shortcite{ta}. They conclude that a supermassive black hole (SMBH) of 
mass $\geq10^{8}M_{\sun}$ can form if either one of the progenitor galaxies contains a `seed' SMBH of mass $\geq10^{7}M_{\sun}$ 
undergoing efficient Bondi type gas accretion over $\sim10^{8}$ years.

\subsubsection{A New Picture of ULIRG Evolution}

It is apparent from this and previous work that there exist serious problems with the S88 picture. Previous results have 
highlighted the ambiguity of whether AGN fraction increases with advancing merger stage. Our results are not consistent with 
ULIRGs being a straightforward transition stage between galaxy mergers and QSOs. Although we find 2/3 warm ULIRGs contain QSOs, 
our results are not consistent with warm ULIRGs being a transition stage between cool ULIRGs and QSOs. 

Our results in conjunction with previous IR spectroscopy and N-body simulations \cite{bar2,mih,mih2,rig} lead us to suggest an 
alternative to the S88 picture. We propose that ULIRGs as a class do not represent a simple transition stage between galaxy 
mergers and QSOs, and that there exists no sequence from `cool' ULIRG to `warm' ULIRG to QSO. Instead we propose that ULIRGs are 
a much more diverse class of object where the time evolution in IR power source and merger morphology is driven {\em solely} by the 
morphologies of the merger progenitors and the local environment. As has been noted by previous authors the presence and size 
of a bulge component in merger progenitors has a fundamental effect on the timing of gas inflow to the galaxy nucleus. Galaxies 
with small or non-existent bulge components would starburst early in the merger, galaxies with a large bulge component would 
starburst late in the merger. It is apparent from our imaging that ultraluminous activity is not mostly confined to late stage 
mergers and that there exist significant numbers of ULIRGs at all stages in galaxy mergers, including $\sim20$\% in systems where 
the galaxies are still separate. This implies that there is a substantial variation in the bulge component over all ULIRG progenitors. 
This morphological diversity amongst the merger progenitors will give rise to a continuum of starburst and AGN power sources as 
a function of merger stage across the ULIRG population. The plausible evidence presented here for diversity in environment and 
hence number of merger progenitors (e.g. pair mergers between field galaxies and multiple mergers between galaxies in compact 
groups) means that this continuum is more complex than may be expected from just pair mergers. It suggests that there may be a 
significant variation in the duration and luminosity of the ULIRG phase depending on the local environment. Pair mergers would 
be ultraluminous in the IR for a shorter period than multiple mergers and have a lower total IR luminosity on average, as there 
can be more starbursts in a multiple merger. It is worth noting that the combination of violent interaction coupled with large 
quantities of gas driven to the centres of mergers would undoubtedly increase the chances of forming an AGN and hence an optical QSO.
We argue however that the S88 picture represents only a subset of the total ULIRG population described by our model.

To illustrate our alternative model, we describe below three points along this continuum of IR power source as a function of merger 
stage. Here we only consider pair mergers, as currently no numerical simulations have been performed of multiple galaxy mergers 
(as may be found in compact galaxy groups rather than in the field) with varying bulge components.

The first point would be a merger between two bulgeless disk/halo galaxies. Starburst activity would be triggered early in the 
interaction while the galaxies are still separate, largely exhausting the gas supplies in both galaxies and leading to a physical merger 
that was not ultraluminous in the IR due to a starburst. The early starburst in the progenitors would make accretion commence early 
in the interaction, possibly leading to an AGN before the galaxies have started to physically merge. The merger would be IR luminous 
for the first $10^{8}$ years or so, followed by a quiescent period before (possibly) ending in an IR luminous AGN phase 
when the galaxies merge. The second point would be a merger between a disk/halo galaxy and a disk/bulge/halo galaxy. A starburst 
would be triggered in the bulgeless galaxy early in the interaction, and will dissipate before the galaxies start to physically merge. 
After about $5\times10^{8}$ years, as the galaxies began to physically merge, a starburst would be triggered in the disk/bulge/halo 
galaxy which had been stabilised against a starburst in the earlier stages of the merger. AGN activity could be triggered both early and 
late in a merger between two such galaxies. The third point would be a merger between 
two disk/bulge/halo galaxies. There would be no early starburst and hence no ULIRG activity whilst the galaxies are still separate. 
Once the galaxies start to coalesce a strong starburst would be triggered. Accretion can only start when the galaxies have 
started to physically merge. IR luminous AGN activity would probably commence shortly before the end of the starburst phase.

This alternative model resolves many of the problems with the S88 picture highlighted by previous observations. Indeed several 
observations which directly contradict the S88 picture are in excellent agreement with our model. For example the study showing 
that the fraction of ULIRGs powered by AGN does not increase with advancing merger stage is in strong agreement with our model 
as our model predicts that no such trend should exist. Under our alternative model, starburst and AGN activity could be observed 
at any point during a merger, depending on the morphologies and local environment of the merger progenitors. As previously noted 
\cite{mih,mih2} this classification scheme would be largely independent of orbital dynamics.

Although it is not possible with imaging data alone to assign our sample to points along this continuum, we can give examples of 
typical morphologies at the points described above. A typical `point 1' object may resemble IRAS 02587-6336. Here the galaxies are 
still widely separated and hence early in the interaction. There is no QSO activity but the object is ultraluminous in the IR. A 
`point 2' object might resemble IRAS 20037-1547. A QSO has been formed in the bulgeless galaxy which formed an AGN early in the 
interaction due to the early starburst. The source IRAS 18580+6527 is a good example of a `point 3' ULIRG, where violent star 
formation has only commenced once the galaxies have started to merge.

An evolutionary track that lasts $10^{8} - 10^{9}$ years (such as the S88 model or our alternative model) is also testable using 
other quantities that vary on similar timescales, or properties that should remain invariant. Observational tests include galaxy 
morphologies, stellar population mixes and Mpc scale environments. In particular, the QSOs in our sample showed negligible evidence 
for preferred timing late in the interaction though the number of QSOs in our sample is small. Increasing the number of ULIRG QSOs
selected purely on the basis of their FIR emission and observed with very high resolution imaging would provide better constraints 
on QSO timing in ULIRGs. Possible theoretical tests include more comprehensive N-body simulations of both pair and multiple galaxy 
mergers, incorporating galaxy morphologies, starburst triggering and QSO formation.

\section{Conclusions}
We have presented a sample of 23 ULIRGs, taken with WFPC2 using the F606W band filter on HST, and compared this to archival 
HST F814W  band data for the same objects. Our conclusions are:

1) There are no isolated normal spirals or ellipticals in the sample; the sources are either galaxy pairs, merging or interacting 
systems, or QSOs. The fraction of observed interacting sources in the sample is 87\%. As a whole the sample is entirely consistent
with ULIRGs being strongly linked with both galaxy interactions and (to a smaller extent) with QSO activity. In three cases, the 
{\em HST} images showed clear signs of interaction which had only been ambiguously detected from the ground. Although the majority 
of the sample are physically merging systems a significant fraction ($\sim40\%$) exist in binary systems or QSOs.

2) Stellar population synthesis models indicate that in most cases the optical emission is dominated by stellar populations with 
ages of several Gyr or less. Although uncertainties due to unconstrained mass normalizations and dereddening vectors are significant, 
it seems likely that any optical starburst triggered by interactions does not contribute significantly to the optical emission. The 
average absolute magnitude for those sources that do not display a QSO is 0.3 magnitudes less bright than the field galaxy magnitude 
in the Cousins {\em I} band.

3) Most of the merging systems in the sample show a number of compact bright `knots' whose colour differs substantially from 
the surrounding regions. Colour maps for nine of the objects show a non-uniform colour structure in all but one case. Observed 
features include blue regions located towards the centres of merging systems, and compact red regions. Although difficult to 
interpret with only imaging data the blue regions are likely to be regions of enhanced star formation and some of the red regions dust 
shrouded starbursts or AGN. 

4) There is no apparent correlation between the IR luminosity and either the optical emission or the merger phase for non-QSO ULIRGs.
Those ULIRGs that contain a QSO are on average no more symmetrical than those that do not, implying that AGN activity does not tend to 
reside in more advanced mergers. The relatively small numbers of ULIRGs with a QSO suggest that this phase is short, and although this 
may be biased by the sample's pre-selected luminosity distribution, the higher IR luminosity QSOs are if anything over-represented compared 
to their relative comoving space density. The relatively small number of physically separate objects also suggests that this phase is 
short and that ULIRGs are dominated by physically merging systems. The QSO host galaxies were found to be either interacting systems or 
ellipticals. Comparisons between our morphologies and previous merger models suggest that ULIRGs can be both pair mergers between field 
galaxies and multiple mergers between galaxies in compact groups.

5) Based on the morphologies of this sample, previous IR spectroscopy, and the results of N-body simulations we propose a new model 
for ULIRG evolution. We propose that ULIRGs are not a simple transition stage between galaxy mergers and QSOs and there exists no 
sequence from `cool' ULIRG to `warm' ULIRG to QSO. Under our model, ULIRGs are an extremely diverse class of object where the time 
evolution of starburst and AGN activity is driven solely by the morphologies of the merger progenitors and the local environment. 
This model suggests that the diversity in ULIRG morphologies and the lack of preferred timing of AGN activity in more advanced 
mergers is due to a continuum of merger progenitor morphologies with varying bulge components. The local environment (field or 
compact group) affects the number of merger progenitors and hence the frequency and luminosity of starburst events. Although the 
combination of violent merging and large central gas concentrations would lead to an increased likelihood of AGN activity and as 
such some ULIRGs would represent such a transition stage, this is only a subset of the ULIRG population as a whole, and there exists 
no automatic evolution to an optical QSO stage after the ultraluminous IR stage.

\section{Acknowledgments}

We would like to thank Matthew McMaster, Senior Data Analyst at the STScI, for invaluable advice on 
creating colour maps, and Leon Lucy for help with the deconvolution algorithms. We would also like to 
thank the referee for his/her very useful comments.

The data presented here were obtained using the NASA/ESA {\em Hubble Space Telescope}, obtained at the 
Space Telescope Science Institute, which is operated by the Association of Universities for Research in 
Astronomy, Inc., under NASA contract NAS 5-2655. The work presented has made use of the NASA/IPAC 
Extragalactic Database (NED), which is operated by the Jet Propulsion Laboratory under contract with 
NASA, and the Digitized Sky Surveys, which were produced at the Space Telescope Science Institute under 
U.S. Government grant NAG W-2166. The images of these surveys are based on photographic data obtained 
using the Oschin Schmidt Telescope on Palomar Mountain and the UK Schmidt Telescope. D.G.F would like 
to acknowledge the award of tuition fees and maintenance grant provided by the Particle Physics and 
Astronomy Research Council. This work was in part supported by PPARC (grant number GR/K98728).

\bsp % ``This paper has been produced using the ...''

\label{lastpage}

\end{document}